\documentclass[a4paper,11pt]{article} \usepackage{jheppub} \usepackage{amsthm,bm,booktabs} 
 \usepackage{mathrsfs}
 \usepackage{xcolor} 
 \usepackage{tikz} 
 \usetikzlibrary{arrows.meta,positioning} 
 \numberwithin{equation}{section} 
 \newtheorem{proposition}{Proposition}[section] 
  \newcommand{\LapH}{\nabla^2_H}
 
 \newcommand{\db}{{\bf d}} 

 \newcommand{\dd}{\mathrm d}

 \newcommand{\Ncal}{\mathcal N}
\newcommand{\PV}{\mathcal P}
 \title{ Spectral recurrences and maximal analyticity: cosmological correlators at one and two loops } 
 \author[1,2]{Bruce A. Bassett}
\author[3,4]{Sergio L. Cacciatori}
\author[3,4]{Ugo Moschella}
\affiliation[1]{Wits MIND Institute and School of Computer Science and Applied Mathematics, University of the Witwatersrand, Johannesburg, South Africa}
\affiliation[2]{School for Data Science and Computational Thinking, University of Stellenbosch, South Africa}
\affiliation[3]{DiSAT, Universit\`a dell'Insubria, Via Valleggio 11, Como}
\affiliation[4]{INFN, Sez. di Milano, Via Celoria 16, 20133 Milano}
\emailAdd{bruce.bassett1@wits.ac.za}
\emailAdd{sergio.cacciatori@uninsubria.it}
\emailAdd{ugo.moschella@uninsubria.it}

\abstract{ We develop a  method for computing de Sitter boundary correlators based on the maximal analyticity of the two-point function and differential--difference equations satisfied by Schwinger--Keldysh diagrams. The vertex integrations are analytically continued through the relevant de Sitter tuboids onto purely imaginary Lobachevskij cycles. Taking the Mehler--Fock transform of the invariant bubble kernels on these Riemannian cycles converts their differential--difference equations into algebraic recurrences in the spectral parameter. The mixed-branch kernel obeys a homogeneous recurrence, whereas the chronological and antichronological kernels satisfy sourced recurrences whose local terms are determined by dimensionally continued tadpoles. We solve these recurrences, constructing explicit particular solutions in terms of generalised hypergeometric functions and identifying the analytic conditions selecting their physical completion. Attachment to the external boundary points then yields an intrinsic spectral representation and a complete solution of the boundary in--in diagrams considered here. Our method is fully coordinate-independent. Specialising to a Poincar\'e chart readily gives expressions that can be compared directly with those obtained by the standard coordinate-dependent approach. Composing two or more bubbles yields a multi-loop chain whose spectral multiplier is determined by the one-loop building blocks and the sum over intermediate Schwinger--Keldysh branches. }
 
\begin{document} \maketitle \flushbottom 

 \section{Introduction} 
 
 The observables of de Sitter cosmology are correlation functions on the future boundary, obtained as late-time limits of bulk correlators by pushing each 
 external point to future infinity along a null direction. They are the de Sitter counterparts of the boundary correlators of AdS holography and are sometimes 
 called celestial correlators of de Sitter space.\footnote{They should not be confused with the celestial amplitudes of flat-space scattering, which live on the 
 celestial sphere of null infinity; for a dictionary between the two, see ref.~\cite{FOSW2025}.}
 
At tree level, these correlators are well understood. Beyond tree level, massive loops remain difficult: the in-in formalism requires a sum over forward and 
backward time branches; the time integrals are oscillatory; ultraviolet divergences must be removed without compromising the boundary limit; and that limit 
cannot naively be interchanged with the loop integrations. 
  
In this paper, we develop a method for computing perturbative de Sitter boundary correlators, applicable in principle at arbitrary loop order. The method is 
based on finite-difference equations satisfied by Schwinger--Keldysh diagrams in the spectral labels of their internal propagators, following our recent 
work~\cite{BCM}. We apply it to the massive one-loop bubble and the two-loop bubble chain. The construction rests on four ingredients, which together address the 
difficulties outlined above.

\paragraph{1. The Bunch--Davies state as an analytic object.}

Complex de Sitter spacetime is essential to our construction, just as complex Minkowski spacetime is essential to standard
Quantum Field Theory, where the spectral condition underlies the analyticity properties of correlation functions~\cite{SW}.
These properties, in turn, underpin fundamental structures of QFT, including its Euclidean formulation, and
provide an essential framework for perturbative renormalization \cite{eg}.

The corresponding analyticity properties in complex  de Sitter spacetime provide our starting point. The primitive domain of analyticity of a two-point function $W(Z,Z')$ satisfying the de Sitter spectral condition \cite{bm} is $\mathcal T_-\times\mathcal T_+$, where $\mathcal T_-$ and $\mathcal T_+$ are the past and future tuboids of complex  de Sitter space~\cite{bm}. For a de Sitter-invariant theory, $W(Z,Z')$ admits an analytic continuation represented by a function $w(z)$ of the single invariant variable $z=Z\cdot Z'$, holomorphic throughout the causal cut plane. 

For the free scalar field, the Bunch--Davies state~\cite{BD}, also called the Euclidean state~\cite{GH}, is characterized by the maximal analyticity of its two-point function~\cite{bgm,bm}. Its boundary values yield all the Schwinger--Keldysh propagators. 

The de Sitter plane waves\footnote{Null vectors $\xi$ of the light cone of the ambient space (the boundary) are often denoted by $P$ in the contemporary literature. We retain $\xi$, following the notation of Gel'fand, Graev and Vilenkin~\cite{GGV5}.} $(Z\cdot\xi)^{-\Delta}$, also introduced in ref.~\cite{bm}, are holomorphic in these tuboids. The maximally analytic free two-point function is a superposition of products of these plane waves, and the external lines of the in-in diagrams are themselves plane waves.

We work directly with perturbative boundary diagrams whose external lines are the plane waves corresponding to the
prescribed asymptotic components of the fields. The late-time limit is thus taken at the level of the external propagators,
before performing the vertex integrations. The identification of these diagrams with the asymptotic coefficients of fully
integrated bulk correlators remains a separate question.

Our construction therefore uses only the maximally analytic two-point function, its boundary values, and the de Sitter plane waves, together with the
time ordering of the in-in contour and explicitly specified integration regions (Section~\ref{sec:definition}).
Neither a coordinate system nor mode functions are needed.
Coordinates are used to describe the integration region and contour
deformation; mode functions are recovered by Fourier transforming the plane waves.
  
\paragraph{2. Integration on the inner Lobachevskij manifolds of the tuboids.} The vertex integrations of the in-in boundary diagrams are deformed, branch by branch, through the tuboids $\mathcal T_+$ and $\mathcal T_-$, up to their inner Lobachevskij manifolds $\pm iH^d$. These are invariant submanifolds of complex  de Sitter spacetime, commonly referred to as Euclidean anti-de Sitter space. 

Propagators joining opposite cycles and propagators joining two points of the same cycle obey different analytic prescriptions. The former are Wightman functions evaluated within their domain of holomorphy, whereas the latter are chronological or antichronological boundary values obtained from opposite sides of the causal cut. The plane waves connecting the internal vertices to the external boundary points remain holomorphic in the corresponding tuboids throughout the deformation. 

The construction therefore remains entirely within complex  de Sitter spacetime. The anti-de Sitter manifold never enters: the cycles $\pm iH^d$ are invariant submanifolds of complex  de Sitter spacetime, rather than anti-de Sitter space obtained by analytic continuation of the de Sitter radius. Although these cycles are geometrically isometric to Euclidean anti-de Sitter space, the functions evaluated on them are values and boundary values of the de Sitter two-point function, not individual anti-de Sitter propagators (Section~\ref{sec:analytic}). 

The relation between de Sitter in-in diagrams and Euclidean anti-de Sitter diagrams was developed systematically by Sleight and Taronna in ref.~\cite{SleightTaronna2021}, building on the Euclidean formulation of de Sitter perturbation theory discussed in ref.~\cite{HMM2011}. Using analytic continuation and Mellin--Barnes representations, they derived general rules expressing de Sitter boundary diagrams as specific linear combinations of EAdS Witten diagrams, with coefficients fixed by the Schwinger--Keldysh prescriptions and consistent factorisation. Their construction provides a general perturbative framework for relating de Sitter boundary correlators to Euclidean anti-de Sitter diagrams. 

Our approach uses the same underlying analytic continuation but retains the maximally analytic de Sitter propagators throughout, without decomposing them into separate EAdS Green functions. Such a decomposition follows from standard connection identities for the Legendre functions. Although useful in other contexts, it introduces additional singularities and discontinuities in the individual terms, which cancel only in the complete de Sitter combination. Working directly with $w_\nu$ preserves its analytic structure throughout the calculation and is particularly suited to the spectral recurrences developed below. 

We also prove the contour deformation. The two vertices are rotated through a common angle,
and each invariant stays on the prescribed side of the causal cut throughout the
deformation, so that each propagator keeps its boundary value. The chronological and
antichronological boundary values are thereby identified separately, without requiring
cancellations between different Schwinger--Keldysh branch assignments.

The contour identity is proved at integer dimension, with an auxiliary analytic regulator
that controls the coincidence singularity: it is first established where all the integrals
converge, and then continued meromorphically in the regulator. A continuation of the
spatial integrations to non-integer dimension is not needed. In four dimensions, the result
is related to the dimensional subtraction used in the rest of the paper by an explicit
finite local term. The argument applies to perturbative boundary diagrams for which the
prescribed asymptotic components of the external fields have already been selected. It is
given here for the bubble; we expect it to extend to higher-loop diagrams, subject to the
corresponding convergence and local-extension conditions.
  
 \paragraph{3. Loops as finite spectral recurrences.} Our method computes loop integrals
by solving finite-difference equations in the spectral labels of their internal lines,
following ref.~\cite{BCM}. The equations follow from local identities, the contiguity
relations of the propagators and Green's identity, and apply directly to the loop
kernels. On the inner Lobachevskij manifolds of the tuboids we derive them for the
complete one-loop bubble. The product of two Wightman functions satisfies a homogeneous
recurrence, whereas the chronological and antichronological products satisfy recurrences
with local sources, determined by the dimensionally continued tadpoles.

On the Lobachevskij manifold the bubble kernels depend only on the geodesic distance
between the two vertices. Following ref.~\cite{BEGMP2010}, we analyse such functions by
the Mehler--Fock transform, the analogue of the Fourier transform for functions of the
distance on hyperbolic space: it expands them in eigenfunctions of the invariant
Laplacian, labelled by a spectral parameter. The transform of a kernel is its spectral
multiplier, the number by which the kernel, acting by convolution, multiplies each
eigenfunction of the Laplacian; the inverse transform reconstructs the kernel. Since the
recurrences involve the Laplacian only through polynomial coefficients, they become
algebraic finite-difference equations for the spectral multipliers.

For the Wightman product, the recurrence and the analyticity of the propagators determine
the multiplier completely. The result is the de Sitter K\"all\'en--Lehmann formula of
ref.~\cite{BEGMP2010}, which we thus rederive by a different method. For the
chronological and antichronological products we solve the sourced recurrences
explicitly: the multiplier combines ordinary generalized hypergeometric functions with
the K\"all\'en--Lehmann multiplier, and we prove that the analytic and asymptotic conditions uniquely
identify the physical solution.

The two-loop bubble chain is the composition of two one-loop bubbles. Its multiplier is
therefore the product of two one-loop multipliers, summed over the Schwinger--Keldysh
branch of the intermediate vertex, and no new two-loop function appears
(Section~\ref{sec:twoloop}). Finally, the recurrences can be transported to the
boundary, where they become finite-difference equations for the boundary correlators
themselves, under the conditions stated in Section~\ref{sec:attachment}.

\paragraph{4. Boundary correlators and renormalization.} The external lines are the
asymptotic plane waves from the outset, and the boundary correlator is obtained by
integrating the two vertices of the bubble kernel over the Lobachevskij manifold, each
against two external plane waves. The Mehler--Fock representation writes the kernel as a
superposition of analytic two-point functions of the principal series. The plane-wave
representation of the analytic two-point function \cite{bgm,bm}, with its prescribed tuboid branches and continuation to the Lobachevskij cycles, expresses each
of them as an integral over the future cone of a product of two plane waves, and so
separates the two vertices. Each vertex integral contains three plane waves, two external
and one internal, and is a triangular invariant in the sense of ref.~\cite{BEGMP2010},
given in closed form by the star--triangle formula. What remains is an integral over the
future cone, which by homogeneity does not depend on the section chosen to perform it, so
that de Sitter covariance is manifest throughout. The final Fourier representation uses a
flat section and spatial momentum variables, to make contact with the more customary
momentum-space methods; there each triangular invariant becomes a triple-$K$ integral.

At one loop, renormalization removes a local contact term. We subtract it with its full
dimension dependence, and the subtracted correlator is then regular at $d=4$, where it is
evaluated. The dimension dependence of the branch phases and of the external plane waves
must be kept, because an evanescent term multiplying an ultraviolet pole leaves a finite
remainder: subtracting the complete contact term and subtracting only its pole differ by
a finite local term. For the principal-series example considered here, the spectral
representation of the subtracted correlator converges.

\paragraph{Summary of results.} The vertices of the in-in diagrams are integrated over the
expanding Poincar\'e patch. We first show that these integrations can be deformed onto the
Lobachevskij cycles of the complex  de Sitter manifold, which are geometrically copies
of Euclidean anti-de Sitter space. The bubble contour identity is proved at fixed integer dimension
with an auxiliary analytic regulator. At $d=4$ its finite part is
matched to dimensional subtraction by an explicit local term. On the cycles we write the
difference--differential equations satisfied by the one-loop bubble kernels for each
Schwinger--Keldysh branch, and solve them. For the Wightman product the solution
reproduces the de Sitter K\"all\'en--Lehmann formula. For the chronological and
antichronological products we obtain an explicit solution in terms of ordinary
generalized hypergeometric functions, and prove its identification with the Mehler--Fock transforms of these products. Inverting the
Mehler--Fock transform gives the bubble kernel for each branch. The boundary correlator
is then expressed as a spectral integral with an attachment evaluated through triangular invariants, and equivalently in Fourier
space, and its subtracted version is evaluated at $d=4$. Finally, the two-loop bubble
chain is expressed through the one-loop multipliers, with its branch dependence fixed by
the Schwinger--Keldysh sum.

 \paragraph{Relation to other work.} The analyticity of quantum fields in complex de Sitter space and the characterization of the Bunch--Davies state were developed in refs.~\cite{bgm,bm}. The original de Sitter K\"all\'en--Lehmann formula and the method of triangular invariants were obtained in ref.~\cite{BEGMP2010}. The relation between in-in perturbation theory and Euclidean constructions was investigated in refs.~\cite{HMM2011,SleightTaronna2021}. More recently, the reconstruction of cosmological correlators from the Euclidean sphere has been studied in ref.~\cite{KatsevichSun2026}, while higher-loop vacuum integrals on the sphere have been investigated in ref.~\cite{LochabDenef2026}. These developments illustrate the usefulness of Euclidean methods for de Sitter perturbation theory. Our approach retains the maximally analytic de Sitter propagators on imaginary Lobachevskij cycles, rather than resolving them into individual EAdS propagators. The spectral recurrence method follows ref.~\cite{BCM}. Its application to the Schwinger--Keldysh bubble makes explicit the distinction between the homogeneous recurrence of the Wightman product and the local sources of the chronological products. Dimensional regularization and the treatment of local terms have also been investigated in refs.~\cite{NowinskiSachs2025,Zhang2026}. Complementary one-loop analyses based on the Bunch--Davies wave function and its differential equations have been developed in ref.~\cite{Benincasa:2024ptf}. Recent spectral and split-representation approaches to massive de Sitter loops and bubble chains provide complementary formulations \cite{DiPietroGorbenkoKomatsu2022, XianyuZhang2023,GraefeSachs2026}. In Appendix~\ref{app:gs}, we examine the discrepancy between our results and those of ref.~\cite{GraefeSachs2026}. We trace it to the omission of a nonvanishing spectral delta-function contribution and an incorrect normalization of the equal-branch propagators. The discrepancy is already present at one loop, but a cancellation between Schwinger--Keldysh branches conceals it for conformally coupled external fields in four spacetime dimensions. Once these errors are corrected, the resulting spectral blocks reproduce our bubble-chain structure.
  
\section{Generalities}\label{sec:definition}

In this section and the next two we define the quantities computed in this paper
intrinsically, without coordinates or mode functions: through the geometry of the
complex  de Sitter manifold, the maximally analytic two-point function and the
Schwinger--Keldysh rules of the in-in formalism, with the integration regions stated
explicitly. Mode functions appear only at the very end, when the results are compared
with momentum-space methods. In Section~\ref{sec:independent-contour} we show that the
vertex integrations can be deformed onto the Lobachevskij cycles, where the quantities
are evaluated.

\subsection{Notation}

\begin{table}[h]
\centering\small
\begin{tabular}{ll}
\toprule
Object & Symbol\\
\midrule
spacetime dimension, and $h$
 & $d$, \ $h=\frac{d-1}2$\\
abbreviation used in appendix~\ref{app:physical-conditions}
 & $a=h-1$\\
real points of $dS_d$
 & $X$, $X'$\\
complex points of $dS^{(c)}_d$
 & $Z=X+iY$\\
points of $H^d$; points of the cycles $\pm iH^d$
 & $Y$; \ $Z=\pm iY$\\
boundary directions (future null cone)
 & $\xi$\\
de Sitter invariant
 & $z=Z\cdot Z'$\\
invariant of two points of $H^d$, and hyperbolic distance
 & $z=Y\cdot Y'=\cosh\rho$\\
conformal time, spatial coordinates, radial coordinate on $H^d$
 & $\eta$, $\bm x$, $y$\\
points of the flat section of the cone, and momenta
 & $\bm k$, \ $\bm p$\\
branch labels of vertices
 & $r,s,t=\pm$\\
spectral label and mass of a field
 & $\nu$, \ $m^2=h^2-\nu^2$\\
boundary dimensions (characteristic exponents)
 & $\Delta=h\pm\nu$\\
two-point function of the invariant, and its falloff coefficients
 & $w_\nu(z)$, \ $a_\Delta$\\
coefficients of the contiguity relation of $w_\nu$
 & $a_\sigma(\nu)=-\frac{\nu-\sigma a}{2\nu}$, \ $\sigma=\pm1$\\
the same for the Mehler--Fock kernel $w_q/w_q(1)$
 & $b_\sigma(q)=\frac{q+\sigma h}{2q}$\\
external lines (plane waves) on the two branches
 & $K^\Delta_\pm(X,\xi)$\\
Laplace--Beltrami operator of $H^d$, and $\mathcal Q$
 & $\nabla^2_H$, \ $\mathcal Q=h^2+\nabla^2_H$\\
branch kernels, and Schwinger--Keldysh kernel
 & $\mathcal K_r$, $r=0,\pm$; \ $\mathcal K$\\
Mehler--Fock label
 & $q=i\lambda$\\
Mehler--Fock transforms of $\mathcal K_r$ and $\mathcal K$
 & $F_r$, \ $F$\\
K\"all\'en--Lehmann function, particular solution, homogeneous coefficient
 & $M$, \ $P$, \ $C_u$\\
\bottomrule
\end{tabular}
\end{table}

 The de Sitter spacetime may be identified with the hyperboloid \begin{equation} dS_d=\{X\in\mathbb R^{1,d}:\ X\cdot X=-1\}, \end{equation} where the ambient Minkowski product has signature $(+,-,\dots,-)$ and we set the de Sitter radius to one. Its complexification is \begin{equation} dS_d^{(c)} =\{Z\in\mathbb C^{d+1}:\ Z\cdot Z=-1\}. \end{equation} Of fundamental importance are the future and past tuboids, which provide the natural domains of holomorphy of the de Sitter plane waves and the primitive domain of analyticity of the Bunch--Davies two-point function \cite{bm}. They are defined by \begin{equation} \mathcal T_\pm =\{Z=X+iY\in dS_d^{(c)}:\ Y\in\pm V_+\}, \end{equation} where $V_+$ is the open future light cone of the ambient Minkowski space.

The invariant scalar product of two complex points is $z(Z,Z')=Z\cdot Z'$. We introduce the complex distance $\rho$ through \begin{equation}\label{eq:complex-distance} Z\cdot Z'=-\cosh\rho. \end{equation} For real timelike separated points, $\rho=\tau\ge0$ is the proper time, and $z=-\cosh\tau\le-1$. For spacelike separated points joined by a geodesic of length $s$, one has $\rho=is$ and $z=-\cos s$, with $0<s\le\pi$. Spacelike separated points with $z>1$ are not joined by a geodesic; for them $\rho=\sigma+i\pi$, with $\sigma>0$ and $z=\cosh\sigma$.

\paragraph{The real integration region.}

The vertices of the in-in diagrams are integrated over the expanding upper 
Poincar\'e patch
\begin{equation}\label{eq:patch}
 dS_d^{up}=\{X\in dS_d:\ X^0+X^d>0\},
\end{equation}
with coordinates $\eta\in(-\infty,0)$ and
$\bm x\in\mathbb R^{d-1}$,
\begin{equation}
 X^0+X^d=-\frac1\eta,\qquad
 X^0-X^d=\eta-\frac{\bm x^2}\eta,\qquad
 X^i=-\frac{x^i}\eta,
\end{equation}
and invariant measure
\begin{equation}
 \dd\mu(X)=\frac{\dd\eta\,\dd^{d-1}x}{(-\eta)^d}. \label{meapp}
\end{equation}

\paragraph{The complex integration cycles.}

The complex integration cycles are usually referred to as Euclidean anti-de Sitter spaces. In fact, they are invariant submanifolds of the complex de Sitter manifold, as their very construction shows. They are the inner boundaries of the tuboids,
\begin{equation}\label{eq:cycles}
 \mathcal H_\pm=\pm iH^d\subset\mathcal T_\pm,\qquad
 H^d=\{Y:\ Y\cdot Y=1,\ Y^0>0\},
\end{equation}
each invariant under the real de Sitter group.

Here, the term ``inner boundary'' refers to the projection onto the imaginary parts. For every $Z=X+iY\in dS_d^{(c)}$, the condition $Z\cdot Z=-1$ implies \begin{equation} X\cdot Y=0,\qquad Y\cdot Y=1+X\cdot X. \end{equation} In the tuboids $\mathcal T_\pm$, the imaginary part $Y$ is timelike. Hence $X$ is spacelike or zero, and $0<Y\cdot Y\le1$, with the upper bound attained exactly when $X=0$. The hyperboloids $\pm H^d$ bound the corresponding projected regions at $Y\cdot Y=1$, and their lifts to complex  de Sitter space are the cycles $\mathcal H_\pm=\pm iH^d$.

However, these are not topological boundaries of the tuboids themselves.
At a point $iY\in\mathcal H_+$, $Y$ is strictly future
timelike, so all sufficiently small perturbations within
$dS_d^{(c)}$ remain in $\mathcal T_+$. Thus $\mathcal H_+$ is a submanifold lying inside the open tuboid.
The same description applies to $\mathcal H_-$ in $\mathcal T_-$.

In the coordinates of the expanding Poincar\'e patch, the cycles are obtained by giving
the conformal time an imaginary value, $\eta=\pm iy$ with $y\in(0,\infty)$, at fixed
spatial coordinates. This gives $Z=\pm iY$ with
\begin{equation}\label{eq:H-coords}
 Y^0+Y^d=\frac1y,\qquad
 Y^0-Y^d=y+\frac{\bm x^2}y,\qquad
 Y^i=\frac{x^i}y,\qquad
 \dd\mu_H(Y)=\frac{\dd y\,\dd^{d-1}x}{y^d}.
\end{equation}

For two points on the same cycle, $Z=\pm iY$ and
$Z'=\pm iY'$, with the same sign, let $\rho$ denote their
hyperbolic distance in the induced Riemannian metric. Then
\begin{equation}
 Z\cdot Z'=-Y\cdot Y'=-\cosh\rho .
\end{equation}
Their invariant scalar product lies on the causal cut,
and the corresponding propagator is a boundary value
$w_\nu(-\cosh\rho\pm i0)$, with the side fixed by the branch.

For points on opposite cycles, $Z=iY$ and $Z'=-iY'$,
one instead has
\begin{equation}
 Z\cdot Z'=Y\cdot Y'\ge1 .
\end{equation}
Their invariant scalar product lies outside the causal cut.
The Wightman function is therefore the regular holomorphic value
$w_\nu(Y\cdot Y')$, including when $Y=Y'$.

A free scalar field of mass $m$ satisfies the Klein--Gordon equation
$(\Box_{dS}+m^2)\varphi=0$. The d'Alembertian extends to a holomorphic operator on the
complexified manifold. On the cycles $\pm iH^d$ the induced metric is the Riemannian
metric of $H^d$, and the equation becomes $(\LapH+m^2)\,\varphi=0$, with
$\LapH=\nabla^a\nabla_a$ the Laplace--Beltrami operator of $H^d$. Reading this equation as
that of an anti-de Sitter field of squared mass $-m^2$ is an unnecessary and somewhat
misleading reinterpretation that we avoid.

\paragraph{The boundary.}

Boundary points are directions $\xi$ of the future null cone
$C^+=\{\xi:\ \xi\cdot\xi=0,\ \xi^0>0\}$.
On its flat section
$\xi(\bm k)=(\frac12(1+\bm k^2),\bm k,\frac12(1-\bm k^2))$, 
\begin{equation} \xi_i=\xi(\bm k_i), \qquad k_{ij}^2=|\bm k_i-\bm k_j|^2.
\label{213}\end{equation}
\begin{equation}\label{eq:Xxi} X\cdot\xi = \frac{\eta^2-(\bm x-\bm k)^2}{2\eta}, \qquad Y\cdot\xi = \frac{y^2+(\bm x-\bm k)^2}{2y}. \end{equation}
and $Z\cdot\xi=\pm i\,Y\cdot\xi$ at $Z=\pm iY$.

\section{The two-point function and its analytic structure}\label{sec:analytic}

Our construction is based on the analytic structure of the so-called Bunch--Davies
two-point function \cite{bgm,bm}. This structure is naturally encoded, in a covariant way,
in the de Sitter plane waves and their holomorphy in the tuboids.

\paragraph{Plane waves and the tuboids.} For a direction $\xi$ on the future null cone and
a complex point $Z=X+iY$, the de Sitter plane waves are
\begin{equation}\label{eq:planewave}
 \psi_\Delta(Z,\xi)=(Z\cdot\xi)^{-\Delta},\qquad \Delta\in\mathbb C.
\end{equation}
Since ${\Im}(Z\cdot\xi)=Y\cdot\xi$, the tuboids are characterised by the sign of this
imaginary part:
\begin{equation}\label{eq:tuboid-planewave}
 \mathcal T_\pm=\left\{Z\in dS_d^{(c)}:\ \pm{\Im}(Z\cdot\xi)>0
 \quad\text{for every future null }\xi\ne0\right\}.
\end{equation}
In these domains the plane waves are holomorphic, the branch of the complex power being
fixed by the half-plane in which $Z\cdot\xi$ lies. The tuboids thus play for de Sitter plane
waves the role that the tubes $\mathbb R^d\pm iV_+$ play for the exponentials $e^{ik\cdot z}$
in Minkowski space.

Each plane wave solves the Klein--Gordon equation $(\Box+m^2)\psi=0$ with
\begin{equation}
 m^2=\Delta(d-1-\Delta);
\end{equation}
writing $\Delta=h\pm\nu$, this gives $m^2=h^2-\nu^2$. Here $\nu$ is an arbitrary complex number. On
the real manifold the plane waves are the boundary values $(X\cdot\xi\pm i0)^{-\Delta}$, which
coincide where $X\cdot\xi>0$ and differ by the phases $e^{\mp i\pi\Delta}$ where $X\cdot\xi<0$.

The two-point function is a superposition of products of plane waves \cite{bm}:
\begin{equation}\label{eq:W-planewaves}
 \begin{aligned}
 W_\nu(Z,Z')&=c_\nu\int_\gamma(Z\cdot\xi)^{-h-\nu}\,(\xi\cdot Z')^{-h+\nu}\,\dd\mu_\gamma(\xi),\\
 &\hspace{2cm} Z\in\mathcal T_-,\qquad Z'\in\mathcal T_+,\qquad m^2=h^2-\nu^2,
 \end{aligned}
\end{equation}
whose two exponents are a dimension $\Delta=h+\nu$ and its shadow $d-1-\Delta=h-\nu$. 
The normalization is fixed by the CCR's:
\begin{equation}\label{eq:cnu-four-tuboids} c_\nu= \frac{\Gamma(h+\nu)\Gamma(h-\nu)} {2^{d+1}\pi^d}\,e^{-i\pi\nu}. 
\end{equation}  Here $\gamma$ is
any section of the future cone that meets (almost) every ray exactly once, with its
induced measure, and $c_\nu$ is fixed by the canonical commutation relations. Since the
integrand is homogeneous of degree $-2h=-(d-1)$ in $\xi$, the integral does not depend on
the choice of $\gamma$; and since Lorentz transformations map sections of the cone into
sections, the expression is de Sitter invariant.
\paragraph{The maximally analytic de Sitter two-point function as a Legendre function.}
 As a function of the invariant $z$ the two-point
function~\eqref{eq:W-planewaves}, is
\begin{equation}\label{eq:w-def}
 w_\nu(z) = \frac{\Gamma(h+\nu)\Gamma(h-\nu)}{(4\pi)^{d/2}\Gamma(\frac d2)}\,
 {}_2F_1\Big(h+\nu,h-\nu;\frac d2;\frac{1-z}{2}\Big).
\end{equation}
It is even in $\nu$ and holomorphic in the plane cut along the causal cut $(-\infty,-1]$.
Actually, this hypergeometric function belongs to the Legendre class:
\begin{equation}\label{eq:w-legendre}
 w_\nu(z) = \frac{\Gamma(h+\nu)\Gamma(h-\nu)}{2(2\pi)^{d/2}}\,
 (z^2-1)^{-(d-2)/4}\,P_{\nu-\frac12}^{-\frac{d-2}{2}}(z),
\end{equation}
where $P_{\nu}^{\mu}$ is the associated Legendre function of the first kind \cite{Bateman1}.
Equation~\eqref{eq:w-legendre} holds as such for real $z>1$. At $z=1$ the zero of $P_{\nu-\frac12}^{-\frac{d-2}{2}}$ compensates the
singularity of the prefactor, and $w_\nu$ is regular. At exceptional parameters the branch singularity at
$z=-1$ may reduce to a pole.

The Legendre form carries information that the hypergeometric form hides. Functions of
the first and of the second kind solve the same equation, the Klein--Gordon equation for
invariant functions, but their analytic structures are different. 

The function of the
first kind, with its prefactor\footnote{Multiplied by the prefactor, the Legendre function becomes a Gegenbauer
function of index $h$:
$(z^2-1)^{-(d-2)/4}P_{\nu-\frac12}^{-\frac{d-2}{2}}(z)
=\frac{2^{-(d-2)/2}}{\Gamma(\frac d2)}\,C^{(h)}_{\nu-h}(z)/C^{(h)}_{\nu-h}(1)$.
For $\nu=i\lambda$ this is, up to normalisation, the spherical function of $H^d$, the
kernel of the Mehler--Fock transform of section~\ref{sec:recurrence}.}, is regular at $z=1$, which in de Sitter space corresponds
to antipodal points, and its singularities lie on the causal cut, beginning at the light
cone $z=-1$: this is exactly the analyticity required by the causal structure of de Sitter
space. 

For the functions of the second kind there is no such compensation: they are themselves
singular at $z=1$, where, with the prefactor, they have the Hadamard singularity
$(z-1)^{-(d-2)/2}$, and they are cut along $(-\infty,1]$. They are unsuitable in de Sitter
space, where they would be singular at antipodal points, but they are the natural
two-point functions of anti-de Sitter space, where $z=1$ is coincidence (see below).
Written only as a hypergeometric series centred at $z=1$, $w_\nu$ shows neither which
solution of the Legendre equation it is, nor its relation to the functions of the second
kind.

 The Wightman function is the boundary value of $w_\nu(Z\cdot Z')$ from $\mathcal T_-\times\mathcal T_+$. For timelike separated points, $z=X\cdot X'<-1$, and \begin{equation}\label{eq:side} W(X,X') = \begin{cases} w_\nu(z+i0), &\text{if $X$ is later than $X'$},\\ w_\nu(z-i0), &\text{if $X'$ is later than $X$}. \end{cases} \end{equation} For spacelike separated points, $z>-1$, and \begin{equation} W(X,X')=W(X',X)=w_\nu(z). \end{equation} Thus the commutator is the discontinuity of $w_\nu$ across the cut and vanishes at spacelike separation.

 Near coincidence, $z\to-1$, the two-point function has the universal Hadamard singularity, for $d>2$, \begin{equation}\label{eq:hadamard} w_\nu(z)\simeq \frac{\Gamma(\frac d2-1)} {4\pi^{d/2}[2(1+z)]^{(d-2)/2}}. \end{equation} The leading term is independent of the mass and determines the leading ultraviolet behaviour of the one-loop bubble. At the antipode, $z=1$, the function is regular, \begin{equation} w_\nu(1)= \frac{\Gamma(h+\nu)\Gamma(h-\nu)} {(4\pi)^{d/2}\Gamma(\frac d2)}. \end{equation}

 At large $|z|$,
\begin{equation}\label{eq:falloffs}
 w_\nu(z) \simeq a_{h-\nu}z^{-(h-\nu)}+a_{h+\nu}z^{-(h+\nu)}, \qquad
 a_{h+\sigma\nu} = \frac{\Gamma(h+\sigma\nu)\Gamma(-\sigma\nu)}{2^{h+\sigma\nu+2}\pi^{h+1}},
 \quad \sigma=\pm1,
\end{equation}
with the powers defined in the plane cut along $(-\infty,-1]$; for integer $\nu$ the two
terms combine into a logarithm, which does not occur for the labels used here. When one
argument tends to future infinity along a null direction $\xi$, each term produces a plane
wave $(Z\cdot\xi)^{-\Delta}$, with $\Delta=h\pm\nu$. These plane waves, with coefficients $a_\Delta$,
provide the external lines of the boundary correlators.

\subsection{De Sitter and anti-de Sitter two-point functions.}

As complex manifolds, the
Complexified de Sitter and anti-de Sitter spacetimes can be identified with the same complex quadric. The de Sitter cycles $\pm iH^d$ are geometrically copies of Lobachevskij space, as is Euclidean anti-de Sitter space~\cite{MoschellaAdS}.  As functions of the invariant variable, the de Sitter and anti-de Sitter two-point functions satisfy the same ordinary differential equation, namely the Legendre--Gegenbauer equation, but differ in their analytic properties and boundary conditions. For an anti-de Sitter field with parameter $\nu$, the two-point functions compatible with the causal structure of anti-de Sitter space~\cite{bem,MoschellaAdS} are represented, as functions of the complex AdS invariant variable $z$, by \begin{equation}\label{eq:wads-def} w^{\rm AdS}_{\pm\nu}(z) = \frac{e^{-i\pi(d-2)/2}}{(2\pi)^{d/2}}\, (z^2-1)^{-(d-2)/4} Q^{\frac{d-2}{2}}_{-\frac12\pm\nu}(z). \end{equation} On the Euclidean section, $z=Y\cdot Y'\geq1$.
 The relation between the two kinds of Legendre function is very well known 
\cite[Eq.~3.3.1(4)]{Bateman1},
\begin{equation}
 Q^\mu_{-\frac12-\nu}(z)=Q^\mu_{-\frac12+\nu}(z)
 +e^{i\pi\mu}\sin(\pi\nu)\,\Gamma\big(\tfrac12+\mu-\nu\big)\Gamma\big(\tfrac12+\mu+\nu\big)\,
 P^{-\mu}_{-\frac12+\nu}(z),
\end{equation}
with $\mu=(d-2)/2$; this  gives, for $z>1$,
\begin{align}
 w_\nu(z)&=\frac{w^{\rm AdS}_{-\nu}(z)-w^{\rm AdS}_\nu(z)}{2\sin(\pi\nu)},\label{eq:wds-wads}\\
 w_\nu(-z\pm i0)&=e^{\mp i\pi h}\,
 \frac{e^{\pm i\pi\nu}\,w^{\rm AdS}_{-\nu}(z)-e^{\mp i\pi\nu}\,w^{\rm AdS}_\nu(z)}{2\sin(\pi\nu)}.
 \label{eq:wds-boundary-ads}
\end{align}

Since the de Sitter invariant is $z$ between opposite cycles and $-z$ on the same cycle,
every propagator on the cycles is a combination of the two anti-de Sitter functions, but
none is an anti-de Sitter propagator. The decomposition introduces singularities that
$w_\nu$ does not have: at $z=1$, where points of opposite cycles are antipodal, and at integer
$\nu$, through the factor $1/\sin(\pi\nu)$. In particular, the mixed-branch kernel, regular at
coincidence and without source in the recurrence of Section~\ref{sec:recurrence}, would
become a combination of singular kernels whose contact terms cancel.

\section{Two-point functions in the four tuboid domains} \label{sametr}\label{sec:four-tuboids} The plane-wave representation of the Wightman function naturally leads to four invariant two-point kernels, corresponding to the four possible assignments of their arguments to the past and future tuboids. All four satisfy the homogeneous Klein--Gordon equation in each argument, without a delta-function source. They are therefore correlation kernels rather than propagators. The ordinary Wightman function is holomorphic in $\mathcal T_-\times\mathcal T_+$ and is given by eq.~\eqref{eq:W-planewaves}. 
The permuted Wightman function is a distinct two-point function given by \begin{equation}\label{eq:W-permuted} \begin{aligned} W_\nu^{+-}(Z,Z') &=W_\nu(Z',Z)\\ &=c_\nu\int_\gamma (Z\cdot\xi)^{-h+\nu} (Z'\cdot\xi)^{-h-\nu} \,\dd\mu_\gamma(\xi), \end{aligned} \end{equation} where $Z'\in\mathcal T_-$ and $Z\in\mathcal T_+$. The permuted Wightman function is holomorphic in $\mathcal T_+\times\mathcal T_-$. 
Its boundary values coincide with those of $W_\nu$ at spacelike separation, by local commutativity. The edge-of-the-wedge theorem then connects the two holomorphic functions by analytic continuation. They are therefore restrictions of the same maximally analytic function.

Two further invariant kernels arise when both arguments belong to the same tuboid. They are defined by \begin{align} W_\nu^{++}(Z,Z') &=e^{i\pi\nu}c_\nu \int_\gamma (Z\cdot\xi)^{-h-\nu} (Z'\cdot\xi)^{-h+\nu} \,\dd\mu_\gamma(\xi), \qquad Z,Z'\in\mathcal T_+, \label{eq:Wpp}\\ W_\nu^{--}(Z,Z') &=e^{i\pi\nu}c_\nu \int_\gamma (Z\cdot\xi)^{-h-\nu} (Z'\cdot\xi)^{-h+\nu} \,\dd\mu_\gamma(\xi), \qquad Z,Z'\in\mathcal T_-. \label{eq:Wmm} \end{align} 
Unlike the ordinary Wightman function, their normalization is not fixed by the canonical commutation relations. We choose it so that the also equal-tuboid kernels are even functions of the spectral parameter $\nu$. 

Their relation to the maximally analytic two-point function can be established by evaluating the integral over a spherical section of the future cone at purely imaginary events $Z=iY$ and $Z'=iY'$, with $Y,Y'\in H^d$. The phases of the two plane waves give \begin{align} W_\nu^{++}(iY,iY') &=e^{i\pi\nu}c_\nu e^{-i\pi h} \int_\gamma (Y\cdot\xi)^{-h-\nu} (Y'\cdot\xi)^{-h+\nu} \,\dd\mu_\gamma(\xi) \nonumber\\ &=e^{i\pi\nu}c_\nu e^{-i\pi h} \frac{2\pi^{d/2}}{\Gamma(d/2)} \,{}_2F_1\left( h+\nu,h-\nu;\frac d2; \frac{1-\cosh\rho}{2} \right). \end{align} 
Comparing with~\eqref{eq:w-def}, we obtain \begin{align} W_\nu^{++}(Z,Z') &=e^{-i\pi h}w_\nu(-Z\cdot Z'), \label{eq:Wpp-w}\\ W_\nu^{--}(Z,Z') &=e^{i\pi h}w_\nu(-Z\cdot Z'). \label{eq:Wmm-w} \end{align} 
The analytic branches are determined by the respective tuboid domains.  The distinction between the two classes of kernels is particularly transparent in their singularity structure. The ordinary and permuted Wightman functions possess the usual coincidence singularity at $Z\cdot Z'=-1$. The equal-tuboid kernels, by contrast, are regular at coincidence and singular at antipodal points, where $Z\cdot Z'=1$. All four kernels satisfy the homogeneous Klein--Gordon equation in each argument. The equal-tuboid kernels are characteristic of the analytic structure of de Sitter quantum field theory. Their singularity structure distinguishes them from the ordinary Wightman functions and reflects the antipodal geometry of de Sitter spacetime.

\section{The one-loop bubble}\label{sec:oneloop}

\subsection{Schwinger--Keldysh propagators and boundary diagrams} 

We now use the boundary values of the maximally analytic two-point function to construct the Schwinger--Keldysh propagators and the corresponding boundary diagrams. 

Each field $f$ has a spectral label $\nu_f$, a mass $m_f^2=h^2-\nu_f^2$, and a corresponding Wightman function $W_{\nu_f}$. The four Schwinger--Keldysh propagators are defined by chronological and antichronological ordering, with $+$ and $-$ denoting the two branches of the Schwinger--Keldysh contour: \begin{align}\label{eq:SK-def} G^f_{++}(X,X') &= \theta(X^0-X'^0)\,W_{\nu_f}(X,X') + \theta(X'^0-X^0)\,W_{\nu_f}(X',X), \nonumber\\ G^f_{--}(X,X') &= \theta(X^0-X'^0)\,W_{\nu_f}(X',X) + \theta(X'^0-X^0)\,W_{\nu_f}(X,X'), \nonumber\\ G^f_{+-}(X,X') &= W_{\nu_f}(X',X), \qquad G^f_{-+}(X,X')=W_{\nu_f}(X,X'). \end{align}
Here $+$ and $-$ label the forward and backward branches of the Schwinger--Keldysh contour and should not be confused with the superscripts used above to distinguish the analytic two-point kernels.

For timelike separated points, eq.~\eqref{eq:side} shows that the chronological propagator $G_{++}$ equals $w(z+i0)$ and the antichronological propagator $G_{--}$ equals $w(z-i0)$, independently of the time ordering. The mixed propagators $G_{+-}$ and $G_{-+}$ are the two oppositely ordered Wightman functions, with the boundary values specified in eq.~\eqref{eq:side}. At spacelike separation, all four propagators coincide.
 The external lines are the boundary values of de Sitter plane
waves,
\begin{equation}\label{eq:K-def}
 K^\Delta_+(X,\xi)=a_\Delta(X\cdot\xi+i0)^{-\Delta},\qquad
 K^\Delta_-(X,\xi)=a_\Delta(X\cdot\xi-i0)^{-\Delta},
\end{equation}
where $a_\Delta$ is given in eq.~\eqref{eq:falloffs}. For each external field we choose an
asymptotic exponent $\Delta_i=h\pm\nu_i$; no conformal-coupling restriction is imposed.

The assignment of outgoing $K^\Delta_+$ to the forward branch and ingoing $K^\Delta_-$ to the
backward branch follows from the operator ordering of the in-in correlator,
\begin{equation}
 \big\langle\Omega\big|\,\bar T\big[e^{\,i\int H_I}\big]\,O_\Delta(\xi)\,
 T\big[e^{-i\int H_I}\big]\big|\Omega\big\rangle .
\end{equation}
The boundary operator is at the turning point of the contour, later than every
interaction vertex. A forward-branch vertex lies to its right and is joined to it by
$W_{\nu_i}(X_{\rm ext},X)$; a backward-branch vertex lies to its left and is joined to it by
$W_{\nu_i}(X,X_{\rm ext})$. As the external point tends to future infinity along the null
direction $\xi$, the contribution of each asymptotic exponent $\Delta$ is
\begin{equation}
 \begin{aligned}
 W_{\nu_i}(X_{\rm ext},X)\big|_\Delta &\simeq R^{-\Delta}K^\Delta_+(X,\xi),\\
 W_{\nu_i}(X,X_{\rm ext})\big|_\Delta &\simeq R^{-\Delta}K^\Delta_-(X,\xi),
 \end{aligned}
\end{equation}
where $|_\Delta$ denotes the contribution with exponent $\Delta$ and $R$ the scale along the ray
of $\xi$. The boundary operator $O_\Delta(\xi)$ is the coefficient of this component.

The mixed internal propagator follows the same ordering:
$G^f_{+-}(X,X')=W_{\nu_f}(X',X)$, with the backward vertex as the first argument and the
forward vertex as the second. Normal analyticity therefore places the backward vertex in
$\mathcal T_-$ and the forward vertex in $\mathcal T_+$. The external plane waves have the
corresponding boundary prescriptions: $K^\Delta_+$ is the boundary value from $\mathcal T_+$, and
$K^\Delta_-$ from $\mathcal T_-$. The opposite assignment would not reproduce the operator ordering
above.

 The correlators of the boundary operators are defined by
the in-in rules, with the external propagators replaced by their asymptotic components
$K^\Delta_\pm$. We include the usual factorials for repeated identical fields in the
interaction density, so that a vertex $v$ on branch $r_v=\pm1$ carries the factor $-ir_vg_v$.
The remaining combinatorics is accounted for by the graph symmetry factor $S$. For a
connected graph $\Gamma$ with vertex set $V$, internal lines $e=(v,v')$ carrying fields $f_e$,
and external lines $i$ attached at vertices $v(i)$, the contribution is
\begin{align}\label{eq:general-rule}
 \left.\Big\langle\prod_i O_{\Delta_i}(\xi_i)\Big\rangle_c\right|_\Gamma
 ={}& S\sum_{\{r_v=\pm1\}}\int_{(dS_d^{up})^{|V|}}\prod_{v\in V}
 \left[(-ir_vg_v)\,\dd\mu(X_v)\right]\nonumber\\
 &\times\prod_{e=(v,v')}G^{f_e}_{r_vr_{v'}}(X_v,X_{v'})
 \prod_i K^{\Delta_i}_{r_{v(i)}}(X_{v(i)},\xi_i).
\end{align}
Two prescriptions complete the definition. At past infinity, the integration contours are
displaced into the tuboids corresponding to their respective branches; this implements
the prescription of the maximally analytic state through the analyticity of the
propagators and external plane waves. Products of Wightman functions with the same
argument order are defined as boundary values of products of holomorphic functions in a
common domain of analyticity \cite{bgm,bm}. Products of chronological or
antichronological propagators require an extension across coincidence, which we define by
dimensional continuation. The local counterterms have real coefficients in the
interaction Hamiltonian and carry the corresponding branch vertex factors.  

\section{Deformation onto the imaginary Lobachevskij cycles}\label{sec:cycleform}

\subsection{The deformation}

\begin{figure}[t]
\centering
\begin{tikzpicture}[>=Stealth,font=\small,ext/.style={circle,draw,fill=white,inner sep=1.3pt},vtx/.style={circle,fill,inner sep=1.5pt}]
\begin{scope}
 \draw[thick] (-1.75,1.35)--(1.75,1.35);
 \node[font=\footnotesize] at (0,1.65) {future boundary};
 \node[ext] (b1) at (-1.35,1.35) {};
 \node[ext] (b2) at (-0.75,1.35) {};
 \node[ext] (b3) at (0.75,1.35) {};
 \node[ext] (b4) at (1.35,1.35) {};
 \node[vtx] (bx) at (-0.55,-0.15) {};
 \node[vtx] (by) at (0.55,-0.15) {};
 \node[below=1pt] at (bx) {$r$};
 \node[below=1pt] at (by) {$s$};
 \draw (b1)--(bx)--(b2);
 \draw (b3)--(by)--(b4);
 \draw (bx) to[out=60,in=120] (by);
 \draw (bx) to[out=-60,in=-120] (by);
 \node at (0,-2.25) {(a) $\sum_{r,s=\pm}$ on $dS_d^{up}$};
\end{scope}
\draw[->,thick] (2.1,0.1)--(3.2,0.1);
\node[align=center,font=\footnotesize] at (2.65,0.75) {deform\\$\eta=\pm iy$};
\begin{scope}[xshift=5.1cm]
 \draw[thick] (-1.75,-1.35)--(1.75,-1.35);
 \node[font=\footnotesize] at (0,-1.65) {$y=0$};
 \node[ext] (c1) at (-1.35,-1.35) {};
 \node[ext] (c2) at (-0.75,-1.35) {};
 \node[ext] (c3) at (0.75,-1.35) {};
 \node[ext] (c4) at (1.35,-1.35) {};
 \node[vtx] (cx) at (-0.55,0.35) {};
 \node[vtx] (cy) at (0.55,0.35) {};
 \draw (c1)--(cx)--(c2);
 \draw (c3)--(cy)--(c4);
 \draw (cx) to[out=60,in=120] node[above,font=\footnotesize]{$u$} (cy);
 \draw (cx) to[out=-60,in=-120] node[below,font=\footnotesize,yshift=1pt]{$v$} (cy);
 \node at (0,-2.25) {(b) kernel $\mathcal K$ on $\pm iH^d$};
\end{scope}
\end{tikzpicture}
\caption{The one-loop bubble before and after deformation. (a) The four branch
assignments on the expanding patch, with the external points on the future boundary.
(b) Their weighted sum on the imaginary Lobachevskij cycles, represented by $\mathcal K$,
eq.~\eqref{eq:weighted-kernel}.}
\label{fig:maps}
\end{figure}
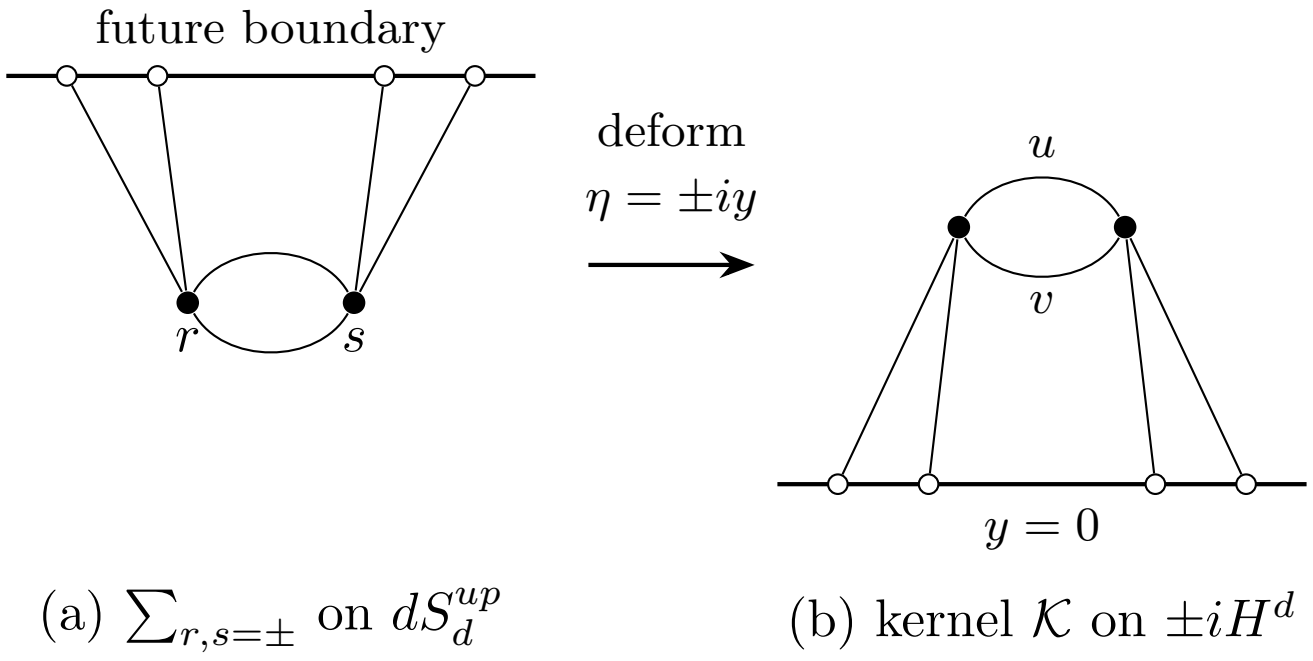

We now describe the deformation of the vertex integrations in 
eq.~\eqref{eq:general-rule}, proved for the bubble in appendix~\ref{sec:independent-contour}, branch by branch,
from the expanding Poincar\'e patch through the corresponding tuboids onto their inner
cycles: the forward-branch vertices onto $iH^d$ and the backward-branch vertices onto $-iH^d$
(figures~\ref{fig:chain} and~\ref{fig:maps}). Since the external lines are plane waves
from the outset, only the interaction vertices are deformed. A related continuation of
in-in diagrams to Euclidean anti-de Sitter space was developed in
ref.~\cite{SleightTaronna2021}.

This deformation is a central step of our construction, and it is not a formality: the
analyticity of the individual propagators does not justify it. For mixed-branch
assignments the relevant invariants remain in domains where the propagators are
holomorphic. For equal-branch assignments they do not. The chronological and
antichronological propagators are boundary values from opposite sides of the causal cut,
and each must stay on its own side while both vertices move: if the two vertices of an
equal-branch line were rotated independently, the invariant would cross the cut and the
integral would change. In addition, the coincidence singularities require the same local
extensions that enter their definition. The problem is therefore branch dependent, and
cannot be reduced to cancellations among different Schwinger--Keldysh assignments.

For the bubble, we establish the deformation by rotating the conformal times of the two
vertices by a common angle, with the sign of their branch, through a continuous family of
contours contained in the tuboids, along which each propagator keeps its prescribed
boundary value. The identity is proved at integer dimension, with an auxiliary analytic
regulator that controls the coincidence singularity: first where all the integrals
converge, and then by meromorphic continuation in the regulator. In four dimensions the
result is related to the dimensionally subtracted quantities used below by an explicit
finite local term, eq.~\eqref{eq:cd-dim-matching}. The complete proof, separately for each
branch assignment, is given in appendix~\ref{sec:independent-contour}.

\begin{figure}[t]
\centering
\resizebox{\textwidth}{!}{
\begin{tikzpicture}[>=Stealth,font=\small]
\begin{scope}[xshift=-6.2cm]
 \fill[gray!12] (-2.2,0)--(2.2,0)--(2.2,1.7)--(-2.2,1.7)--cycle;
 \fill[gray!12] (-2.2,0)--(2.2,0)--(2.2,-1.7)--(-2.2,-1.7)--cycle;
 \draw[very thick] (-2.2,0)--(2.2,0);
 \node[fill=white,inner sep=1pt,font=\footnotesize] at (0,0) {real edge $dS_d$};
 \draw[very thick,dashed] (-2.2,1.7)--(2.2,1.7);
 \draw[very thick,dashed] (-2.2,-1.7)--(2.2,-1.7);
 \node[above,font=\footnotesize] at (0,1.7) {$iH^d$: $X=0$, $Y\cdot Y=1$};
 \node[below,font=\footnotesize] at (0,-1.7) {$-iH^d$};
 \node at (-1.6,0.85) {$\mathcal T_+$};
 \node at (-1.6,-0.85) {$\mathcal T_-$};
 \draw[->,thick] (0.9,0.25)--(0.9,1.55);
 \node[right,font=\footnotesize] at (0.9,0.9) {$+$ branch};
 \draw[->,thick] (0.9,-0.25)--(0.9,-1.55);
 \node[right,font=\footnotesize] at (0.9,-0.9) {$-$ branch};
 \node at (0,-2.75) {(a)};
\end{scope}
\begin{scope}
 \draw[->] (-3.4,0)--(0.8,0) node[right]{${\rm Re}\,\eta$};
 \draw[->] (0,-2.2)--(0,2.3) node[above]{${\Im}\,\eta$};
 \draw[very thick,->] (-3.2,0.18)--(-0.1,0.18);
 \draw[very thick,->] (-0.1,-0.18)--(-3.2,-0.18);
 \node[above] at (-1.9,0.2) {$+$ branch};
 \node[below] at (-1.9,-0.2) {$-$ branch};
 \draw[very thick,dashed,<-] (0.12,0.05)--(0.12,2.0);
 \draw[very thick,dashed,->] (0.12,-0.05)--(0.12,-2.0);
 \node[right] at (0.15,1.6) {$\eta=iy$};
 \node[right] at (0.15,-1.6) {$\eta=-iy$};
 \draw[->,gray] (-1.3,0.45) arc[start angle=160,end angle=105,radius=1.4];
 \draw[->,gray] (-1.3,-0.45) arc[start angle=200,end angle=255,radius=1.4];
 \node at (-1.4,-2.6) {(b)};
\end{scope}
\begin{scope}[xshift=6.3cm,yshift=2.1cm,box/.style={draw,rounded corners,align=center,minimum width=5.4cm,inner sep=4pt}]
 \node[box] (l) at (0,0) {In-in diagram on $dS_d^{up}$:\\ branch assignments at each vertex};
 \node[box] (a) at (0,-1.7) {Cycles $\pm iH^d\subset\mathcal T_\pm$:\\ $w_\nu$ and its boundary values};
 \node[box] (t) at (0,-3.4) {Mehler--Fock transform $F(\lambda)$:\\ recurrence and solution};
 \draw[->,thick] (l)--node[right]{\ contour deformation} (a);
 \draw[->,thick] (a)--node[right]{\ spectral recurrence} (t);
 \node at (0,-4.6) {(c)};
\end{scope}
\end{tikzpicture}}
\caption{(a) The tuboids $\mathcal T_\pm$, schematically, with their common real edge and the
inner cycles $\pm iH^d$. These cycles are boundaries only in the imaginary-part projection,
not topological boundaries of the tuboids. (b) The contour deformation in complex conformal
time: $\eta=iy$ for the forward branch and $\eta=-iy$ for the backward branch, with $y>0$.
(c) From the in-in diagram to its cycle kernel and Mehler--Fock transform.}
\label{fig:chain}
\end{figure}
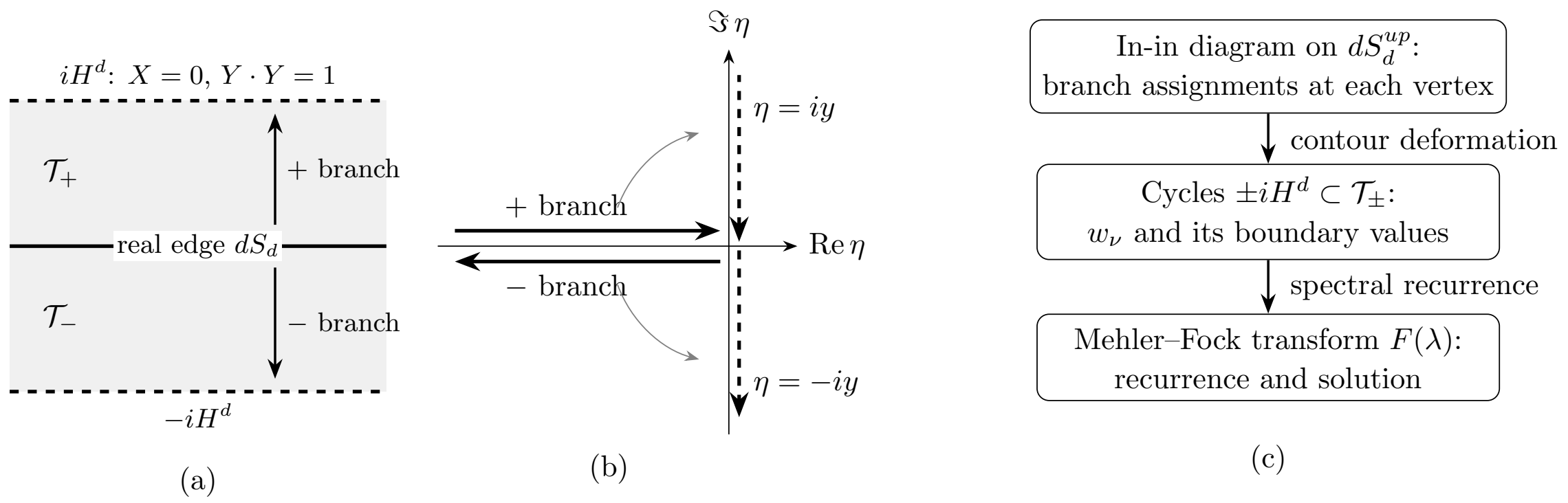

\subsection{The bubble on the Lobachevskij cycles}

For a vertex on branch $r=\pm1$, the deformation gives
\begin{equation}\label{eq:vertex-deformation}
 \int_{dS_d^{up}}\dd\mu(X)\,F(X)\;\longrightarrow\;
 (-ir)^{1-d}\int_{H^d}\dd\mu_H(Y)\,F(riY),
\end{equation}
with $-ir=e^{-ir\pi/2}$; the phase is that of $\dd\eta/(-\eta)^d$ at $\eta=riy$ in the
coordinates~\eqref{eq:H-coords}. Including the factor $-ir$ of the interaction, each
vertex carries the weight
\begin{equation}\label{eq:vertex-weight}
 \varpi_r=(-ir)^{2-d}=e^{ir\pi(d-2)/2}.
\end{equation}
Away from coincidence, the continued propagators are
\begin{equation}\label{eq:G-cycles}
 \begin{aligned}
 G_{++}(iY,iY') &=w(-Y\cdot Y'+i0), &\qquad G_{+-}(iY,-iY') &=w(Y\cdot Y'),\\
 G_{--}(-iY,-iY') &=w(-Y\cdot Y'-i0), &\qquad G_{-+}(-iY,iY') &=w(Y\cdot Y').
 \end{aligned}
\end{equation}
The mixed-branch propagators are evaluated where the two-point function is holomorphic;
the equal-branch propagators are boundary values from the two sides of the causal cut.
The external lines become $K^\Delta_\pm(\pm iY,\xi)=a_\Delta e^{\mp i\pi\Delta/2}(Y\cdot\xi)^{-\Delta}$, with no
boundary prescription, since $Y\cdot\xi>0$.

For the bubble, consider two interaction vertices,
\begin{equation}
 g_L\,\phi_1\phi_2\chi_a\chi_b\ \text{at}\ X,\qquad g_R\,\phi_3\phi_4\chi_a\chi_b\ \text{at}\ X',
\end{equation}
and let $u,v$ be the spectral labels of the internal fields $\chi_a,\chi_b$. At each vertex the
weight combines with the two external lines into a phase and a branch-independent factor:
\begin{align}
 \varpi_r K_r^{\Delta_1}(riY,\xi_1)K_r^{\Delta_2}(riY,\xi_2) &= e^{ir\theta_L}f_L(Y),\\
 \varpi_s K_s^{\Delta_3}(siY',\xi_3)K_s^{\Delta_4}(siY',\xi_4) &= e^{is\theta_R}f_R(Y'),
\end{align}
where
\begin{equation}
 \theta_L=\frac{\pi}{2}(d-2-\Delta_1-\Delta_2),\qquad
 \theta_R=\frac{\pi}{2}(d-2-\Delta_3-\Delta_4),\qquad
 \Theta=\theta_L+\theta_R,
\end{equation}
and
\begin{align}\label{legs}
 f_L(Y) &= a_{\Delta_1}a_{\Delta_2}(Y\cdot\xi_1)^{-\Delta_1}(Y\cdot\xi_2)^{-\Delta_2},\\
 f_R(Y') &= a_{\Delta_3}a_{\Delta_4}(Y'\cdot\xi_3)^{-\Delta_3}(Y'\cdot\xi_4)^{-\Delta_4}.
\end{align}
By eq.~\eqref{eq:general-rule}, deformed onto the cycles, the contribution of the branch
pair $(r,s)$ is therefore
\begin{equation}\label{eq:def-oneloop}
 \mathcal C_{rs}=e^{i(r\theta_L+s\theta_R)}\int_{H^d}\!\dd\mu_H(Y)\int_{H^d}\!\dd\mu_H(Y')\;
 f_L(Y)\,G^a_{rs}(riY,siY')\,G^b_{rs}(riY,siY')\,f_R(Y').
\end{equation}
With $Y\cdot Y'=\cosh\rho$, the products of propagators are the three branch kernels
\begin{equation}\label{eq:branch-kernels}
 \begin{aligned}
 \mathcal K_0(\rho) &= w_u(\cosh\rho)\,w_v(\cosh\rho),\\
 \mathcal K_\pm(\rho) &= w_u(-\cosh\rho\pm i0)\,w_v(-\cosh\rho\pm i0):
 \end{aligned}
\end{equation}
the two mixed assignments give $\mathcal K_0$, the equal-branch assignments the chronological
and antichronological products $\mathcal K_\pm$. Summing over branches, the bubble contribution
to the connected four-point function is
\begin{equation}\label{eq:bubble-sum}
 \mathcal B_{12|34}=Sg_Lg_R\sum_{r,s=\pm}\mathcal C_{rs}
 =Sg_Lg_R\int_{H^d}\!\dd\mu_H(Y)\int_{H^d}\!\dd\mu_H(Y')\,
 f_L(Y)\,\mathcal K(Y,Y')\,f_R(Y'),
\end{equation}
where the Schwinger--Keldysh kernel
\begin{equation}\label{eq:weighted-kernel}
 \mathcal K=2\cos(\theta_L-\theta_R)\,\mathcal K_0+e^{i\Theta}\mathcal K_++e^{-i\Theta}\mathcal K_-
\end{equation}
combines the three branch kernels, with phases fixed by the external dimensions. With the
coupling convention of eq.~\eqref{eq:general-rule}, $S=1$ for distinct internal fields and
$S=\frac12$ for two identical internal lines. For the slow components of conformally
coupled external fields, $\Delta_i=(d-2)/2$, both angles vanish and
\begin{equation}\label{eq:wightman-kernel}
 \mathcal K^{\rm conf}=2\mathcal K_0+\mathcal K_++\mathcal K_- .
\end{equation}

\paragraph{Local singularities.} The mixed kernel $\mathcal K_0$ is regular at every finite
$\rho$: at $\rho=0$ the corresponding complex de Sitter points are antipodal rather than
coincident. The ultraviolet singularity is therefore carried entirely by the equal-branch
kernels. The Hadamard form~\eqref{eq:hadamard} gives
\begin{equation}
 w_u(-\cosh\rho\pm i0)\simeq e^{\mp i\pi(d-2)/2}\,
 \frac{\Gamma(\frac d2-1)}{4\pi^{d/2}\rho^{d-2}},
\end{equation}
independently of the internal spectral label at leading order. Since
$\Theta=\pi(d-2)-\pi\Sigma/2$, with $\Sigma=\sum_{i=1}^4\Delta_i$, the leading singular part of the
Schwinger--Keldysh kernel is
\begin{equation}\label{eq:kernel-singular}
 \mathcal K(\rho)\simeq 2\cos\Big(\frac{\pi\Sigma}{2}\Big)
 \left[\frac{\Gamma(\frac d2-1)}{4\pi^{d/2}}\right]^2\rho^{4-2d},
\end{equation}
with a coefficient independent of the internal masses. For the slow components of
conformally coupled external fields, $\Sigma=2(d-2)$, and
$\cos(\pi\Sigma/2)=\cos[\pi(d-2)]=1+O(\epsilon^2)$ for $d=4-2\epsilon$.

Near coincidence $(\sinh\rho)^{d-1}\dd\rho\simeq\rho^{d-1}\dd\rho$, so the radial integrand behaves as
$\rho^{3-d}$, and for $d=4-2\epsilon$ the divergent radial integral is proportional to
$\int_0\dd\rho\,\rho^{-1+2\epsilon}=1/(2\epsilon)+\text{finite}$. 
We normalize the invariant delta distribution by
$\int_{H^d}\dd\mu_H(Y')\,\delta_H(Y,Y')f(Y')=f(Y)$.
\begin{equation}\label{eq:kernel-local-pole}
 \left.\mathcal K\right|_{\rm pole}=\frac{\cos\Theta_4}{8\pi^2\epsilon}\,\delta_{H^d},
 \qquad d=4-2\epsilon,\qquad \Theta_4=\Theta|_{d=4}.
\end{equation}
If $\cos\Theta_4=0$, the pole cancels in $\mathcal K$. The $O(\epsilon)$ part of the branch phases
may nevertheless multiply the $1/\epsilon$ singularity and leave a finite local contribution,
so their $d$ dependence must be retained until after subtraction.

\paragraph{Convergence at the boundary.} Besides the ultraviolet singularities at
coincident vertices, the integrals may also diverge at the conformal boundary. When one
vertex approaches a boundary point away from the external insertions, each external line
attached to it contributes $y^{\Delta_i}$ and each internal line decays at least as
$y^{\alpha_e}$, with $\alpha_e=h-|{\rm Re}\,\nu_e|$; the integral converges if
\begin{equation}
 \sum_{i\ {\rm attached}}{\rm Re}\,\Delta_i+\sum_{e\ {\rm incident}}\alpha_e>d-1 .
\end{equation}
For the slow components of two conformally coupled external lines and two principal-series
internal lines in $d=4$, the left-hand side is $5$, and the condition holds. Absolute
convergence in all boundary regions, including the approach to the external insertions
and simultaneous approaches of both vertices, is proved in
Appendix~\ref{sec:independent-contour}.

\section{Difference--differential equations for the Schwinger--Keldysh bubble}\label{sec:recurrence}

The branch kernels $\mathcal K_r$, $r=0,\pm$, and the Schwinger--Keldysh kernel $\mathcal K$ are the
basic objects of the rest of the paper. Following the ideas and methods introduced in
ref.~\cite{BCM}, we derive the equations they satisfy: difference--differential equations
for the kernels, and finite-difference equations for their Mehler--Fock transforms. They follow from the contiguity relations of the de Sitter two-point function, together
with the Klein--Gordon equation and Green's identity, and relate the kernels in which one
internal spectral label is shifted by $\pm1$.

On invariant functions of $z=\cosh\rho$ we write
\begin{equation}
 \mathcal Q=h^2+\LapH,\qquad \LapH=(z^2-1)\partial_z^2+d\,z\,\partial_z ,
\end{equation}
so that the Klein--Gordon equation for a label $u$, with $m_u^2=h^2-u^2$, reads
$(\mathcal Q-u^2)w_u=0$. The recurrence coefficients are
\begin{equation}\label{eq:p-pm}
 p_u^\pm(t)=\big[(h-1\pm u+v)^2-t\big]\big[(h-1\pm u-v)^2-t\big].
\end{equation}
The sources involve the dimensionally continued tadpoles \cite{BCM}
\begin{equation}\label{eq:ss-tadpole}
 T_u(d)=\frac{\Gamma(1-\frac d2)}{(4\pi)^{d/2}}
 \frac{\Gamma(h+u)\Gamma(h-u)}{\Gamma(\frac12+u)\Gamma(\frac12-u)}
\end{equation}
through the combination
\begin{equation}\label{eq:Su}
 S_u(t)=2T_v+\frac{T_u}{(h-1)^2-u^2}\big(u^2+v^2-(h-1)^2-t\big).
\end{equation}

The mixed-branch kernel $\mathcal K_0$ is regular at coincidence, whereas each equal-branch
propagator satisfies
\begin{equation}\label{eq:ss-green}
 (\mathcal Q-u^2)\,w_u(-z+ri0)=s_r(d)\,\delta_H,\qquad s_r(d)=-e^{-ir\pi(d-2)/2},\qquad r=\pm .
\end{equation}
The single-label recurrences of ref.~\cite{BCM} therefore give
\begin{align}
 p_u^-(\mathcal Q)\mathcal K_{0;u+1,v}-p_u^+(\mathcal Q)\mathcal K_{0;u-1,v} &=0, \label{eq:ss-mixed}\\
 p_u^-(\mathcal Q)\mathcal K_{r;u+1,v}-p_u^+(\mathcal Q)\mathcal K_{r;u-1,v}
 &=2u(d-2)\,s_r(d)\,S_u(\mathcal Q)\,\delta_H,\qquad r=\pm. \label{eq:ss-same}
\end{align}
In each recurrence only one internal label is shifted; the equations obtained by
exchanging $u$ and $v$ hold as well. The source is local, a combination of $\delta_H$ and
$\mathcal Q\delta_H$ with coefficients fixed by the tadpoles. The phase $s_r(d)$ comes from the
singular boundary value of one propagator; the tadpole is the regular coincidence
coefficient of the other and carries no additional phase.

Since the coefficients of the Schwinger--Keldysh kernel~\eqref{eq:weighted-kernel} are
independent of the internal labels, the three branch recurrences combine linearly into a
recurrence for $\mathcal K$. The mixed-branch kernel contributes no source, so only the
equal-branch phases enter the right-hand side. Using
\begin{equation}
 e^{i\Theta}s_+(d)+e^{-i\Theta}s_-(d)=-2\cos\Big[\Theta-\frac{\pi(d-2)}{2}\Big],
\end{equation}
we obtain
\begin{align}\label{eq:ss-full}
 &p_u^-(\mathcal Q)\mathcal K_{u+1,v}-p_u^+(\mathcal Q)\mathcal K_{u-1,v}\nonumber\\
 &\qquad=-4u(d-2)\cos\Big[\Theta-\frac{\pi(d-2)}{2}\Big]S_u(\mathcal Q)\,\delta_H .
\end{align}
For the slow components of conformally coupled external fields, $\Theta=0$. These
identities hold as meromorphic distributional relations in $d$ and in the spectral
labels, with the coincidence extensions fixed by dimensional continuation.

\paragraph{Proof away from coincidence.} The contiguity relations of the associated
Legendre functions \cite{Bateman1} give
\begin{align}
 &\sum_{\sigma=\pm1}a_\sigma(\nu)\,w_{\nu+\sigma}(z)=z\,w_\nu(z), \label{eq:contiguity}\\
 &\sum_{\sigma=\pm1}(h-\sigma\nu)\,a_\sigma(\nu)\,w_{\nu+\sigma}(z)=(1-z^2)\,\partial_z w_\nu(z),
 \label{eq:derivative-contiguity}
\end{align}
with
\begin{equation}
 a_\sigma(\nu)=-\frac{\nu-\sigma(h-1)}{2\nu}.
\end{equation}
These identities also hold for the boundary values on the causal cut. Let $f=w_u$ and
$g=w_v$ be the two propagators of a branch kernel. For the equal-branch boundary values,
the contiguity relations give
\begin{equation}\label{eq:ss-ladder}
 f_{u+1}=\frac{(u+h)\,zf+(z^2-1)f'}{u-h+1},\qquad
 f_{u-1}=\frac{(u-h)\,zf-(z^2-1)f'}{u+h-1};
\end{equation}
for the Wightman functions of the mixed-branch kernel the relations have the opposite
overall signs. The Klein--Gordon equations,
\begin{equation}
 f''=-\frac{dz\,f'+m_u^2f}{z^2-1},\qquad g''=-\frac{dz\,g'+m_v^2g}{z^2-1},
\end{equation}
reduce every derivative of $fg$ to a linear combination of $fg$, $f'g$, $fg'$ and $f'g'$.
Substituting eq.~\eqref{eq:ss-ladder} into the recurrence and using the Klein--Gordon
equations, the coefficients of all four combinations cancel identically. This proves the
recurrence away from coincidence, for both the mixed- and the equal-branch kernels.

\paragraph{The contact term.} The mixed-branch kernel is regular at $\rho=0$ and produces
no contact term. Near coincidence, an equal-branch propagator of label $j$ behaves as
\begin{align}
 w_j(-\cosh\rho+ri0)={}T_j\Big(1-\frac{m_j^2}{2d}\rho^2+O(\rho^4)\Big)
 +c_r\,\rho^{2-d}\Big(1-\frac{m_j^2+\chi_d}{2(4-d)}\rho^2+O(\rho^4)\Big),
 \label{eq:ss-local-expansion}
\end{align}
with
\begin{equation}
 c_r=e^{-ir\pi(d-2)/2}\,\frac{\Gamma(\frac d2-1)}{4\pi^{d/2}},\qquad
 \chi_d=\frac{(d-1)(2-d)}{3}.
\end{equation}
The singular part has radial flux $(2-d)\,\Omega_{d-1}c_r=s_r(d)$, with
$\Omega_{d-1}=2\pi^{d/2}/\Gamma(\frac d2)$, which fixes the phase and normalisation of the
delta-function source in eq.~\eqref{eq:ss-green}.

We apply Green's identity to the shifted products outside a ball of radius $\rho_0$ around
coincidence. The contributions containing two singular factors cancel in the recurrence,
and the remaining surface terms, apart from those with one singular and one regular
factor, vanish for $0<{\rm Re}\,d<2$. The recurrence is therefore first established there,
as an identity of distributions, and then extended meromorphically in $d$, by the same
dimensional continuation that defines the coincidence products in
section~\ref{sec:oneloop}. The surviving terms contain the singular part of one
propagator and the regular coefficient $T_j$ of the other. Using
\begin{equation}
 T_{u+1}=\frac{u+h}{u-h+1}\,T_u,\qquad T_{u-1}=\frac{u-h}{u+h-1}\,T_u,
\end{equation}
their sum is $2u(d-2)\,s_r(d)\,S_u(\mathcal Q)\,\delta_H$, which proves
eq.~\eqref{eq:ss-same}. The recurrence~\eqref{eq:ss-full} for $\mathcal K$ follows by
linearity.

\subsection{The Mehler--Fock transform}\label{sec:MF-transform}

Following ref.~\cite{BEGMP2010}, we now introduce the generalised Mehler--Fock
transforms $F_r$, $r=0,\pm$, of the three branch kernels $\mathcal K_r$.  
The Mehler--Fock kernel is the two-point function $w_q$, normalised to one at $z=1$:
\begin{equation}\label{eq:phi-def}
 \varphi_q(z)=\frac{w_q(z)}{w_q(1)}
 =2^{\frac d2-1}\,\Gamma\Big(\frac d2\Big)\,(z^2-1)^{-\frac{d-2}{4}}\,P^{-\frac{d-2}{2}}_{-\frac12+q}(z),
 \qquad \varphi_q(1)=1 
 \end{equation}

\begin{equation}\label{eq:transform-convention}
 F_r(u,v;q)=\Omega_{d-1}\int_1^\infty\dd z\,(z^2-1)^{\frac{d-2}{2}}\,\varphi_q(z)\,\mathcal K_r(z),
 \qquad q=i\lambda .
\end{equation}
In radial formulas, $\varphi_q(\rho)$ abbreviates $\varphi_q(\cosh\rho)$.\\
The integrals are defined first in their convergence domains, and then continued
meromorphically.
The passage from the kernel recurrences to their transforms rests on two facts. First, the
Mehler--Fock kernel is an eigenfunction of $\mathcal Q$,
\begin{equation}\label{eq:MF-eigenvalue}
 \mathcal Q\,\varphi_q=q^2\,\varphi_q ,
\end{equation}
with eigenvalue $-\lambda^2$ on the physical line $q=i\lambda$; this is the Klein--Gordon equation
of $w_q$. Second, on functions of $z$,
\begin{equation}
 \LapH f=(z^2-1)^{-\frac{d-2}{2}}\,\partial_z\big[(z^2-1)^{\frac d2}\,\partial_zf\big],
\end{equation}
so that $\mathcal Q$ is formally symmetric with respect to the measure
$(z^2-1)^{(d-2)/2}\dd z$ of eq.~\eqref{eq:transform-convention}, and integration by parts
replaces $\mathcal Q$ by $q^2=-\lambda^2$. For
real $3<d<4$ and principal-series labels this is legitimate. The shifted transforms
exist, since the convergence strip $|{\rm Re}\,u|<h$ contains ${\rm Re}\,u=\pm1$ when $h>1$.
The boundary term at infinity vanishes: with $z=\cosh\rho$, the shifted kernels decay at
least as $e^{-(2h-1)\rho}$, the Mehler--Fock kernel as $e^{-h\rho}$, and the boundary factor
grows as $e^{2h\rho}$, leaving $O(e^{-(h-1)\rho})$.

We may therefore pair the distributional recurrences~\eqref{eq:ss-mixed}
and~\eqref{eq:ss-same} directly with the Mehler--Fock kernel. With the
normalisation~\eqref{eq:transform-convention} the transform of $\delta_H$ is one, and we
obtain
\begin{align}
 p_u^-(-\lambda^2)F_0(u+1,v;i\lambda)-p_u^+(-\lambda^2)F_0(u-1,v;i\lambda) &=0,
 \label{eq:spectral-recurrence-0}\\
 p_u^-(-\lambda^2)F_r(u+1,v;i\lambda)-p_u^+(-\lambda^2)F_r(u-1,v;i\lambda)
 &=2u(d-2)\,s_r(d)\,S_u(-\lambda^2),\qquad r=\pm,
 \label{eq:spectral-recurrence-r}
\end{align}
The corresponding recurrences in $v$ hold as well. The transform of the
Schwinger--Keldysh kernel~\eqref{eq:weighted-kernel} is, by linearity,
\begin{equation}
 F=2\cos(\theta_L-\theta_R)\,F_0+e^{i\Theta}F_++e^{-i\Theta}F_- ,
\end{equation}
and satisfies
\begin{align}\label{eq:spectral-recurrence-d}
 &p_u^-(-\lambda^2)\,F(u+1,v;i\lambda)-p_u^+(-\lambda^2)\,F(u-1,v;i\lambda)\nonumber\\
 &\qquad=-4u(d-2)\cos\Big[\Theta-\frac{\pi(d-2)}{2}\Big]S_u(-\lambda^2).
\end{align}

The spectral recurrences are obtained first where the transforms are defined, and then
extend meromorphically in the dimension and in the spectral labels. At $d=4$ the
ultraviolet pole must be subtracted first; the subtracted expression is then regular and
is evaluated at $d=4$.

\section{Solving the recurrences: the Mehler--Fock transforms}\label{sec:solution}

We begin with the homogeneous recurrence~\eqref{eq:spectral-recurrence-0} for the
transform $F_0$ of the mixed-branch kernel. The Mehler--Fock integral of $\mathcal K_0$ is
precisely the one from which Bros, Epstein, Gaudin, Moschella and Pasquier
\cite{BEGMP2010} first obtained the de Sitter K\"all\'en--Lehmann formula, which has since
been rederived and widely used, not always with reference to its original derivation.

Here we obtain the same formula from the recurrence. The homogeneous recurrence
determines $F_0$ as a product of Gamma functions, up to a function periodic in the
internal spectral labels. The analyticity and growth of the defining integral remove this
periodic freedom, and a single residue, determined by the large-distance behaviour of the
two-point functions, fixes the remaining dependence on the spectral variable. The
K\"all\'en--Lehmann formula thus follows from the recurrence, rather than being used as
an input.

The same method then applies to the chronological and antichronological products. Their
finite-difference equations contain the local sources of section~\ref{sec:recurrence},
and the defining integrals provide the analytic conditions that identify the solutions.

We work initially at real $3<d<4$ and generic principal-series labels, and then continue
the resulting identities meromorphically. The analytic estimates and algebraic
identities needed in the derivation are collected in
appendix~\ref{app:physical-conditions}.

\subsection{The homogeneous recurrence and the K\"all\'en--Lehmann formula}\label{sec:KL-recurrence}

We write the spectral variable as $q=i\lambda$. On the physical line $q$ is purely
imaginary; in the derivation below it is continued to complex values. The
transform~\eqref{eq:transform-convention} of the mixed-branch kernel then reads
\begin{equation}\label{eq:Wightman-transform}
 F_0(u,v;q)=(2\pi)^{d/2}\int_1^\infty\dd z\,(z^2-1)^{(d-2)/4}\,
 P_{-\frac12+q}^{-\frac{d-2}{2}}(z)\,\mathcal K_0(z).
\end{equation}
The transform~\eqref{eq:Wightman-transform} satisfies the homogeneous recurrence of
section~\ref{sec:recurrence},
\begin{equation}\label{eq:homogeneous}
 p_u^-(q^2)\,F_0(u+1,v;q)-p_u^+(q^2)\,F_0(u-1,v;q)=0,
\end{equation}
where, at $t=q^2$, the coefficients~\eqref{eq:p-pm} factorise as
\begin{equation}
 p_u^\pm(q^2)=\prod_{\sigma,\tau=\pm1}\big(h-1\pm u+\sigma v+\tau q\big).
\end{equation}
Equivalently,
\begin{equation}\label{eq:homogeneous-ratio}
 \frac{F_0(u+1,v;q)}{F_0(u-1,v;q)}
 =\prod_{\sigma,\tau=\pm1}\frac{h-1+u+\sigma v+\tau q}{h-1-u+\sigma v+\tau q}.
\end{equation}
The corresponding recurrence with $u$ and $v$ exchanged holds as well.
The linear factors are reproduced by shifts of Gamma functions. A solution symmetric in
the three spectral labels is
\begin{equation}\label{eq:N}
 N(u,v,q)=\prod_{\epsilon_1,\epsilon_2,\epsilon_3=\pm1}
 \Gamma\Big(\frac{h+\epsilon_1u+\epsilon_2v+\epsilon_3q}{2}\Big).
\end{equation}
By $\Gamma(z+1)=z\Gamma(z)$, $N$ satisfies the homogeneous recurrences in both $u$ and $v$,
and it is even in all three spectral labels.

The recurrences alone do not determine the solution uniquely. The quotient
\begin{equation}
 E(u,v;q)=\frac{F_0(u,v;q)}{N(u,v,q)}
\end{equation}
has period two in each internal label,
\begin{equation}
 E(u+2,v;q)=E(u,v;q),\qquad E(u,v+2;q)=E(u,v;q).
\end{equation}
For the transform~\eqref{eq:Wightman-transform} this periodic freedom is removed by
analyticity and growth. The defining integral is holomorphic on complete period strips in
$u$ and $v$ and satisfies the corresponding uniform bounds, and a Fourier-mode argument
then shows that the periodic factor is independent of both internal labels; the proof and
the required estimates are given in appendix~\ref{app:KL-uniqueness}. Hence
\begin{equation}\label{eq:F-KN}
 F_0(u,v;q)=K(q)\,N(u,v,q).
\end{equation}

The recurrences have thus determined the full dependence on the internal labels, and only
the spectral factor $K(q)$ remains. It is fixed by a pole of the defining integral: the
large-distance behaviour of the three factors in eq.~\eqref{eq:Wightman-transform} produces
a simple pole on the divisor
\begin{equation}\label{eq:KL-pole}
 q=h+u+v,
\end{equation}
whose residue is determined by the leading large-distance coefficients of the two
Wightman functions and of the Mehler--Fock kernel. Comparison with the corresponding
residue of $N$ gives
\begin{equation}\label{eq:K-factor}
 K(q)=\frac{1}{32\pi^{(d+3)/2}\,\Gamma(h)\Gamma(h+q)\Gamma(h-q)};
\end{equation}
the residue calculation, including its normalisation, is given in
appendix~\ref{app:KL-uniqueness}. Therefore $F_0=M$, with
\begin{equation}\label{eq:M}
 M(u,v;q)=\frac{\displaystyle\prod_{\epsilon_1,\epsilon_2,\epsilon_3=\pm1}
 \Gamma\Big(\frac{h+\epsilon_1u+\epsilon_2v+\epsilon_3q}{2}\Big)}
 {32\pi^{(d+3)/2}\,\Gamma(h)\Gamma(h+q)\Gamma(h-q)} ,
\end{equation}
and Mehler--Fock inversion gives the de Sitter K\"all\'en--Lehmann formula.

\subsection{The sourced recurrences}\label{sec:particular}

We now turn to the chronological and antichronological products. In terms of $q$, their
recurrences~\eqref{eq:spectral-recurrence-r} read
\begin{equation}\label{eq:sourced-recurrence}
 p_u^-(q^2)F_r(u+1,v;q)-p_u^+(q^2)F_r(u-1,v;q)=s_r(d)\,j_d(u),\qquad r=\pm,
\end{equation}
with
\begin{equation}\label{eq:jd}
 j_d(u)=2u(d-2)\,S_u(q^2).
\end{equation}
To construct a particular solution, define
\begin{equation}\label{eq:H}
 B_{\sigma\tau}=\frac{h+u+\sigma v+\tau q}{2},\qquad A_{\sigma\tau}=B_{\sigma\tau}-h+1,\qquad
 \mathcal H(u)=\prod_{\sigma,\tau=\pm1}\frac{\Gamma(B_{\sigma\tau})}{\Gamma(A_{\sigma\tau})} .
\end{equation}
Since
\begin{equation}
 \frac{\mathcal H(u+2)}{\mathcal H(u)}=\frac{p_{u+1}^+(q^2)}{p_{u+1}^-(q^2)},
\end{equation}
division by $\mathcal H$ reduces the sourced recurrence to a first difference, and summing
the tail defines
\begin{equation}\label{eq:P-sum}
 P(u,v;q)=-\mathcal H(u)\sum_{n=0}^{\infty}
 \frac{j_d(u+2n+1)}{p_{u+2n+1}^-(q^2)\,\mathcal H(u+2n+2)} .
\end{equation}
Uniform asymptotics of the Gamma-function ratios show that the series converges
absolutely, and locally normally, for ${\rm Re}\,d>2$ at generic spectral parameters. It
therefore defines a holomorphic function away from its singular divisors, with
meromorphic continuation in $u$, $v$ and $q$, and shifting $u$ by two in the series shows
that $P$ satisfies the recurrence with source $j_d$.

The series can be summed explicitly. With $b_u=(u+1)/2$, separating the three tadpole
contributions gives
\begin{equation}\label{eq:P}
 P(u,v;q)=-\frac{(u+1)(d-2)}{8\displaystyle\prod_{\sigma,\tau=\pm1}B_{\sigma\tau}}
 \left[2T_v\,\mathbf F_1-T_{u+1}\,\mathbf F_2
 +\frac{q^2-v^2}{(u+1)^2-(h-1)^2}\,T_{u+1}\,\mathbf F_3\right],
\end{equation}
with
\begin{align}
 \mathbf F_1&={}_6F_5\left[\begin{matrix}A,\ b_u+1,\ 1\\ B+1,\ b_u\end{matrix};1\right],
 \nonumber\\
 \mathbf F_2&={}_8F_7\left[\begin{matrix}A,\ b_u+1,\ \frac{u+h+1}{2},\ \frac{u+h+2}{2},\ 1\\[2pt]
 B+1,\ b_u,\ \frac{u-h+2}{2},\ \frac{u-h+3}{2}\end{matrix};1\right],
 \label{eq:F123}\\
 \mathbf F_3&={}_8F_7\left[\begin{matrix}A,\ b_u+1,\ \frac{u+h+1}{2},\ \frac{u+h}{2},\ 1\\[2pt]
 B+1,\ b_u,\ \frac{u-h+3}{2},\ \frac{u-h+4}{2}\end{matrix};1\right],
 \nonumber
\end{align}
where $A$ and $B+1$ denote the corresponding sets of four parameters. The three series
have excesses $2d-4$, $d-2$ and $d$, and therefore converge absolutely for ${\rm Re}\,d>2$.
At exceptional parameter values, eq.~\eqref{eq:P} is understood by meromorphic
continuation from the series~\eqref{eq:P-sum}: apparent singularities of the individual
hypergeometric functions need not be singularities of $P$.

The particular solution $P$ is not even in $u$, whereas the original two-point function
is even. Since $S_u$ is even in $u$, the source $j_d$ is odd, and since $p^\pm_{-u}=p^\mp_u$,
the function $P(-u,v;q)$ satisfies the same sourced recurrence as $P(u,v;q)$.

Their difference therefore solves the homogeneous recurrence~\eqref{eq:homogeneous}, so
by subsection~\ref{sec:KL-recurrence} it is $M$ times a function of period two in $u$
and $v$, and the odd part of $P$ must be cancelled by such a homogeneous contribution. Introduce
\begin{equation}\label{eq:C}
 C_u(v,q)=\frac{\cos(\pi v)-\cos[\pi(h-u)]\cos(\pi q)}{\sin[\pi(h-u)]} .
\end{equation}
The recurrence solution obeys
\begin{equation}\label{eq:odd-identity}
 \frac{P(u,v;q)-P(-u,v;q)}{M(u,v;q)}=C_{-u}(v,q)-C_u(v,q),
\end{equation}
as proved in appendix~\ref{app:physical-conditions}. Since $M$ is even in $u$, this is
equivalent to
\[
 P(u,v;q)+C_u(v,q)M(u,v;q)=P(-u,v;q)+C_{-u}(v,q)M(-u,v;q),
\]
so that $P+C_uM$ is even in $u$.

\subsection{Identifying the transforms of the chronological and antichronological products}\label{sec:sourced-completion}

The finite-difference equations alone do not determine a unique meromorphic function.
Although $P$ fixes the inhomogeneous part and the reflection identity fixes its odd
homogeneous part, an even homogeneous contribution remains a priori undetermined. The
defining Mehler--Fock integrals supply the additional information required to eliminate
this freedom: analyticity and growth in the complete spectral strips, the
contiguity relations, the simultaneous large-mass behaviour and the ultraviolet
normalisation.

Appendix~\ref{app:physical-conditions} proves that the transforms of the chronological
and antichronological products and the explicit recurrence solution belong to the same
analytic class. Both satisfy the two sourced internal-label recurrences and the
contiguity relation in the transform label, obtained by applying
eq.~\eqref{eq:contiguity} under the defining integral. The difference between them is
therefore a homogeneous solution. The strip bounds and the internal contiguity relations
reduce this difference to
\[
 M(u,v;q)\big[A_0(q)+A_1(q)\cos(\pi u)\cos(\pi v)\big].
\]
The simultaneous large-mass behaviour at $u=v=i\mu$ eliminates the second term, while the
contiguity in $q$, evenness and the $q$-strip bound imply $A_0(q)=A\cos(\pi q)$. Finally,
equality of the ultraviolet coefficients gives $A=0$. Consequently, the transforms of the
chronological and antichronological products are
\begin{equation}\label{eq:result}
 F_r(u,v;q)=s_r(d)\,P(u,v;q)+e^{-ir\pi h}\big[\cos(\pi q)-ir\,C_u(v,q)\big]M(u,v;q),
 \qquad r=\pm1.
\end{equation}
The equality holds initially in the common convergence domain and extends
meromorphically in the dimension and spectral parameters. The first term accounts for
the local source, while the second is the uniquely selected homogeneous term.
Continuation of the inverse transform is understood with the contour prescription stated
in section~\ref{sec:inversion}.

Together with $F_0=M$, eq.~\eqref{eq:result} determines the Mehler--Fock transform of the
Schwinger--Keldysh kernel~\eqref{eq:weighted-kernel}:
\begin{equation}
 F=2\cos(\theta_L-\theta_R)\,F_0+e^{i\Theta}F_++e^{-i\Theta}F_- .
\end{equation}
On the physical spectral line $q=i\lambda$, this becomes
\begin{align}\label{eq:F-weighted}
 F(u,v;i\lambda)={}&2\sin(\Theta-\pi h)\big[P(u,v;i\lambda)+C_u(v,i\lambda)M(u,v;i\lambda)\big]
 \nonumber\\
 &+2\big[\cos(\theta_L-\theta_R)+\cos(\Theta-\pi h)\cosh(\pi\lambda)\big]M(u,v;i\lambda).
\end{align}
The external dimensions enter only through the branch angles, while $P$, $C_u$ and $M$
depend only on the internal spectral labels and on $\lambda$.

For the slow components of conformally coupled external fields, $\theta_L=\theta_R=0$, and
therefore
\begin{align}\label{eq:F-conf}
 F^{\rm conf}(u,v;i\lambda)={}&-2\sin(\pi h)\big[P(u,v;i\lambda)+C_u(v,i\lambda)M(u,v;i\lambda)\big]
 \nonumber\\
 &+2\big[1+\cos(\pi h)\cosh(\pi\lambda)\big]M(u,v;i\lambda).
\end{align}

\subsection{Mehler--Fock inversion}\label{sec:inversion}

With the normalisation of eq.~\eqref{eq:transform-convention}, the inverse Mehler--Fock
transform reconstructs the Schwinger--Keldysh kernel as
\begin{equation}\label{eq:inverse-transform}
 \mathcal K(z)=2^{\frac d2-1}\,\Gamma\Big(\frac d2\Big)\,(z^2-1)^{-\frac{d-2}{4}}
 \int_0^\infty\dd\lambda\,p_d(\lambda)\,P_{-\frac12+i\lambda}^{-\frac{d-2}{2}}(z)\,F(u,v;i\lambda),
 \qquad z>1,
\end{equation}
where
\begin{equation}\label{eq:plancherel-density}
 p_d(\lambda)=\frac{2\,\Gamma(h+i\lambda)\Gamma(h-i\lambda)}
 {(4\pi)^{d/2}\,\Gamma(\frac d2)\,\Gamma(i\lambda)\Gamma(-i\lambda)} ,
\end{equation}
that is, $$\mathcal K(z)=\int_0^\infty\dd\lambda\,p_d(\lambda)\,\varphi_{i\lambda}(z)\,F(u,v;i\lambda)$$ in terms of the
Mehler--Fock kernel~\eqref{eq:phi-def}. The same formula, with $F$ replaced by $F_r$,
reconstructs the branch kernels $\mathcal K_r$.
For principal internal labels, the inversion of $F_0=M$ converges absolutely. The
chronological, antichronological and Schwinger--Keldysh kernels require instead an Abel
prescription: the $\lambda$-integral in eq.~\eqref{eq:inverse-transform} is replaced by
\begin{equation}\label{eq:Abel-inversion}
 \lim_{\epsilon\downarrow0}\int_0^\infty\dd\lambda\,e^{-\epsilon\lambda}\,p_d(\lambda)\,
 P_{-\frac12+i\lambda}^{-\frac{d-2}{2}}(z)\,F(u,v;i\lambda),
\end{equation}
understood distributionally, and pointwise wherever the limit exists. Meromorphic
continuation in the dimension and in the spectral labels is performed after this
prescription has been specified. At $d=4$, the ultraviolet pole must be subtracted first;
the subtracted expression is then evaluated at $d=4$.

The inversion is first restricted to principal-series internal labels. Continuing to labels of the 
complementary series requires deforming the spectral contour and including the
residues of the poles that cross it. In particular, for positive complementary labels with
$u+v>h$, the poles at $q=u+v-h-2n>0$, with $n=0,1,2,\ldots$, produce discrete terms \cite{BEGMP2010}, and the integral along the
principal axis alone is then not the full continued representation.

\section{The boundary correlator}\label{sec:attachment}

We now attach the external plane waves to the bubble kernel. The evaluated spectral formula defines the dimensional continuation
used below; appendix~\ref{sec:independent-contour} identifies it at $d=4$ with the analytically regulated cycle integral, up to the
explicit finite local conversion to dimensional subtraction.

The evaluation uses the method of triangular invariants of ref.~\cite{BEGMP2010}, where the
de Sitter K\"all\'en--Lehmann formula was derived by combining the plane-wave
representation with integration over the Lobachevskij cycles, which reduces the relevant
integrals to triangular invariants evaluated by the star--triangle formula. The same
mechanism applies here to the Mehler--Fock representation of the Schwinger--Keldysh
kernel.

Inserting the inverse Mehler--Fock representation~\eqref{eq:inverse-transform} into
eq.~\eqref{eq:bubble-sum}, we obtain
\begin{equation}\label{eq:attached-spectral}
 \mathcal B_{12|34}(d)=Sg_Lg_R\int_0^\infty\dd\lambda\,p_d(\lambda)\,F(u,v;i\lambda)\,\mathcal A_\lambda,
\end{equation}
where
\begin{equation}\label{eq:spectral-attachment}
 \mathcal A_\lambda=2^{\frac d2-1}\,\Gamma\Big(\frac d2\Big)\int_{H^d}\dd\mu_H(Y)\int_{H^d}\dd\mu_H(Y')\,
 f_L(Y)\,f_R(Y')\,(z^2-1)^{-\frac{d-2}{4}}P_{-\frac12+i\lambda}^{-\frac{d-2}{2}}(Y\cdot Y').
\end{equation}
Here $f_L$ and $f_R$ are defined in eq.~\eqref{legs}, $F$ is the transform~\eqref{eq:F-weighted}
with its full dependence on $d$ retained, and $p_d$ is the Plancherel
density~\eqref{eq:plancherel-density}.

The spectral and vertex integrations are interchanged at fixed dimension using the heat regulator in appendix~\ref{sec:independent-contour}.
The evaluated expression is then continued meromorphically in $d$, with the inversion prescription of section~\ref{sec:inversion}.

\subsection{Evaluation by triangular invariants}\label{sec:triangular-invariants}

The crucial observation is that the plane-wave representation~\eqref{eq:W-planewaves} of
the two-point function, specialized to the purely imaginary Lobachevskij cycles, writes
the Mehler--Fock kernel as a superposition of products of two plane waves, one depending
only on $Y$ and the other only on $Y'$. Explicitly,
\begin{align}\label{eq:attachment-plane-waves}
 &2^{\frac d2-1}\,\Gamma\Big(\frac d2\Big)\,(z^2-1)^{-\frac{d-2}{4}}\,
 P_{-\frac12+i\lambda}^{-\frac{d-2}{2}}(Y\cdot Y')\nonumber\\
 &\qquad=\frac1{\Omega_{d-1}}\int_{S^{d-1}}\dd\omega(\xi)\,(Y\cdot\xi)^{-h-i\lambda}\,(Y'\cdot\xi)^{-h+i\lambda}.
\end{align}
Substitution into eq.~\eqref{eq:spectral-attachment} factorises the two integrations over
$H^d$ into triangular invariants of the type evaluated in ref.~\cite{BEGMP2010},
\begin{equation}\label{eq:triangular-invariant}
 \mathcal T_{\Delta_1\Delta_2\Delta}(\xi_1,\xi_2,\xi)
 =\int_{H^d}\dd\mu(Y)\,(Y\cdot\xi_1)^{-\Delta_1}(Y\cdot\xi_2)^{-\Delta_2}(Y\cdot\xi)^{-\Delta},
\end{equation}
and gives
\begin{equation}\label{eq:attachment-two-triangles}
 \mathcal A_\lambda=\frac{\prod_{i=1}^4a_{\Delta_i}}{\Omega_{d-1}}\int_\gamma\dd\mu_\gamma(\xi)\,
 \mathcal T_{\Delta_1\Delta_2\Delta}(\xi_1,\xi_2,\xi)\,
 \mathcal T_{\Delta_3\Delta_4,2h-\Delta}(\xi_3,\xi_4,\xi),\qquad \Delta=h+i\lambda .
\end{equation}
The triangular invariant is given by the star--triangle formula
\begin{equation}\label{eq:attachment-star-triangle}
 \mathcal T_{\Delta_1\Delta_2\Delta}(\xi_1,\xi_2,\xi)
 =\frac{c_d(\Delta_1,\Delta_2,\Delta)}
 {(\xi_1\cdot\xi_2)^{\frac{\Delta_1+\Delta_2-\Delta}{2}}\,
 (\xi_2\cdot\xi)^{\frac{\Delta_2+\Delta-\Delta_1}{2}}\,
 (\xi\cdot\xi_1)^{\frac{\Delta+\Delta_1-\Delta_2}{2}}},
\end{equation}
where
\begin{align}\label{eq:attachment-three-point-coefficient}
 c_d(\Delta_1,\Delta_2,\Delta)={}&
 \frac{2^{\frac{\Delta_1+\Delta_2+\Delta}{2}-1}\pi^h}{\Gamma(\Delta_1)\Gamma(\Delta_2)\Gamma(\Delta)}\,
 \Gamma\Big(\frac{\Delta_1+\Delta_2+\Delta-2h}{2}\Big)\nonumber\\
 &\times\Gamma\Big(\frac{\Delta_1+\Delta_2-\Delta}{2}\Big)
 \Gamma\Big(\frac{\Delta_2+\Delta-\Delta_1}{2}\Big)
 \Gamma\Big(\frac{\Delta+\Delta_1-\Delta_2}{2}\Big).
\end{align}
At the vertex $Y'$ the same formula holds with $(\Delta_1,\Delta_2,\xi_1,\xi_2)$ replaced by
$(\Delta_3,\Delta_4,\xi_3,\xi_4)$ and $\Delta$ by its shadow $2h-\Delta=h-i\lambda$.
Applying eq.~\eqref{eq:attachment-star-triangle} at the two vertices evaluates both
integrations over the Lobachevskij cycles. What remains is the integral over the cone
direction $\xi$ of the plane-wave representation, in which each external direction
$\xi_i$ appears through a power $(\xi_i\cdot\xi)^{-\beta_i}$, with
\begin{align}\label{eq:attachment-exponents}
 \beta_1&=\frac{h+i\lambda+\Delta_1-\Delta_2}{2}, & \beta_2&=\frac{h+i\lambda+\Delta_2-\Delta_1}{2},
 \nonumber\\
 \beta_3&=\frac{h-i\lambda+\Delta_3-\Delta_4}{2}, & \beta_4&=\frac{h-i\lambda+\Delta_4-\Delta_3}{2}.
\end{align}
These satisfy $\sum_{i=1}^4\beta_i=2h=d-1$, so the remaining four-point cone integral,
\begin{equation}\label{eq:F4-def}
 F_4(\beta_i;\xi_i)=\int_\gamma\dd\mu_\gamma(\xi)\prod_{i=1}^4(\xi_i\cdot\xi)^{-\beta_i},
\end{equation}
is independent of the section $\gamma$, which may be any section of the future cone meeting
almost every ray exactly once. It is a conformally covariant four-point function of the
boundary directions, with weight $\beta_i$ at $\xi_i$: up to a standard prefactor, a function
of their two cross-ratios ($F_4$ is not the Appell
hypergeometric function). We obtain
\begin{align}\label{eq:attachment-boundary}
 \mathcal A_\lambda={}&\frac{\prod_{i=1}^4a_{\Delta_i}}{\Omega_{d-1}}\,
 c_d(\Delta_1,\Delta_2,h+i\lambda)\,c_d(\Delta_3,\Delta_4,h-i\lambda)\nonumber\\
 &\times(\xi_1\cdot\xi_2)^{-\frac{\Delta_1+\Delta_2-h-i\lambda}{2}}\,
 (\xi_3\cdot\xi_4)^{-\frac{\Delta_3+\Delta_4-h+i\lambda}{2}}\,
 F_4(\beta_i;\xi_i).
\end{align}

This calculation shows the advantage of the coordinate-independent formulation. De Sitter
covariance is manifest throughout, the two integrations over the Lobachevskij cycles are
evaluated directly by the method of triangular invariants, and what remains is the
conformally covariant four-point function $F_4$ on the cone. No choice of coordinates is
required.

\paragraph{Convergence of the spectral representation.} The evaluated spectral representation is absolutely convergent
for real $3<d<4$, principal-series internal labels and distinct boundary directions in the domain
\begin{align}\label{eq:attachment-convergence-domain}
 {\rm Re}(\Delta_1+\Delta_2)&>h, & |{\rm Re}(\Delta_1-\Delta_2)|&<h,\nonumber\\
 {\rm Re}(\Delta_3+\Delta_4)&>h, & |{\rm Re}(\Delta_3-\Delta_4)|&<h,
\end{align}
the same inequalities as in the fixed-dimensional contour theorem. These conditions ensure the absolute
convergence of the two triangular integrals and imply $0<{\rm Re}\,\beta_i<h$, so that $F_4$
is locally integrable at each insertion and bounded uniformly in real $\lambda$. By
Stirling's formula, each of the two coefficients $c_d$ decays as $e^{-\pi\lambda/2}$, and
\begin{equation}\label{eq:attachment-large-lambda}
 |\mathcal A_\lambda|\le C\,(1+\lambda)^{{\rm Re}\,\Sigma-2h-3}\,e^{-\pi\lambda}.
\end{equation}
The transform $F(u,v;i\lambda)$ is bounded, since the plane-wave representation bounds the
normalized spherical function by its value at $\lambda=0$ and $p_d(\lambda)=O(\lambda^{2h})$. The spectral integrand
therefore decays as $(1+\lambda)^{{\rm Re}\,\Sigma-3}e^{-\pi\lambda}$ at large $\lambda$; at $\lambda=0$,
$p_d(\lambda)=O(\lambda^2)$ and the other factors are generically regular.

\subsection{Difference equations for the boundary correlator}\label{sec:boundary-recurrence}

The kernel recurrence carries over to the boundary correlator. Its source involves the
contact integral
\begin{equation}\label{eq:contact-def}
 \mathcal D_d(\xi_1,\xi_2,\xi_3,\xi_4)=\int_{H^d}\dd\mu(Y)\,f_L(Y)\,f_R(Y),
\end{equation}
a single vertex with all four external lines, which converges for ${\rm Re}\,\Sigma>2h$ and
${\rm Re}(\Sigma-2\Delta_i)>0$ for each $i$, and is defined elsewhere by meromorphic continuation.

The factor $f_L(Y)$ is invariant under simultaneous transformations of $Y$, $\xi_1$ and $\xi_2$,
so the vertex Laplacian acts as the boundary Casimir: 
\begin{equation}\label{eq:casimir-vertex}
 \LapH f_L=\mathcal C_{12}\,f_L ,
\end{equation}
where $\mathcal C_{12}$ is the quadratic Casimir of $SO(1,d)$ acting simultaneously on $\xi_1$ and $\xi_2$,
normalised so that it equals $\Delta(\Delta-2h)$ on a scalar of dimension $\Delta$, the eigenvalue of
$\LapH$ on the plane wave $(Y\cdot\xi)^{-\Delta}$. Integrating by parts in $Y$, with boundary terms
vanishing in the domain~\eqref{eq:attachment-convergence-domain}, the operator $\mathcal Q$ acting
on the kernel in eq.~\eqref{eq:bubble-sum} becomes $h^2+\mathcal C_{12}$, which does not act on
$f_R$, and the source $\delta_H$ produces the contact integral. Eq.~\eqref{eq:ss-full} thus gives,
for the boundary correlator $\mathcal B_{u,v}$ with internal labels $(u,v)$,
\begin{align}\label{eq:boundary-recurrence}
 &p_u^-\big(h^2+\mathcal C_{12}\big)\,\mathcal B_{u+1,v}-p_u^+\big(h^2+\mathcal C_{12}\big)\,\mathcal B_{u-1,v}
 \nonumber\\
 &\qquad=-4u(d-2)\cos\Big[\Theta-\frac{\pi(d-2)}{2}\Big]\,Sg_Lg_R\,
 S_u\big(h^2+\mathcal C_{12}\big)\,\mathcal D_d ,
\end{align}
together with the corresponding recurrence in $v$, first in the convergence domain and
elsewhere by continuation. 

These difference--differential equations combine shifts of the
internal labels with the boundary Casimir in place of $\mathcal Q$.
One may equivalently use $\mathcal C_{34}$, since the two Casimirs
agree on the invariant correlator. The recurrence is consistent with the spectral representation. In
eq.~\eqref{eq:attachment-two-triangles}, the triangular invariant
$\mathcal T_{\Delta_1\Delta_2\Delta}(\xi_1,\xi_2,\xi)$ is invariant under simultaneous transformations of $\xi_1$,
$\xi_2$ and $\xi$, so $\mathcal C_{12}$ acts on it as the Casimir in $\xi$, which equals
$\Delta(\Delta-2h)=-(h^2+\lambda^2)$ by homogeneity. Hence $h^2+\mathcal C_{12}$ acts on $\mathcal A_\lambda$ as $-\lambda^2$, the
eigenvalue of $\mathcal Q$ in the Mehler--Fock transform, and eq.~\eqref{eq:boundary-recurrence} is
the spectral recurrence~\eqref{eq:spectral-recurrence-d} multiplied by $p_d(\lambda)\mathcal A_\lambda$ and
integrated over $\lambda$.

\subsection{Four dimensions}\label{sec:4d-limit}

We set $d=4-2\epsilon$, keeping the coordinate-independent evaluation. Near $d=4$ the
Schwinger--Keldysh kernel~\eqref{eq:weighted-kernel} has the pole of
eq.~\eqref{eq:kernel-local-pole},
\begin{equation}\label{eq:kernel-laurent}
 \mathcal K_d=\frac{c_{-1}}{\epsilon}\,\delta_H+\mathcal K_{d,\mathrm{sub}},\qquad
 c_{-1}=\frac{\cos\Theta_4}{8\pi^2},
\end{equation}
where the subscript $d$ indicates the dependence on the dimension. Integrated against the
external plane waves, the local term $\delta_H$ sets $Y'=Y$ and gives the contact integral
$ \mathcal D_d(\xi_1,\xi_2,\xi_3,\xi_4)$. 
The dimension-dependent contact subtraction is therefore
\begin{equation}\label{eq:attached-local-pole}
 \frac{Sg_Lg_Rc_{-1}}{\epsilon}\,\mathcal D_d .
\end{equation}
This is not just the Laurent pole of the boundary correlator: its expansion also contains
a finite term.

For definiteness, we hold the internal spectral labels and the chosen external exponents
fixed, continue their normalisations with $d$, and keep the coupling prefactor fixed; a
different continuation, or a dimension-dependent coupling and renormalisation-scale
convention, changes the finite local term and must be specified separately. Assuming
that $\mathcal D_d$ is regular at $d=4$,
\begin{equation}
 \mathcal D_{4-2\epsilon}=\mathcal D_4-2\epsilon\,\mathcal D'_4+O(\epsilon^2),\qquad
 \mathcal D'_4=\left.\frac{\dd\mathcal D_d}{\dd d}\right|_{d=4}.
\end{equation}
We compare two prescriptions:
\begin{align}\label{eq:attached-renormalized}
 \mathcal B_{\rm dim}&=\lim_{\epsilon\to0}\Big[\mathcal B_{12|34}(d)-\frac{Sg_Lg_Rc_{-1}}{\epsilon}\,\mathcal D_d\Big],
 \nonumber\\
 \mathcal B_{\rm pole}&=\lim_{\epsilon\to0}\Big[\mathcal B_{12|34}(d)-\frac{Sg_Lg_Rc_{-1}}{\epsilon}\,\mathcal D_4\Big].
\end{align}
They differ by a finite local term,
\begin{equation}\label{eq:finite-contact}
 \mathcal B_{\rm dim}-\mathcal B_{\rm pole}=2Sg_Lg_Rc_{-1}\,\mathcal D'_4
 =\frac{Sg_Lg_R\cos\Theta_4}{4\pi^2}\,\mathcal D'_4 .
\end{equation}
More generally, replacing $c_{-1}$ in the dimension-dependent subtraction by
$c(\epsilon)=c_{-1}+\epsilon c_1+O(\epsilon^2)$ changes the right-hand side to
$Sg_Lg_R(2c_{-1}\mathcal D'_4-c_1\mathcal D_4)$. If $\mathcal D_d$ has additional poles, its full
Laurent expansion must be retained.

The physical local counterterm can also be displayed directly. With the factorial
convention of section~\ref{sec:oneloop}, a quartic counterterm with real coefficient
$\delta g$ contributes
\begin{equation}\label{eq:quartic-counterterm}
 \mathcal B_{\rm ct}=2\,\delta g\cos\Big[\frac\pi2(d-2-\Sigma)\Big]\mathcal D_d .
\end{equation}
At $d=4$ the cosine equals $-\cos\Theta_4$, so the pole coefficient
$\delta g=Sg_Lg_R/(16\pi^2\epsilon)$ cancels the bubble pole. Its full dimensional phase, any scale
factors and any finite part must be retained, and this prescription need not coincide in
its finite part with the kernel subtraction~\eqref{eq:kernel-laurent}.

For the dimension-dependent kernel subtraction, the spectral representation gives
\begin{equation}\label{eq:ren-spectral}
 \mathcal B_{\rm dim}=Sg_Lg_R\lim_{\epsilon\to0}\int_0^\infty\dd\lambda\,
 p_d(\lambda)\Big[F(u,v;i\lambda)-\frac{c_{-1}}{\epsilon}\Big]\mathcal A_\lambda,
\end{equation}
with $F$ retaining its full dependence on $d$, the Abel prescription understood. A finite quartic counterterm changes the result by
a multiple of $\mathcal D_4$, so no scheme-independent quantity is defined here. Note that
$\mathcal D_4$, although it comes from a local bulk vertex, is not a contact term on the boundary: it is a non-trivial function of the four boundary directions.

\section{Fourier representation of the boundary correlator in Poincar\'e coordinates}\label{sec:fourier-representation}

We now Fourier transform the boundary variables in a flat section of the cone, obtaining a triple-$K$ representation for comparison
with momentum-space calculations. Related results are discussed in the forthcoming work of Moschella, Sleight and Taronna~\cite{MST}.

We use the flat coordinates $(\eta,\bm x)$ of the Poincar\'e patch and the corresponding flat
section of the cone, parametrised as in eq.~\eqref{213} by $\bm k\in\mathbb R^{d-1}$, with measure
$\dd\mu_\gamma(\xi(\bm k))=\dd^{d-1}k$; we put $\xi_i=\xi(\bm k_i)$, so that
$\xi_i\cdot\xi_j=k_{ij}^2/2$ with $k_{ij}=|\bm k_i-\bm k_j|$. For the boundary values
$\psi_\Delta^\pm(X,\xi)=(X\cdot\xi\pm i0)^{-\Delta}$ of the plane waves~\eqref{eq:planewave}, we
take the Fourier transform with respect to $\bm x$, at fixed conformal time $\eta<0$ and fixed
boundary point $\xi(\bm k)$:
\begin{equation}\label{eq:plane-wave-Fourier-definition}
 \widetilde\psi_\Delta^\pm(\eta,\bm p;\bm k)=\int_{\mathbb R^{d-1}}\dd^{d-1}x\,e^{i\bm p\cdot\bm x}\,
 \psi_\Delta^\pm(X(\eta,\bm x),\xi(\bm k)).
\end{equation}
Writing $\alpha=\Delta-h$, the Hankel representations of ref.~\cite{BEM2010} give
\begin{align}\label{eq:plane-wave-Fourier-Hankel}
 \widetilde\psi_\Delta^-(\eta,\bm p;\bm k)&=\frac{i\pi(2\pi)^h}{\Gamma(\Delta)}\,(-\eta)^h\,e^{i\bm p\cdot\bm k}\,
 p^\alpha H_{-\alpha}^{(1)}(-p\eta),\nonumber\\
 \widetilde\psi_\Delta^+(\eta,\bm p;\bm k)&=-\frac{i\pi(2\pi)^h}{\Gamma(\Delta)}\,(-\eta)^h\,e^{i\bm p\cdot\bm k}\,
 p^\alpha H_{-\alpha}^{(2)}(-p\eta),
\end{align}
with $p=|\bm p|$, and Fourier inversion yields
\begin{equation}\label{eq:plane-wave-Fourier-inversion}
 \psi_\Delta^\pm(X(\eta,\bm x),\xi(\bm k))=\int_{\mathbb R^{d-1}}\frac{\dd^{d-1}p}{(2\pi)^{d-1}}\,
 e^{-i\bm p\cdot\bm x}\,\widetilde\psi_\Delta^\pm(\eta,\bm p;\bm k).
\end{equation}
Each tuboid boundary value thus contains only one kind of Hankel function, with no admixture of the other
solution: $\psi_\Delta^-$
contains $H^{(1)}$ and $\psi_\Delta^+$ contains $H^{(2)}$. The Bunch--Davies mode prescription is therefore not an additional choice
imposed on the Fourier modes, but the spatial Fourier expression of the analyticity
already built into the de Sitter plane waves~\cite{BEM2010,bm}; combined in the two-point
function, these plane waves give the maximally analytic state.

On the imaginary Lobachevskij cycles, with coordinates $(y,\bm x)$, $y>0$, as in
eq.~\eqref{eq:H-coords}, which correspond to $\eta=\pm iy$, the decaying continuations of
the Hankel functions are Macdonald functions:
\begin{equation}\label{eq:fourier-plane-wave}
 \int_{\mathbb R^{d-1}}\dd^{d-1}k\,e^{i\bm p\cdot\bm k}\,(Y(y,\bm x)\cdot\xi(\bm k))^{-\Delta}
 =\frac{2(2\pi)^h}{\Gamma(\Delta)}\,e^{i\bm p\cdot\bm x}\,y^h\,p^{\Delta-h}K_{\Delta-h}(py).
\end{equation}
This identity follows by Schwinger parametrisation of $(Y\cdot\xi(\bm k))^{-\Delta}$, with
$Y\cdot\xi(\bm k)=(y^2+|\bm x-\bm k|^2)/(2y)$. Formulae are first understood in a
convergence domain, and elsewhere by continuation as Fourier distributions.

\subsection{The boundary correlator in the Fourier space of the flat section}

We Fourier transform the triangular invariant~\eqref{eq:triangular-invariant} with respect
to its three boundary points $\bm k_1$, $\bm k_2$, $\bm k$. Inserting eq.~\eqref{eq:fourier-plane-wave}
for each plane wave, the integration over the spatial coordinate $\bm x$ of the vertex
$Y=Y(y,\bm x)$ gives momentum conservation, and the integration over $y$ leaves the power
$y^{3h-d}=y^{h-1}$. The triangular invariant thus becomes a triple-$K$ integral, the
momentum-space form of a conformal three-point function~\cite{BMS2013}:
\begin{align}\label{eq:triangle-triple-K}
 &\int\dd^{d-1}k_1\,\dd^{d-1}k_2\,\dd^{d-1}k\;e^{i(\bm p_1\cdot\bm k_1+\bm p_2\cdot\bm k_2+\bm p\cdot\bm k)}\,
 \mathcal T_{\Delta_1\Delta_2\Delta}(\xi(\bm k_1),\xi(\bm k_2),\xi(\bm k))\nonumber\\
 &\qquad=\frac{8(2\pi)^{5h}\,p_1^{\Delta_1-h}\,p_2^{\Delta_2-h}\,p^{\Delta-h}}{\Gamma(\Delta_1)\Gamma(\Delta_2)\Gamma(\Delta)}\,
 \delta^{(d-1)}(\bm p_1+\bm p_2+\bm p)\,\mathcal J_{\Delta_1\Delta_2\Delta}(p_1,p_2,p),
\end{align}
with
\begin{equation}\label{eq:fourier-triple-K}
 \mathcal J_{\Delta_1\Delta_2\Delta}(p_1,p_2,p)=\int_0^\infty\dd y\,y^{h-1}
 K_{\Delta_1-h}(p_1y)\,K_{\Delta_2-h}(p_2y)\,K_{\Delta-h}(py).
\end{equation}
For non-vanishing momenta, $p_1,p_2,p>0$, $\mathcal J_{\Delta_1\Delta_2\Delta}$ converges at $y=0$ if
$h>|{\rm Re}(\Delta_1-h)|+|{\rm Re}(\Delta_2-h)|+|{\rm Re}(\Delta-h)|$, and at infinity the Macdonald
functions decay exponentially; elsewhere it is defined by continuation.

The function $\mathcal A_\lambda$ has two representations on the flat section. Write
$\Delta=h+i\lambda$ and $\widetilde\Delta=2h-\Delta=h-i\lambda$. In position space, the
result~\eqref{eq:attachment-boundary}, manifestly $SO(1,d)$-invariant and homogeneous of
degree $-\Delta_i$ in each $\xi_i$, becomes an expression in which these properties are no
longer manifest:
\begin{align}\label{eq:attachment-flat}
 \mathcal A_\lambda={}&\frac{2^{h+\Sigma/2}\prod_{i=1}^4a_{\Delta_i}}{\Omega_{d-1}}\,
 c_d(\Delta_1,\Delta_2,\Delta)\,c_d(\Delta_3,\Delta_4,\widetilde\Delta)\nonumber\\
 &\times(k_{12}^2)^{-\frac{\Delta_1+\Delta_2-\Delta}{2}}\,
 (k_{34}^2)^{-\frac{\Delta_3+\Delta_4-\widetilde\Delta}{2}}
 \int_{\mathbb R^{d-1}}\dd^{d-1}k\,\prod_{i=1}^4|\bm k-\bm k_i|^{-2\beta_i}.
\end{align}
In momentum space, its Fourier transform in the four boundary points is obtained from
eq.~\eqref{eq:attachment-two-triangles}, by applying eq.~\eqref{eq:triangle-triple-K} to both
triangular invariants. The integration over $\bm k$ identifies their internal momenta with
the exchanged momentum $\bm Q=\bm p_1+\bm p_2=-(\bm p_3+\bm p_4)$, $Q=|\bm Q|$, and the powers
$Q^{\pm i\lambda}$ cancel:
\begin{gather}\label{eq:fourier-A}
 \widetilde{\mathcal A}_\lambda=(2\pi)^{d-1}\,\delta^{(d-1)}\Big(\sum_i\bm p_i\Big)
 \prod_{i=1}^4\mathcal N_i(p_i)\,
 \frac{4(2\pi)^{2h} \mathcal J_{\Delta_1\Delta_2\Delta}(p_1,p_2,Q)\,\mathcal J_{\Delta_3\Delta_4\widetilde\Delta}(p_3,p_4,Q)}{\Omega_{d-1}\Gamma(\Delta)\Gamma(\widetilde\Delta)}\,
,
 \nonumber\\
 \mathcal N_i(p_i)=\frac{2(2\pi)^h a_{\Delta_i}}{\Gamma(\Delta_i)}\,p_i^{\Delta_i-h}.
\end{gather}

Equating the two representations~\eqref{eq:attachment-flat} and~\eqref{eq:fourier-A}, the
coefficients $a_{\Delta_i}$ and the factor $\Omega_{d-1}$ cancel, and we obtain
\begin{align}\label{eq:shadow-fourier}
 &\int\prod_{i=1}^4\dd^{d-1}k_i\,e^{i\sum_i\bm p_i\cdot\bm k_i}\,
 (k_{12}^2)^{-\frac{\Delta_1+\Delta_2-\Delta}{2}}\,(k_{34}^2)^{-\frac{\Delta_3+\Delta_4-\widetilde\Delta}{2}}
 \int_{\mathbb R^{d-1}}\dd^{d-1}k\,\prod_{i=1}^4|\bm k-\bm k_i|^{-2\beta_i}\nonumber\\
 &\quad=\frac{64\,(2\pi)^{8h}\,\delta^{(d-1)}\big(\sum_i\bm p_i\big) \prod_{i=1}^4p_i^{\Delta_i-h}\;
 \mathcal J_{\Delta_1\Delta_2\Delta}(p_1,p_2,Q)\,\mathcal J_{\Delta_3\Delta_4\widetilde\Delta}(p_3,p_4,Q)}
 {2^{h+\Sigma/2}\,c_d(\Delta_1,\Delta_2,\Delta)\,c_d(\Delta_3,\Delta_4,\widetilde\Delta)\,
 \Gamma(\Delta)\Gamma(\widetilde\Delta)\prod_{i=1}^4\Gamma(\Delta_i)}
.
\end{align}
The left-hand side is the shadow representation of a conformal partial wave with exchanged
dimension $\Delta$: the integral over $\bm k$ of two conformal three-point structures, with
dimensions $\Delta$ and its shadow $\widetilde\Delta$. Eq.~\eqref{eq:shadow-fourier} shows that in momentum
space the shadow integral reduces to momentum conservation, and the partial wave
factorises into a product of two triple-$K$ integrals at fixed exchanged momentum $Q$.

The Fourier transform of the boundary correlator finally follows from
eq.~\eqref{eq:attached-spectral}. With the Plancherel density~\eqref{eq:plancherel-density},
and the Abel prescription of section~\ref{sec:inversion} understood,
\begin{align}\label{eq:fourier-attached-bubble}
 \widetilde{\mathcal B}_{12|34}={}&(2\pi)^{d-1}\,\delta^{(d-1)}\Big(\sum_i\bm p_i\Big)\,
 Sg_Lg_R\prod_{i=1}^4\mathcal N_i(p_i)\,\frac{2}{\pi^2}\int_0^\infty\dd\lambda\,
 \lambda\sinh(\pi\lambda)\,F(u,v;i\lambda)\nonumber\\
 &\times\mathcal J_{\Delta_1\Delta_2\Delta}(p_1,p_2,Q)\,\mathcal J_{\Delta_3\Delta_4\widetilde\Delta}(p_3,p_4,Q),
\end{align}
a spectral superposition of conformal partial waves, each factorised into two triple-$K$
integrals.

For the exchanged dimensions $\Delta=h\pm i\lambda$ the convergence condition of the triple-$K$
integrals reduces to $h>|{\rm Re}(\Delta_1-h)|+|{\rm Re}(\Delta_2-h)|$ and
$h>|{\rm Re}(\Delta_3-h)|+|{\rm Re}(\Delta_4-h)|$, a domain strictly contained in the
domain~\eqref{eq:attachment-convergence-domain} of the position-space representation;
outside it, the formulas are understood by meromorphic continuation, with distributional
extensions and local subtractions where required. Interchanging the vertex
integrals with the spectral integration requires, in addition, regulated bounds for the
whole integrand, not merely the convergence of each vertex integral.

\subsection{The contact term and the subtraction}
Fourier transforming the contact integral~\eqref{eq:contact-def}, a single vertex with four
plane waves, in its four boundary points, and inserting eq.~\eqref{eq:fourier-plane-wave}
for each plane wave, gives
\begin{equation}\label{eq:fourier-contact}
 \widetilde{\mathcal D}_d=(2\pi)^{d-1}\,\delta^{(d-1)}\Big(\sum_i\bm p_i\Big)
 \prod_{i=1}^4\mathcal N_i(p_i)\int_0^\infty\dd y\,y^{2h-1}\prod_{i=1}^4K_{\Delta_i-h}(p_iy).
\end{equation}
The dimension-dependent subtraction is therefore
\begin{equation}\label{eq:fourier-subtraction}
 \widetilde{\mathcal B}_{\rm dim}=\lim_{\epsilon\to0}\Big[\widetilde{\mathcal B}_{12|34}(d)
 -\frac{Sg_Lg_Rc_{-1}}{\epsilon}\,\widetilde{\mathcal D}_d\Big],\qquad d=4-2\epsilon ;
\end{equation}
since both terms carry the same momentum-conserving delta function, this amounts to
replacing $F(u,v;i\lambda)\mapsto F(u,v;i\lambda)-c_{-1}/\epsilon$ in eq.~\eqref{eq:fourier-attached-bubble}
before $\epsilon\to0$, with the regulated spectral prescription understood. The external
exponents, their normalisations and the coupling factors must follow the same
dimensional continuation as in the coordinate-independent subtraction: replacing them
prematurely by their four-dimensional values can discard finite contact terms.

\subsection{Four dimensions}

For the slow components of conformally coupled external fields in four dimensions,
\begin{equation}
 h=\frac32,\qquad \Delta_i=1,\qquad \Delta_i-h=-\frac12,\qquad a_1=\frac1{8\pi^2}.
\end{equation}
Using $K_{1/2}(z)=\sqrt{\pi/(2z)}\,e^{-z}$, each external factor becomes
\begin{equation}\label{eq:fourier-conformal-external}
 \mathcal N_i(p_i)\,y^{3/2}K_{1/2}(p_iy)=\frac{y}{2p_i}\,e^{-p_iy}.
\end{equation}
Put $E_L=p_1+p_2$ and $E_R=p_3+p_4$, and define
\begin{equation}\label{eq:fourier-Laplace-K}
 \mathcal I_\lambda(E,Q)=\int_0^\infty\frac{\dd y}{\sqrt y}\,e^{-Ey}\,K_{i\lambda}(Qy).
\end{equation}
After the dimension-dependent subtraction, the boundary correlator is
\begin{equation}\label{eq:fourier-four-dimensional-weight}
 \widetilde{\mathcal B}_{\rm dim}=(2\pi)^3\,\delta^{(3)}\Big(\sum_i\bm p_i\Big)\,
 \frac{Sg_Lg_R}{8\pi^2p_1p_2p_3p_4}\int_0^\infty\dd\lambda\,\lambda\sinh(\pi\lambda)\,
 F_{\rm sub}(u,v;i\lambda)\,\mathcal I_\lambda(E_L,Q)\,\mathcal I_\lambda(E_R,Q),
\end{equation}
where, with the full dimensional dependence of $F$ retained,
\[
 F_{\rm sub}(u,v;i\lambda)=\lim_{\epsilon\to0}
 \left[\left.F(u,v;i\lambda)\right|_{d=4-2\epsilon}
       -\frac{c_{-1}}{\epsilon}\right].
\]
The external continuation is the one fixed in section~\ref{sec:4d-limit}.
Finite phase contributions are already included in this multiplier;
any additional finite quartic counterterm contributes separately
through $\widetilde{\mathcal D}_4$. For the same external components, the contact integral is elementary:
\begin{equation}\label{eq:fourier-contact-four}
 \widetilde{\mathcal D}_4=(2\pi)^3\,\delta^{(3)}\Big(\sum_i\bm p_i\Big)\,
 \frac{1}{16p_1p_2p_3p_4(p_1+p_2+p_3+p_4)} .
\end{equation}
The exchanged momentum $Q$ remains arbitrary throughout. A zero-transfer limit requires a
separate uniform analysis, and cannot be obtained by setting $Q=0$ in the individual
Macdonald functions of imaginary order.

{ 
\section{The two-loop bubble chain}\label{sec:twoloop}

Consider now the two-bubble chain with four external lines, a two-loop diagram, generated
by the three vertices
\begin{equation}
 g_L\,\phi_1\phi_2\chi_a\chi_b,\qquad g_\chi\,\chi_a\chi_b\chi_c\chi_d,\qquad g_R\,\phi_3\phi_4\chi_c\chi_d,
\end{equation}
where the internal fields $\chi_a,\chi_b$ of the left bubble have spectral labels $(u_L,v_L)$ and
those of the right bubble, $\chi_c,\chi_d$, have labels $(u_R,v_R)$. The three vertices carry
independent Schwinger--Keldysh branch indices, giving eight branch assignments. The
external factors $f_L$, $f_R$ and the endpoint phases $\theta_L$, $\theta_R$ are those of
section~\ref{sec:cycleform}. With the graph symmetry factor $S$, the two-loop boundary
correlator is
\begin{equation}\label{eq:chain-attachment}
 \mathcal B^{(2)}_{12|34}=Sg_Lg_\chi g_R\int_{H^d}\dd\mu(Y)\int_{H^d}\dd\mu(Y')\,
 f_L(Y)\,\Gamma_{\rm SK}(Y,Y')\,f_R(Y').
\end{equation}

\subsection{Composition of the two bubbles}

Let $V$ be the central vertex in the cycles. For branch indices $r,s,t=\pm1$, define
\begin{equation}\label{eq:chain-internal-kernel}
 \Gamma_{\rm int}^{rst}(Y,Y')=\varpi_s\int_{H^d}\dd\mu(V)\,
 \mathcal K_L^{rs}(Y,V)\,\mathcal K_R^{st}(V,Y'),
\end{equation}
where $\mathcal K_b^{rs}$, $b=L,R$, is the branch kernel of section~\ref{sec:cycleform} with the
labels of bubble $b$: the mixed-branch kernel $\mathcal K_0$ if $r\ne s$, and the equal-branch
kernel $\mathcal K_r$ if $r=s$. The Schwinger--Keldysh kernel of the chain is
\begin{equation}\label{eq:chain-weighted-kernel}
 \Gamma_{\rm SK}(Y,Y')=\sum_{r,s,t=\pm1}e^{ir\theta_L+it\theta_R}\,\Gamma_{\rm int}^{rst}(Y,Y').
\end{equation}
Each endpoint vertex, together with its two external lines, contributes to the phase
$e^{ir\theta_L}$ or $e^{it\theta_R}$, while the central vertex contributes its weight $\varpi_s$,
included in $\Gamma_{\rm int}^{rst}$. This weight and the source phases of eq.~\eqref{eq:ss-green}
take a simple form in terms of the angle
\begin{equation}\label{eq:chain-central-phase}
 \vartheta=\frac{\pi(4-d)}{2},
\end{equation}
which vanishes at $d=4$:
\begin{equation}
 \varpi_s=-e^{-is\vartheta},\qquad s_\pm(d)=e^{\pm i\vartheta}.
\end{equation}
The Mehler--Fock transform of the convolution in eq.~\eqref{eq:chain-internal-kernel} is the
product of the transforms of the two bubbles. We denote the transform of $\Gamma_{\rm SK}$ by
$F^{(2)}$, with $q=i\lambda$ on the physical line as in section~\ref{sec:solution}. All
compositions below are first understood in a common regulated domain, with the
prescribed regulators and coincidence extensions in place.

\subsection{The one-loop blocks}

Let
\begin{equation}
 A(u,v;q)=P(u,v;q)+C_u(v,q)\,M(u,v;q),\qquad c=\cos(\pi q),
\end{equation}
so that $c=\cosh(\pi\lambda)$ on the physical line, and write $A_b$, $M_b$, $F_{s,b}$ for $A$, $M$, $F_s$
evaluated at the labels $(u_b,v_b)$ of bubble $b=L,R$. Since $e^{-is\pi h}=is\,e^{is\vartheta}$,
eq.~\eqref{eq:result} takes the form
\begin{equation}\label{eq:chain-branch-decomposition}
 F_{s,b}=e^{is\vartheta}\big[A_b+is\,c\,M_b\big],\qquad s=\pm1,
\end{equation}
so that $A_b$ and $M_b$ are recovered from the two branches as
\begin{equation}\label{eq:chain-A-definition}
 A_b=\frac12\sum_{s=\pm1}e^{-is\vartheta}F_{s,b},\qquad
 2ic\,M_b=\sum_{s=\pm1}s\,e^{-is\vartheta}F_{s,b}.
\end{equation}
By eqs.~\eqref{eq:homogeneous} and~\eqref{eq:sourced-recurrence}, $M$ satisfies the homogeneous
recurrence, and $A$ the sourced recurrence with the phase-free source~\eqref{eq:jd}:
\begin{equation}\label{eq:chain-block-recurrences}
 p_u^-(q^2)\,A(u+1,v;q)-p_u^+(q^2)\,A(u-1,v;q)=j_d(u),
\end{equation}
together with the corresponding relations in $v$.

\subsection{The Mehler--Fock transform of the chain}

Once the branch $s$ of the central vertex is fixed, the left bubble depends only on the
branch $r$ of the left endpoint and the right bubble only on the branch $t$ of the right
endpoint, so the sum over $r$ and $t$ factorises. At the left endpoint, the branch $r=s$ gives the equal-branch
transform $F_{s,L}$ with phase $e^{is\theta_L}$, and the branch $r=-s$ the mixed-branch transform
$M_L$ with phase $e^{-is\theta_L}$; likewise at the right endpoint. With the central weight
$\varpi_s$,
\begin{equation}\label{eq:chain-sewing}
 F^{(2)}=-\sum_{s=\pm1}e^{-is\vartheta}
 \big[e^{is\theta_L}F_{s,L}+e^{-is\theta_L}M_L\big]
 \big[e^{is\theta_R}F_{s,R}+e^{-is\theta_R}M_R\big],
\end{equation}
which contains all eight branch assignments. Substituting
eq.~\eqref{eq:chain-branch-decomposition}, with
\begin{equation}\label{eq:chain-phase-combinations}
 \Psi=\theta_L+\theta_R+\vartheta,\qquad \omega=\theta_L-\theta_R,
\end{equation}
we get
\begin{equation}\label{eq:chain-expanded}
 F^{(2)}=-2\cos\Psi\,A_LA_R+2\big(c\sin\Psi-\cos\omega\big)\big(A_LM_R+M_LA_R\big)
 +2(c^2-1)\cos\Psi\,M_LM_R .
\end{equation}
The three terms compose, respectively, the sourced parts of the two bubbles, a sourced
part with a homogeneous one, and the homogeneous parts alone. The last term solves the homogeneous recurrences in all four internal labels, so the
recurrences alone would leave its coefficient undetermined; here it is fixed by the branch
sum. Since $A_b=P_b+C_{u_b}(v_b,q)\,M_b$, eqs.~\eqref{eq:P}, \eqref{eq:C}
and~\eqref{eq:M} express $F^{(2)}$ as a finite combination of generalised hypergeometric
functions, Gamma functions and trigonometric factors: the composition requires no
spectral function beyond the one-loop blocks.

\paragraph{Conformally coupled external fields}

For the slow components of conformally coupled external fields, $\Delta_i=(d-2)/2$ and
$\theta_L=\theta_R=0$, so that $\Psi=\vartheta$ and $\omega=0$. Eq.~\eqref{eq:chain-expanded} reduces to
\begin{equation}\label{eq:chain-slow-components}
 F^{(2)}=-2\cos\vartheta\,A_LA_R+2\big(c\sin\vartheta-1\big)\big(A_LM_R+M_LA_R\big)
 +2(c^2-1)\cos\vartheta\,M_LM_R,
\end{equation}
while the one-loop transforms, eq.~\eqref{eq:F-conf} at the labels $(u_b,v_b)$, are
\begin{equation}\label{eq:chain-weighted-one-loop}
 F_b^{\rm conf}=2\cos\vartheta\,A_b+2\big(1-c\sin\vartheta\big)M_b .
\end{equation}
For $\cos\vartheta\ne0$, eq.~\eqref{eq:chain-slow-components} can be written in the factorised form
\begin{equation}\label{eq:chain-factorised}
 F^{(2)}=-\frac{F_L^{\rm conf}F_R^{\rm conf}}{2\cos\vartheta}
 +\frac{2(c-\sin\vartheta)^2}{\cos\vartheta}\,M_LM_R ,
\end{equation}
whose second term, proportional to $M_LM_R$, solves the homogeneous recurrences in all four
internal labels.

At $d=3$, where $\cos\vartheta=0$ and $\sin\vartheta=1$, the factorised form
does not apply, while eq.~\eqref{eq:chain-slow-components} remains regular and gives
\begin{equation}\label{eq:chain-exceptional-phase}
 F^{(2)}=2(c-1)\big(A_LM_R+M_LA_R\big).
\end{equation}
These specialisations rely on $\theta_L=\theta_R=0$; for general external components the full
expression~\eqref{eq:chain-expanded} must be retained.

The transform $F^{(2)}$ is invariant under reflection or interchange of the two internal
labels within either bubble, and under the simultaneous exchange $L\leftrightarrow R$,
$\theta_L\leftrightarrow\theta_R$. For real $d$, principal-series internal labels, real $\lambda$ and real
endpoint phases, $A_b$ and $M_b$ are real on the physical line, and hence so is $F^{(2)}$; for
complex external exponents, complex conjugation also acts on the endpoint phases.

These formulae define the regulated composition. Near $d=4$, renormalisation must be
performed on the dimensionally continued branch sum, with the full $d$-dependent branch
phases retained: the one-loop subdivergences and any remaining local divergence must be
subtracted before $\epsilon\to0$.

\subsection{The boundary correlator of the chain}\label{sec:chain-external-attachment}

The two endpoints of the chain are integrated against the external plane waves exactly as
for the one-loop bubble. Inserting the Mehler--Fock representation of $\Gamma_{\rm SK}$ into
eq.~\eqref{eq:chain-attachment} gives
\begin{equation}\label{eq:chain-attached-spectral}
 \mathcal B^{(2)}_{12|34}(d)=Sg_Lg_\chi g_R\int_0^\infty\dd\lambda\,p_d(\lambda)\,
 F^{(2)}(\lambda)\,\mathcal A_\lambda,
\end{equation}
with $q=i\lambda$ and $\mathcal A_\lambda$ as in eq.~\eqref{eq:attachment-boundary}. No further vertex
integration is needed: the spectral representation of section~\ref{sec:attachment} applies
unchanged, with the one-loop transform $F$ replaced by $F^{(2)}$, and with the Abel
prescription and the continuations in the dimension and in the spectral labels
understood as for one loop. In the convergence domain~\eqref{eq:attachment-convergence-domain},
$F^{(2)}(\lambda)=O(1)$ as $\lambda\to\infty$, by the boundedness of the one-loop transforms, so the
convergence argument following eq.~\eqref{eq:attachment-large-lambda} applies without
modification.
\subsection{Four dimensions}\label{sec:chain-four-dimensional}

Set $d=4-2\epsilon$ and $\kappa=(16\pi^2)^{-1}$. We keep the internal spectral labels and the
coupling prefactor fixed, and continue the external normalisation factors with $d$; the
external exponents are also held fixed, except where the slow conformal continuation
$\Delta_i=(d-2)/2$ is explicitly considered. For generic values of the parameters,
\begin{equation}\label{eq:chain-finite-blocks}
 \widehat A_b(d)=A_b(d)-\frac{\kappa}{\epsilon},\qquad
 \widehat A_{b,4}=\lim_{\epsilon\to0}\widehat A_b(d),\qquad
 M_{b,4}=M_b(d)\big|_{d=4}
\end{equation}
are finite: the residue of the pole of $A_b$ is fixed by the universal coincidence
singularity, while the mixed-branch product has no ultraviolet pole.

We implement the local quartic counterterm of eq.~\eqref{eq:quartic-counterterm}, keeping
the full $d$-dependence of its branch-sum factor $2\cos[\frac\pi2(d-2-\Sigma)]$, not only its value
at $d=4$. For each bubble this amounts to the replacements
\begin{equation}\label{eq:chain-branch-subtraction}
 F_{s,b}\longmapsto F_{s,b}-e^{is\vartheta}\frac{\kappa}{\epsilon},\qquad
 A_b\longmapsto\widehat A_b(d),\qquad M_b\longmapsto M_b .
\end{equation}
To relate this to the subtraction of section~\ref{sec:4d-limit}, consider the transform
$F$ of eq.~\eqref{eq:F-weighted} for the external lines of the chain and the internal labels
of bubble $b$. Since $\Theta-\pi h=\Psi-3\pi/2$, it reads
\begin{equation}\label{eq:chain-contracted-multiplier}
 F_b=2\cos\Psi\,A_b+2\big(\cos\omega-c\sin\Psi\big)M_b ,
\end{equation}
and its pole subtraction is $2\kappa\cos\Psi/\epsilon$. Since
$2\cos[\frac{\pi}{2}(d-2-\Sigma)]=-2\cos\Psi$, the subtraction~\eqref{eq:chain-branch-subtraction} is
generated by a real local quartic counterterm, and implements the physical-phase
prescription discussed after eq.~\eqref{eq:finite-contact}; its finite part need not
coincide with that obtained by subtracting only the fixed residue in
eq.~\eqref{eq:kernel-laurent}.

Substituting eq.~\eqref{eq:chain-branch-subtraction} into eq.~\eqref{eq:chain-expanded} gives,
at finite $\epsilon$,
\begin{align}\label{eq:chain-subtracted-multiplier}
 F^{(2)}_{\rm sub}={}&F^{(2)}+\frac{\kappa}{\epsilon}\big(F_L+F_R\big)-\frac{2\kappa^2\cos\Psi}{\epsilon^2}
 \nonumber\\
 ={}&-2\cos\Psi\,\widehat A_L\widehat A_R+2\big(c\sin\Psi-\cos\omega\big)
 \big(\widehat A_LM_R+M_L\widehat A_R\big)+2(c^2-1)\cos\Psi\,M_LM_R .
\end{align}
The terms linear in $\kappa/\epsilon$ are the insertions of the one-loop counterterm: in
$(\kappa/\epsilon)F_L$ the right bubble is replaced by its counterterm, and conversely. Once these
subdivergences are subtracted, the remaining divergence $2\kappa^2\cos\Psi/\epsilon^2$ is independent
of $\lambda$, hence the transform of a multiple of $\delta_H$, and the last term of the first line
is the corresponding overall local counterterm, a quartic vertex of the external fields.
It cannot be replaced by the two single-bubble insertions, and with it no further pole
remains. Since $\cos\Psi$ is kept with its full dependence on $d$, its expansion generates the
associated simple-pole and finite terms.

Let $\mathcal B_{12|34,b}(d)$ denote the spectral representation~\eqref{eq:chain-attached-spectral}
with $F^{(2)}$ replaced by the one-loop transform $F_b$, that is, the one-loop boundary
correlator with the internal labels of bubble $b$, multiplied by $g_\chi$. The renormalised
boundary correlator of the chain is then
\begin{align}\label{eq:chain-renormalized-attachment}
 \mathcal B^{(2)}_{12|34,\rm ren}
 ={}&\lim_{\epsilon\to0}\Big[\mathcal B^{(2)}_{12|34}(d)
 +\frac{\kappa}{\epsilon}\big(\mathcal B_{12|34,L}(d)+\mathcal B_{12|34,R}(d)\big)
 -Sg_Lg_\chi g_R\frac{2\kappa^2\cos\Psi}{\epsilon^2}\,\mathcal D_d\Big]\nonumber\\
 ={}&Sg_Lg_\chi g_R\int_0^\infty\dd\lambda\,p_4(\lambda)\,F^{(2)}_{\rm ren}(\lambda)\,\mathcal A_{\lambda,4},
\end{align}
with the regulated limit and the continuations understood as in
section~\ref{sec:4d-limit}. Since $\vartheta\to0$ as $d\to4$, $\Psi_4=\Theta_4$, and
\begin{align}\label{eq:chain-finite-general}
F^{(2)}_{\rm ren}(\lambda)
={}&
-2\cos\Theta_4\,
\widehat A_{L,4}\widehat A_{R,4}
+
2\bigl(c\sin\Theta_4-\cos\omega_4\bigr)
\bigl(
\widehat A_{L,4}M_{R,4}
+
M_{L,4}\widehat A_{R,4}
\bigr)
\nonumber\\
&+
2(c^2-1)\cos\Theta_4\,
M_{L,4}M_{R,4},
\qquad
c=\cosh(\pi\lambda).
\end{align}
All quantities on the right-hand side are finite one-loop functions. Here $\widehat 
A_{b,4}=\widehat A_4(u_b,v_b;i\lambda)$ and $M_{b,4}=M_4(u_b,v_b;i\lambda)$.

\paragraph{Evaluation of the finite blocks.} The dependence of the one-loop blocks on 
$d$ is suppressed except when a limit is taken. The finite block $\widehat A_{b,4}$ has an
explicit recurrence representation, which follows from the tail
representation~\eqref{eq:P-sum}. With $t_j=\frac14-j^2$, the value at $d=4$ of the
denominator $(h-1)^2-j^2$ in eq.~\eqref{eq:Su}, define
\begin{equation}\label{eq:chain-finite-tadpole}
 \tau_j=\kappa\Big\{t_j\Big[\gamma_E-1-\log(4\pi)+\psi\Big(\frac32+j\Big)+\psi\Big(\frac32-j\Big)\Big]-1\Big\},
\end{equation}
where $\psi$ is the digamma function. Thus $\tau_j=\operatorname{FP}_{\epsilon=0}T_j-\kappa$. The corresponding finite source is
\begin{equation}\label{eq:chain-finite-source}
 j_{\rm ren}(x;v,q)=4x\Big[2\tau_v+\frac{\tau_x}{t_x}\Big(x^2+v^2-\frac14-q^2\Big)
 -\kappa\Big(\frac14+x^2-v^2-q^2\Big)\Big].
\end{equation}
With a subscript or superscript $4$ denoting evaluation at $d=4$,
\begin{equation}\label{eq:chain-finite-A-series}
 \widehat A_4(u,v;q)=C_u^{(4)}(v,q)\,M_4(u,v;q)
 -\mathcal H_4(u)\sum_{n=0}^{\infty}\frac{j_{\rm ren}(u+2n+1;v,q)}
 {p^-_{u+2n+1,4}(q^2)\,\mathcal H_4(u+2n+2)},
\end{equation}
and $\widehat A_{b,4}=\widehat A_4(u_b,v_b;q)$, with \(q=i\lambda\), $\mathcal H$, $C_u$ and $M$ given in
eqs.~\eqref{eq:H}, \eqref{eq:C} and~\eqref{eq:M}. Indeed, by
eq.~\eqref{eq:chain-block-recurrences}, a constant solves the recurrence with source
$p_x^--p_x^+$, so subtracting $\kappa/\epsilon$ from the tail solution replaces its source by
$j_d(x)-\frac{\kappa}{\epsilon}(p_x^--p_x^+)$; the tail sum of the second term reproduces the constant,
up to a boundary term that vanishes for ${\rm Re}\,d>3$. The finite part of the resulting
source is eq.~\eqref{eq:chain-finite-source}. For generic parameters the summands behave as
$O(n^{-3}\log n)$, and on compact parameter sets near $d=4$ the subtracted tail admits a
common summable majorant, so the limit may be taken term by term. At exceptional divisors
the expression is defined by meromorphic continuation. Equation~\eqref{eq:chain-finite-A-series}
thus evaluates the finite block, with the same identification as the one-loop solution
of section~\ref{sec:solution}.

\paragraph{Conformally coupled external fields.} For the slow conformal continuation $\Delta_i(d)=(d-2)/2$,
define
\begin{equation}\label{eq:chain-finite-one-loop-conf}
 F_{b,\rm ren}^{\rm conf}=\lim_{\epsilon\to0}\Big[F_b^{\rm conf}-\frac{2\kappa\cos\vartheta}{\epsilon}\Big]
 =2\big(\widehat A_{b,4}+M_{b,4}\big).
\end{equation}
These finite one-loop transforms agree with the kernel subtraction of
section~\ref{sec:4d-limit}, since $\cos\vartheta=1+O(\epsilon^2)$. Performing the subtractions in
eq.~\eqref{eq:chain-factorised} before $\epsilon\to0$ gives
\begin{equation}\label{eq:chain-finite-conformal}
 F^{(2)}_{\rm ren}(\lambda)=-\frac12\,F_{L,\rm ren}^{\rm conf}(\lambda)\,F_{R,\rm ren}^{\rm conf}(\lambda)
 +2\cosh^2(\pi\lambda)\,M_{L,4}(\lambda)\,M_{R,4}(\lambda).
\end{equation}
The term $M_LM_R$ thus survives renormalisation, with the finite coefficient
$2\cosh^2(\pi\lambda)$. Importantly, comparing with the results the recent bubble-chain construction of Ref.~\cite{GraefeSachs2026}, we realized that this 
term is absent there. As shown in Appendix~\ref{app:gs}, the difference originates in extending the split representation of the equal-branch propagator from 
positive spectral label to the full real $\tilde\nu'$-axis. Under the even extension, the coefficient of the delta-function contribution is 
$\coth(\pi|\tilde\nu'|)$. 
The resulting term is therefore even and does not vanish against the even spectral density. Restoring this contribution, together with the corresponding 
branch-projection phase, gives the corrected same-branch blocks and reproduces precisely the $M_LM_R$ term in eq.~\eqref{eq:chain-finite-conformal}.

\paragraph{Comparison of subtraction prescriptions.} With the single-bubble counterterms
included, the remaining freedom is, as in section~\ref{sec:4d-limit}, the continuation in
$d$ of the overall counterterm. Write it as $-\kappa^2G(\epsilon)/\epsilon^2$, with
\begin{equation}\label{eq:chain-contact-expansion}
 G(\epsilon)=2Sg_Lg_\chi g_R\cos\Psi(4-2\epsilon)\,\mathcal D_{4-2\epsilon}
 =G_0+\epsilon G_1+\epsilon^2G_2+O(\epsilon^3),
\end{equation}
and let $\mathcal B^{(2)}_{\rm dim}$ and $\mathcal B^{(2)}_{\rm pole}$ denote the results of subtracting the full
expression $\kappa^2G(\epsilon)/\epsilon^2$ and only its pole part $\kappa^2(G_0/\epsilon^2+G_1/\epsilon)$. They differ by
\begin{equation}\label{eq:chain-finite-contact-difference}
 \mathcal B^{(2)}_{\rm dim}-\mathcal B^{(2)}_{\rm pole}=-\kappa^2G_2
 =-4Sg_Lg_\chi g_R\,\kappa^2\,\partial_d^2\big[\cos\Psi(d)\,\mathcal D_d\big]\Big|_{d=4}.
\end{equation}
For fixed external exponents, $\Psi(d)=\frac\pi2(d-\Sigma)$, and the difference becomes
\begin{equation}\label{eq:chain-finite-contact-fixed}
 \kappa^2Sg_Lg_\chi g_R\Big[\pi^2\cos\Theta_4\,\mathcal D_4
 +4\pi\sin\Theta_4\,\partial_d\mathcal D_d\big|_{d=4}
 -4\cos\Theta_4\,\partial_d^2\mathcal D_d\big|_{d=4}\Big];
\end{equation}
along the slow conformal continuation it is
$\kappa^2Sg_Lg_\chi g_R\big[\pi^2\mathcal D_4-4\,\partial_d^2\mathcal D_d|_{d=4}\big]$, where the derivatives include
the $d$-dependence of the external exponents.

These formulas assume that the contact integral is regular at $d=4$; if it is not, its full
Laurent expansion must be retained. Finite changes in the one-loop counterterms also
modify the one-loop counterterm contributions to the chain, while a finite overall
counterterm shifts the result by a multiple of $\mathcal D_4$. Any renormalisation scale or
$d$-dependent coupling convention must be retained consistently throughout the same
expansion.

\subsection{Fourier representation}\label{sec:chain-fourier}

With the conventions of section~\ref{sec:fourier-representation}, and $Q=|\bm p_1+\bm p_2|$, the
spectral representation~\eqref{eq:chain-attached-spectral} becomes
\begin{align}\label{eq:chain-fourier-attached}
 \widetilde{\mathcal B}^{(2)}_{12|34}={}&(2\pi)^{d-1}\,\delta^{(d-1)}\Big(\sum_i\bm p_i\Big)\,
 Sg_Lg_\chi g_R\prod_{i=1}^4\mathcal N_i(p_i)\,\frac{2}{\pi^2}\int_0^\infty\dd\lambda\,
 \lambda\sinh(\pi\lambda)\,F^{(2)}(\lambda)\nonumber\\
 &\times\mathcal J_{\Delta_1\Delta_2\Delta}(p_1,p_2,Q)\,\mathcal J_{\Delta_3\Delta_4\widetilde\Delta}(p_3,p_4,Q),
\end{align}
with $\Delta=h+i\lambda$, $\widetilde\Delta=h-i\lambda$, the normalisations $\mathcal N_i$ of eq.~\eqref{eq:fourier-A} and the
triple-$K$ integrals~\eqref{eq:fourier-triple-K}. All branch dependence is in $F^{(2)}$, and
the Fourier transformation introduces no additional phase. Renormalisation replaces
$F^{(2)}$ by $F^{(2)}_{\rm sub}$ before $\epsilon\to0$, with the dimensional continuation of
section~\ref{sec:chain-four-dimensional}; the one-loop counterterms and the overall
counterterm are transformed in the same way, the latter through the contact
integral~\eqref{eq:fourier-contact}.

For the slow components of conformally coupled external fields at $d=4$, with
$E_L=p_1+p_2$, $E_R=p_3+p_4$ and the function $\mathcal I_\lambda(E,Q)$ of
eq.~\eqref{eq:fourier-Laplace-K}, the renormalised result is
\begin{equation}\label{eq:chain-fourier-four}
 \widetilde{\mathcal B}^{(2)}_{12|34,\rm ren}=(2\pi)^3\,\delta^{(3)}\Big(\sum_i\bm p_i\Big)\,
 \frac{Sg_Lg_\chi g_R}{8\pi^2p_1p_2p_3p_4}\int_0^\infty\dd\lambda\,\lambda\sinh(\pi\lambda)\,
 F^{(2)}_{\rm ren}(\lambda)\,\mathcal I_\lambda(E_L,Q)\,\mathcal I_\lambda(E_R,Q),
\end{equation}
with $F^{(2)}_{\rm ren}$ given by eq.~\eqref{eq:chain-finite-conformal}. A further finite local
counterterm shifts the coefficient of the delta function by a multiple of
$1/[16p_1p_2p_3p_4(E_L+E_R)]$, in accordance with eq.~\eqref{eq:fourier-contact-four}.

Equations~\eqref{eq:chain-fourier-attached} and~\eqref{eq:chain-fourier-four} are understood
with the regulated spectral prescription, and under the convergence conditions of the
triple-$K$ integrals stated in section~\ref{sec:fourier-representation}. They hold for $Q>0$;
the limit $Q\to0$ cannot be obtained by direct substitution, and requires the separate
analysis mentioned at the end of section~\ref{sec:fourier-representation}.

\section{Conclusions}
\label{sec:conclusions}

In this work, we have developed a framework for perturbative quantum field theory in de Sitter space based on three interconnected ingredients: maximal analyticity, the branchwise deformation of Schwinger--Keldysh vertex integrations onto imaginary Lobachevskij cycles, and finite spectral recurrence relations. Together, these ingredients turn the evaluation of loop amplitudes into a recursive problem in the spectral parameters of the internal propagators and provide a direct route to the corresponding boundary correlation functions.

The contour deformation proved here for the bubble boundary diagram, defined on the real contour by its tilt, allows one to retain the maximally analytic de Sitter propagators throughout the calculation, without decomposing them into AdS building blocks. The resulting recurrence relations are therefore derived directly at the level of de Sitter amplitudes and preserve their analytic structure. The Lobachevskij cycles are not merely a technical device: they provide a natural setting in which the algebraic properties of the propagators become transparent and can be exploited systematically.

The main result is the construction of finite spectral recurrences relating amplitudes whose internal spectral parameters differ by discrete shifts. For products of Wightman functions, these recurrences are homogeneous. For equal-branch causal products, they acquire a source term determined by the local singularity structure of the propagators. For these sourced recurrences, we construct an explicit meromorphic solution and prove that complete-strip analyticity, spectral
contiguity, simultaneous large-mass behaviour and ultraviolet normalisation uniquely identify it with the chronological and antichronological defining transforms.
The same relations determine the loop amplitudes recursively and, after attachment to the external boundary points, become differential--difference equations for the corresponding boundary correlation functions.

We have illustrated the method through several nontrivial applications. For the one-loop bubble, the recurrence admits an explicit meromorphic solution and yields the de Sitter K\"all\'en--Lehmann representation as a direct consequence of the propagator algebra. The attachment procedure then produces the associated position-space and Fourier-space boundary correlators, with dimensional regularisation and local counterterms treated at the level of the full Schwinger--Keldysh branch sum. For the two-bubble, two-loop chain, the composition is determined entirely by the one-loop spectral blocks. In particular, the complete branch sum fixes a finite contribution quadratic in the homogeneous blocks, rather than leaving it as an undetermined homogeneous term.

Finally, the two-loop analysis clarifies a discrepancy that we have identified with the recent bubble-chain construction of Ref.~\cite{GraefeSachs2026}. 
The complete Schwinger--Keldysh branch sum fixes the additional homogeneous $M_LM_R$ contribution in eq.~\eqref{eq:chain-finite-conformal}. Its relation to the
treatment of the equal-branch split representation in Ref.~\cite{GraefeSachs2026} is derived in detail in Appendix~\ref{app:gs}.

These results suggest that finite spectral recurrences provide a natural organising principle for perturbation theory in de Sitter space. The examples studied 
here should be viewed not only as explicit calculations, but also as evidence for a broader framework in which loop amplitudes are characterised by their 
behaviour under discrete shifts of the internal spectral parameters. Extending this framework to more general multiloop topologies, higher-point boundary
correlators and interacting fields with spin constitutes a natural direction for future work.

 \acknowledgments

AI tools (notably GPT-6 Astra and Claude Fable 5.1) were used
to assist with analytical calculations and with  consistency
and numerical checks. The authors take full responsibility
for the work.


}
\newpage 

\appendix
\section{Details of the recurrence solution}\label{app:physical-conditions}

We collect here the analytic estimates and algebraic identities used in
section~\ref{sec:solution}. Unless otherwise stated, we work at real
\[
 3<d<4,\qquad h=\frac{d-1}{2},\qquad a=h-1\in\big(0,\tfrac12\big),
\]
with generic principal-series spectator labels. The identities obtained below extend
meromorphically, while the estimates are used only in the domains in which they are
proved.

\subsection{Analyticity, growth and the K\"all\'en--Lehmann residue}\label{app:KL-uniqueness}

Consider the Mehler--Fock transform~\eqref{eq:Wightman-transform}, and the Mehler--Dirichlet
representation
\begin{align}\label{eq:app-KL-Mehler}
 &w_u(\cosh\rho)=\frac{\sqrt{2/\pi}\,\Gamma(h+u)\Gamma(h-u)}{2(2\pi)^{d/2}\,\Gamma(h)}\,
 (\sinh\rho)^{-(d-2)}\int_0^\rho f_\rho(t)\cosh(ut)\,\dd t
\end{align}
where $f_\rho(t) =(\cosh\rho-\cosh t)^a$ is absolutely continuous on $[0,\rho]$. For $u=x+iy\ne0$, one
integration by parts gives
\[
 \Big|\int_0^\rho f_\rho(t)\cosh(ut)\,\dd t\Big|\le\frac{e^{|x|\rho}}{|u|}\,(\cosh\rho-1)^a ,
\]
and with uniform Stirling asymptotics on $|x|\le1$,
\begin{equation}\label{eq:app-KL-w-bound}
 \big|w_{x+iy}(\cosh\rho)\big|\le C|y|^{d-3}e^{-\pi|y|}\,e^{|x|\rho}\,(\sinh\rho)^{-(d-2)}(\cosh\rho-1)^a .
\end{equation}

For fixed $v$ and $q$, the bound reduces the integral in eq.~\eqref{eq:Wightman-transform} to
$|y|^{d-3}e^{-\pi|y|}$ times a convergent radial integral (its integrand behaves as $\rho^{d-2}$ at
the origin and as $(1+\rho)^2e^{-(h-1)\rho}$ at infinity). Hence
\begin{equation}\label{eq:app-KL-F-bound}
 \big|F_0(x+iy,v;q)\big|\le C|y|^{d-3}e^{-\pi|y|},\qquad |x|\le1,
\end{equation}
for $|y|$ sufficiently large, uniformly in $x$. Since $|\cosh(ut)|\le\cosh(t\,{\rm Re}\,u)$, the integrand of eq.~\eqref{eq:Wightman-transform}
decays at large $\rho$ as $e^{(|{\rm Re}\,u|-h)\rho}$, uniformly on compact subsets of the strip
$|{\rm Re}\,u|<h$, so $F_0(u,v;q)$ is holomorphic there. The same holds with $u$ and $v$
exchanged.

For the Gamma product~\eqref{eq:N}, uniform Gamma asymptotics give
\begin{equation}\label{eq:app-KL-N-bound}
 \frac{1}{|N(x+iy,v,q)|}\le C|y|^{6-2d}e^{2\pi|y|},\qquad |x|\le1 .
\end{equation}
Hence the quotient $E(u,v;q)=F_0(u,v;q)/N(u,v,q)$ satisfies
\begin{equation}\label{eq:app-KL-E-bound}
 |E(x+iy,v;q)|\le C|y|^{3-d}e^{\pi|y|}.
\end{equation}
Since $1/N$ is entire, $E$ is holomorphic on the strip $|{\rm Re}\,u|<h$, which contains a full period because $h > 1$; since $F_0$ and $N$
satisfy the same homogeneous recurrence~\eqref{eq:homogeneous}, $E$ is two-periodic in $u$.
It therefore extends to an entire two-periodic function. Its Fourier coefficients
\begin{equation}
 e_n(v;q)=\frac12\int_{-1}^{1}E(x+iy,v;q)\,e^{-i\pi n(x+iy)}\,\dd x
\end{equation}
are independent of $y$, by Cauchy's theorem on a rectangle of width two, whose vertical sides
cancel by periodicity. By eq.~\eqref{eq:app-KL-E-bound},
$|e_n|\le C|y|^{3-d}e^{\pi|y|+\pi ny}$. For $n\ge1$ we let $y\to-\infty$: the bound vanishes
exponentially for $n\ge2$, and as $|y|^{3-d}$ for $n=1$, since $d>3$. For $n\le-1$ we let
$y\to+\infty$ in the same way. Hence $e_n=0$ for $n\ne0$, so $E$ does not depend on $u$, and
repeating the argument in $v$ gives
\begin{equation}\label{eq:app-KL-KN}
 F_0(u,v;q)=K(q)\,N(u,v,q).
\end{equation}

It remains to determine $K(q)$, for which one pole suffices. At large $\rho$, the integrand of
eq.~\eqref{eq:Wightman-transform} is a sum of exponentials given by the large-distance
powers~\eqref{eq:falloffs} of its three factors, the Mehler--Fock kernel being proportional
to $w_q$ by eq.~\eqref{eq:w-legendre}. The term proportional to $a_{h+u}a_{h+v}\,e^{-(h+u+v-q)\rho}$
produces a simple pole at $q=h+u+v$, where no other term is singular. With the measure
$\Omega_{d-1}(\sinh\rho)^{d-1}\dd\rho$, its residue is
\begin{equation}\label{eq:app-KL-integral-residue}
 \underset{q=h+u+v}{\rm Res}\,F_0(u,v;q)
 =-\frac{\Gamma(h+u+v)\,\Gamma(h+u)\Gamma(h+v)\,\Gamma(-u)\Gamma(-v)}
 {16\pi^{(d+3)/2}\,\Gamma(2h+u+v)} .
\end{equation}
$N$ has the same pole, with residue
\begin{equation}\label{eq:app-KL-N-residue}
 \underset{q=h+u+v}{\rm Res}\,N=-2\,\Gamma(h)\,\Gamma(h+u+v)\,\Gamma(h+u)\Gamma(h+v)\,
 \Gamma(-u)\Gamma(-v)\,\Gamma(-u-v).
\end{equation}
Since $F_0=K(q)N$, the ratio of the residues gives
\begin{equation}\label{eq:app-KL-K}
 K(q)=\frac{1}{32\pi^{(d+3)/2}\,\Gamma(h)\,\Gamma(h+q)\Gamma(h-q)} .
\end{equation}
With eq.~\eqref{eq:N}, this gives $F_0=M$.

\subsection{Properties of the particular solution $P(u,v;q)$}\label{app:particular-identities}

\paragraph{Recurrence in the second label.} The series~\eqref{eq:P-sum} was constructed to
solve the sourced recurrence in $u$, and is not symmetric in $u$ and $v$. It nevertheless
satisfies the recurrence in $v$,
\begin{equation}\label{eq:app-second-sourced-recurrence}
 p_v^-(q^2)\,P(u,v+1;q)-p_v^+(q^2)\,P(u,v-1;q)=j_d(v),
\end{equation}
with the source~\eqref{eq:jd} for the roles of $u$ and $v$ exchanged. Dividing
eq.~\eqref{eq:P-sum} by $M$, the two-periodicity of $M/\mathcal H$ replaces
$\mathcal H(u)/\mathcal H(u+2n+2)$ by $M(u)/M(u+2n+2)$, so that
\begin{equation}\label{eq:A-tail-M}
 \frac{P}{M}=-\sum_{n\ge0}A(u+2n+1,v;q),\qquad A(u,v;q)=\frac{j_d(u)}{p_u^-(q^2)\,M(u+1,v;q)} .
\end{equation}
With $B(u,v;q)=j_d(v)/[p_v^-(q^2)M(u,v+1;q)]$, Gamma and tadpole shifts give
\begin{equation}\label{eq:A-second-label-certificate}
 A(u,v+1;q)-A(u,v-1;q)=B(u+1,v;q)-B(u-1,v;q),
\end{equation}
an identity that becomes rational after multiplication by $M(u+1,v+1;q)$, and holds
separately for the coefficients of $T_u$ and $T_v$. Summed over $n$, the right-hand side
telescopes, with an endpoint $O(N^{-d})+O(N^{4-2d})$ that vanishes for ${\rm Re}\,d>2$:
\[
 \frac{P(u,v+1;q)}{M(u,v+1;q)}-\frac{P(u,v-1;q)}{M(u,v-1;q)}=B(u,v;q).
\]
Multiplying by $p_v^-(q^2)M(u,v+1;q)=p_v^+(q^2)M(u,v-1;q)$ gives
eq.~\eqref{eq:app-second-sourced-recurrence}.

\paragraph{Contiguity in the labels.} The recurrences shift one label at a time, and leave
the solution undetermined up to $M$ times functions periodic in each label separately. The
contiguity relations instead mix the labels, tying a shift of $u$ to a shift of $v$ or of the
Mehler--Fock label $q$, and so correlate these periodic factors. In
subsection~\ref{app:sourced-uniqueness} we show that the transforms $F_r$ coincide with the
full solution~\eqref{eq:result}. Their difference solves the homogeneous recurrences, so it is
$M$ times functions periodic in each label, and the contiguity relations restrict these
functions, provided that both the transforms and the full solution satisfy them. We prove
this here.

For the defining integrals, the relations follow from the contiguity
relation~\eqref{eq:contiguity}, which holds at every argument, including the boundary values
on the cut. Both propagators of a branch kernel have the same argument, $z$ for the mixed
kernel and $-z\pm i0$ for the equal-branch kernels, so eq.~\eqref{eq:contiguity} applied to
either of them multiplies the kernel by the same factor, $z$ or $-z$. Multiplication by $z$ is
in turn traded for a shift of the Mehler--Fock label, since the kernel, normalised as
$\varphi_q=w_q/w_q(1)$, satisfies
\begin{equation}\label{eq:A-q-contiguity-spherical}
 \sum_{\sigma=\pm1}b_\sigma(q)\,\varphi_{q+\sigma}(z)=z\,\varphi_q(z),\qquad b_\sigma(q)=\frac{q+\sigma h}{2q}.
\end{equation}
Hence the contiguity sums of each transform in $u$, $v$ and $q$ coincide, up to the sign of
the argument:
\begin{align}
 \sum_{\sigma=\pm1}a_\sigma(u)\,F_0(u+\sigma,v;q)=\sum_{\sigma=\pm1}a_\sigma(v)\,F_0(u,v+\sigma;q)
 &=\sum_{\sigma=\pm1}b_\sigma(q)\,F_0(u,v;q+\sigma),\label{eq:A-q-contiguity-mixed}\\
 \sum_{\sigma=\pm1}a_\sigma(u)\,F_\pm(u+\sigma,v;q)=\sum_{\sigma=\pm1}a_\sigma(v)\,F_\pm(u,v+\sigma;q)
 &=-\sum_{\sigma=\pm1}b_\sigma(q)\,F_\pm(u,v;q+\sigma),\label{eq:A-q-contiguity-equal}\, 
\end{align}
 the two chains being the transforms of $z\mathcal K_0$ and of $-z\mathcal K_r$.

The full solution~\eqref{eq:result} is a combination of $P$, $M$, $\cos(\pi q)M$ and $C_uM$, with
coefficients independent of the labels. Since $M=F_0$, $M$ satisfies
eq.~\eqref{eq:A-q-contiguity-mixed}; we denote by $Z$ the common value of its three sums. Since
$C_{u\pm1}(v,q)=C_u(v\pm1,q)=-C_u(v,q\pm1)$ and $\cos\pi(q\pm1)=-\cos(\pi q)$, multiplying $M$ by
either factor reverses the sign of the last member of eq.~\eqref{eq:A-q-contiguity-mixed}, so
$\cos(\pi q)M$ and $C_uM$ satisfy eq.~\eqref{eq:A-q-contiguity-equal}. It remains to show that $P$
satisfies the two equalities of eq.~\eqref{eq:A-q-contiguity-equal}.

 To prove them, we write $P=ME$. Dividing the recurrences for $P$ by $p_u^-M(u+1,v;q)$ and
$p_v^-M(u,v+1;q)$, and using eq.~\eqref{eq:homogeneous} for $M$, gives two differences of $E$ in
closed form,
\begin{equation}\label{eq:A-E-increments}
 E(u+1,v)-E(u-1,v)=A(u,v;q),\qquad E(u,v+1)-E(u,v-1)=B(u,v;q).
\end{equation}
A third combination of values of $E$ is written, by eq.~\eqref{eq:A-tail-M}, as a sum over $n$
whose summand is the difference of an explicit function at two neighbouring points, so the
sum telescopes and the combination too has a closed form. The contiguity relation then
reduces to a finite identity.

\emph{First equality.} With $P=ME$, eq.~\eqref{eq:A-E-increments} reduces it to
\begin{equation}\label{eq:A-uv-reduced}
 Z\,\big[E(u+1,v)-E(u,v+1)\big]=a_-(u)\,M(u-1,v;q)\,A(u,v;q)-a_-(v)\,M(u,v-1;q)\,B(u,v;q).
\end{equation}
By eq.~\eqref{eq:A-tail-M}, the difference on the left is a sum over $n$ which telescopes, since
\begin{equation}\label{eq:A-first-kind-certificate}
 A(u+1,v+1;q)-A(u+2,v;q)=D_1(u,v;q)-D_1(u+2,v;q),
\end{equation}
with
\begin{equation}\label{eq:A-D1}
 D_1(u,v;q)=\frac{(d-2)\big[T_v/(v-a)-T_u/(u-a)\big]}{\big[(a-u+v)^2-q^2\big]\,M(u+1,v;q)};
\end{equation}
hence $E(u+1,v)-E(u,v+1)=D_1(u,v;q)$, the term at infinity vanishing for $3<d<4$. Both
eq.~\eqref{eq:A-first-kind-certificate} and eq.~\eqref{eq:A-uv-reduced} reduce to polynomial
identities by $\Gamma(x+1)=x\Gamma(x)$ and $({x-a})T_{x+1}=({x+h})T_x$, which hold by direct expansion.

\emph{Second equality.} Written for $P$, the second equality of
eq.~\eqref{eq:A-q-contiguity-equal} has four terms. By eq.~\eqref{eq:A-tail-M}, each of them is a
sum over the same lattice $t=u+2n+1$, the shifts $u\pm1$ moving the summand to $t\pm1$ and the
shifts $q\pm1$ changing its Mehler--Fock label. Expressing all $M$ factors through $M(t,v;q)$
and $M(u+1,v;q)$, and the tadpoles at $t\pm1$ through $T_t$, the difference of the two sides
becomes
\[
 -M(u+1,v;q)\sum_{n\ge0}\frac{f(t_n)}{M(t_n,v;q)},\qquad f(t)=f_U(t)\,T_t+f_V(t)\,T_v,
\]
with $f_U$ and $f_V$ rational. The summand is a difference,
\begin{equation}\label{eq:A-q-contiguity-certificate}
 \frac{f(t)}{M(t,v;q)}=\mathcal B(t)-\mathcal B(t+2),\qquad
 \mathcal B(t)=\frac{Q_U(t)\,T_t+Q_V(t)\,T_v}{M(t,v;q)},
\end{equation}
with
\begin{equation}\label{eq:A-QUV}
 Q_V(t)=-\frac{2a(d-2)(t-u-1)\,L_V(t)}{u\,p_u^+(q^2)\,p_{t-1}^-(q^2)},\qquad
 Q_U(t)=-\frac{a(d-2)(t-u-1)\,L_U(t)}{u\,p_u^+(q^2)\,p_{t-1}^-(q^2)\,(t+a-1)(t+a)},
\end{equation}
where $L_V(t)=2u(t-1)(u^2+v^2-a^2-q^2)+(a^2-v^2-q^2)^2-4v^2q^2-u^4+4u^2v^2$ and, with
$X_0=a^2+q^2$, $Y_0=a^2q^2$, $U_0=u^2+v^2-X_0$ and $s=t-1$, $L_U(t)=c_0+c_1s+c_2s^2+c_3s^3$,
\begin{align*}
 c_0&=(X_0-v^2)\big\{(X_0-u^2-v^2)^2-4u^2v^2\big\}-4Y_0(X_0-v^2-2u^2),\\
 c_1&=-4u\big\{(X_0-v^2)(X_0+v^2-u^2)-2Y_0\big\},\\
 c_2&=-U_0^2+4(Y_0-u^2v^2),\\
 c_3&=-4uU_0 .
\end{align*}
Eq.~\eqref{eq:A-q-contiguity-certificate} holds identically: after reduction with
$\Gamma(x+1)=x\Gamma(x)$ and $T_{x+1}=\frac{x+h}{x-a}T_x$, it becomes a polynomial identity, which
holds by direct expansion. The sum therefore reduces to its boundary terms,
$\mathcal B(u+1)-\lim_{t\to\infty}\mathcal B(t)$, and both vanish: the first because $Q_U$ and $Q_V$
contain the factor $t-u-1$, the second because $\mathcal B(t)=O(t^{2-d})+O(t^{4-2d})$ on the
lattice, which tends to zero for ${\rm Re}\,d>2$. The difference of the two sides is
therefore zero, which proves the second equality for $P$.  Together with the first equality,
and with the relations for $\cos(\pi q)M$ and $C_uM$ shown above, it follows that the full
solution~\eqref{eq:result} satisfies eq.~\eqref{eq:A-q-contiguity-equal}.

\paragraph{Poles of $P$.} In the strip $|{\rm Re}\,u|<h$, $P$ has only two poles, at $u=a$ and
$u=a-1$, both simple, with residues
\begin{align}
 \underset{u=a}{\rm Res}\,P(u,v;q)&=-\frac{M(a,v;q)}{\pi}\,\big[\cos(\pi v)+\cos(\pi q)\big],
 \label{eq:app-first-P-residue}\\
 \underset{u=a-1}{\rm Res}\,P(u,v;q)&=\frac{M(a-1,v;q)}{\pi}\,\big[\cos(\pi v)-\cos(\pi q)\big].
 \label{eq:app-second-P-residue}
\end{align}

\paragraph{The odd part.} Define
\begin{equation}
 D(u,v;q)=\frac{P(u,v;q)-P(-u,v;q)}{M(u,v;q)}-\big[C_{-u}(v,q)-C_u(v,q)\big].
\end{equation}
The residues above cancel its possible poles, so $D$ is entire, odd and two-periodic in $u$,
and its growth leaves only
\begin{equation}\label{eq:A-D-mode}
 D(u,v;q)=\beta(v,q)\,\sin(\pi u).
\end{equation}
The product $MD$ satisfies the first equality of eq.~\eqref{eq:A-q-contiguity-equal}: $P(u,v;q)$
does, as just shown; $P(-u,v;q)$ does, since $a_\sigma(-u)=a_{-\sigma}(u)$; and $C_{\pm u}M$ do. With
$\sin[\pi(u\pm1)]=-\sin(\pi u)$, this gives
\[
 \sum_{\sigma=\pm1}a_\sigma(v)\,M(u,v+\sigma;q)\,\big[\beta(v+\sigma,q)+\beta(v,q)\big]=0 .
\]
Evenness, two-periodicity and the strip bound leave $\beta(v,q)=b_0(q)+k(q)\cos(\pi v)$, for which
the last equation reads $2Zb_0(q)=0$; hence $b_0=0$, since $Z$ is not identically zero. The
second equality of eq.~\eqref{eq:A-q-contiguity-equal} then makes $k$ one-periodic in $q$, since
$k$ does not depend on $u$ and $v$, and its exponential type, smaller than $2\pi$, excludes every
non-constant Fourier mode. Hence
\begin{equation}\label{eq:A-B-mode}
 \beta(v,q)=k(d)\cos(\pi v).
\end{equation}
The coefficient $k(d)$ is fixed at $u=1/2$, $v=a$, $q=3/2$, where $P$ and $M$ have a common
pole. For $0<a<1/2$,
\[
 \underset{q=3/2}{\rm Res}\,M=\frac{\Gamma(a)}{16\pi^{a+2}(a+3/2)},\qquad
 \underset{q=3/2}{\rm Res}\,P=-\frac{\Gamma(a)}{8\pi^{a+2}(a+3/2)},
\]
while $P(-1/2,a;q)$ is regular there. The limiting odd-part ratio is therefore $-2$, equal to
the limit of $C_{-1/2}-C_{1/2}$, so that $k(d)\cos(\pi a)=0$ and $k(d)=0$. Consequently
\begin{equation}\label{eq:app-complete-odd-identity}
 \frac{P(u,v;q)-P(-u,v;q)}{M(u,v;q)}=C_{-u}(v,q)-C_u(v,q).
\end{equation}

\subsection{The complete solution: uniqueness.} \label{app:sourced-uniqueness} 
We now show that the full solution~\eqref{eq:result} has the properties used in the
uniqueness argument below.

\begin{proposition}[Properties of the full solution]\label{prop:A-candidate-admissibility}
Let $3<d<4$, and let
\begin{equation}\label{eq:A-candidate-definition}
 \widehat F_r(u,v;q)=s_r(d)\,P(u,v;q)+e^{-ir\pi h}\big[\cos(\pi q)-irC_u(v,q)\big]M(u,v;q),
 \qquad r=\pm1,
\end{equation}
denote the right-hand side of eq.~\eqref{eq:result}. Then $\widehat F_r$ is holomorphic in the
strips $|{\rm Re}\,u|<h$ and $|{\rm Re}\,v|<h$ and, with the labels not varied in each bound fixed
in compact subsets of the principal series,
\begin{align}
 \widehat F_r(x+iy,v;q)&=O\big((1+|y|)^{d-4}\big),\qquad |x|\le1,\label{eq:A-candidate-u-bound}\\
 \widehat F_r(u,x+iy;q)&=O\big((1+|y|)^{d-4}\big),\qquad |x|\le1,\label{eq:A-candidate-v-bound}\\
 \widehat F_r(u,v;x+iy)&=O\big((1+|y|)^{d-4}\big),\qquad |x|\le1,\label{eq:A-candidate-q-bound}\\
 \widehat F_r(i\mu,i\mu;q)&=O(\mu^{d-4})=o(\mu^{d-3}),\qquad \mu\to+\infty.\label{eq:A-candidate-simultaneous-bound}
\end{align}
It satisfies the two sourced recurrences and the contiguity
relations~\eqref{eq:A-q-contiguity-equal}, is even in $u$ and in $v$ and symmetric under
$u\leftrightarrow v$, and its leading ultraviolet coefficient coincides with that of the
transform $F_r$.
\end{proposition}

\begin{proof}
\emph{Recurrences, contiguity and evenness.} The sourced recurrences follow from the
construction of the tail in eq.~\eqref{eq:P-sum}, which solves
eq.~\eqref{eq:sourced-recurrence}, and from eq.~\eqref{eq:app-second-sourced-recurrence}. The
contiguity relations~\eqref{eq:A-q-contiguity-equal} were established in
subsection~\ref{app:particular-identities}. By the odd-part
identity~\eqref{eq:app-complete-odd-identity}, $\widehat F_r$ is even in $u$; it is even in $v$,
since $P$, $C_u$ and $M$ depend on $v$ only through $v^2$ and $T_v=T_{-v}$.

\emph{Holomorphy.} In the strip $|{\rm Re}\,u|<h$, $P$ has only the two simple poles at $u=a$ and
$u=a-1$, with the residues~\eqref{eq:app-first-P-residue}
and~\eqref{eq:app-second-P-residue}. They cancel the residues of $C_uM$ at the same points,
where $\sin\pi(h-u)$ vanishes, in the combination~\eqref{eq:A-candidate-definition}. In $v$,
the terms of the tail are holomorphic for $|{\rm Re}\,v|<h$, where it converges normally.
Hence $\widehat F_r$ is holomorphic in both strips.

\emph{Bounds.} Put $a=h-1$ and let $Y$ be the large imaginary part.
Reflection of the tadpole gives
\[
 T_z=\frac{\Gamma(1-d/2)}{(4\pi)^{d/2}}
 \frac{\cos\pi z}{\sin\pi(h-z)}
 \frac{\Gamma(h+z)}{\Gamma(1-h+z)}.
\]
On the lattice $z=u+2n+1$, the trigonometric quotient is independent
of $n$. For $u=x+iY$, $|x|\le1$, the tail in eq.~\eqref{eq:P-sum}
is consequently bounded by
$CY^{4a}\sum_{n\ge0}(n+Y)^{d-5-4a}=O(Y^{d-4})$.
For a large imaginary part in $v$ or $q$, its weight is bounded by
\[
 \left|\frac{\mathcal H(u)}{p_z^-(q^2)\mathcal H(z+1)}\right|
 \le \frac{CY^{4a}}{((n+1)^2+Y^2)^{2a+2}}.
\]
The source is bounded by
$C\{(n+1)Y^{d-2}+(n+1)^{d-1}+Y^2(n+1)^{d-3}\}$
in the $v$ strip; in the $q$ strip the first term is replaced by
$C(n+1)$. Summation gives $P=O(Y^{d-4})$ in both cases.
Stirling's formula gives the same bounds for the homogeneous
completion, proving eqs.~\eqref{eq:A-candidate-u-bound}--\eqref{eq:A-candidate-q-bound}.

On the diagonal $u=v=i\mu$, retain the cancellation between the
two tadpole terms: for $n\le\mu$,
$T_{i\mu+2n+1}/T_{i\mu}=[(i\mu+2n+1)/(i\mu)]^{d-2}
[1+O(\mu^{-1})]$, and hence $j_d(i\mu+2n+1)=O((n+1)\mu^{d-2})$.
The tail weight is bounded by
$C\mu^{2a}/[(n+1)^{2a+2}(n+\mu)^{2a+2}]$.
Thus this part is $O(\mu^{d-4})$; the remaining tail is
$O(\mu^{-1})$, and the homogeneous completion is exponentially
small. This proves eq.~\eqref{eq:A-candidate-simultaneous-bound}.
All bounds are uniform on compact sets of the generic spectator
parameters specified in the proposition.

\emph{Ultraviolet coefficient.} Set $q=i\lambda$ and
$R_u=\cos(\pi u)/\sin[\pi(h-u)]$. Rescaling the tail by
$t=2n/\lambda$ gives
\[
 \lambda^{4-d}P(u,v;i\lambda)\longrightarrow
 \frac{(d-2)\Gamma(1-d/2)R_u}{(4\pi)^{d/2}}
 \int_0^\infty\frac{t^{d-3}\,\dd t}{(1+t^2)^{d-2}}
 =-C_dR_u.
\]
The preceding bounds control the scaled contributions from
$n<\delta\lambda$ and $n>L\lambda$ by
$C\delta^{d-2}+o(1)$ and $CL^{2-d}$, respectively, justifying
this limit. Independently, Stirling's formula gives
\[
 \lambda^{4-d}\cosh(\pi\lambda)M\longrightarrow m_d,
 \qquad m_d=\frac{2^{3-2d}\pi^{(3-d)/2}}{\Gamma(h)}
 =-C_d\cos(\pi h).
\]
With $\theta=\pi(h-u)$, one also has
$\lambda^{4-d}C_uM\to-m_d\cot\theta$. Substitution in
\eqref{eq:A-candidate-definition}, using
$\cos\pi u-\cos\pi h\cos\theta=\sin\pi h\sin\theta$, gives
$\lambda^{4-d}\widehat F_r\to e^{-ir\pi(d-2)}C_d$.
This fixes the candidate's ultraviolet coefficient independently
of its identification with the defining transform.

\emph{Symmetry.} Let $S(u,v)=\widehat F_r(u,v;q)-\widehat F_r(v,u;q)$. By
eq.~\eqref{eq:app-second-sourced-recurrence}, $\widehat F_r(v,u;q)$ satisfies the same two
sourced recurrences as $\widehat F_r(u,v;q)$, so $S$ solves the homogeneous ones and $S=M\,e$,
with $e$ two-periodic in $u$ and $v$ and entire, since $\widehat F_r$ is holomorphic in both
strips. By the bounds~\eqref{eq:A-candidate-u-bound} and~\eqref{eq:A-candidate-v-bound} and
eq.~\eqref{eq:app-KL-N-bound}, $|e|\le C|y|^{2-d}e^{2\pi|y|}$ in each strip, which, as in
subsection~\ref{app:KL-uniqueness}, excludes the Fourier modes with $|n|\ge2$; with evenness,
this leaves the modes $1$, $\cos(\pi u)$, $\cos(\pi v)$ and $\cos(\pi u)\cos(\pi v)$. The first
equality of eq.~\eqref{eq:A-q-contiguity-equal}, which both functions satisfy, removes
$\cos(\pi u)$ and $\cos(\pi v)$. The two remaining modes are symmetric, while $e$ is
antisymmetric, because $S$ is and $M$ is symmetric. Hence $e=0$, and $\widehat F_r$ is symmetric
under $u\leftrightarrow v$.
\end{proof}

For real $3<d<4$ and fixed generic principal-series $v,q$, $F_r(u,v;q)$ is holomorphic for $|\operatorname{Re}u|<h$ and satisfies \begin{equation}\label{eq:app-mass-strip} F_r(x+iy,v;q) = O(|y|^{d-3}), \qquad |x|\le1, \end{equation} uniformly in $x$. This follows from the Mehler--Dirichlet representation by contour deformation and one integration by parts. The same result holds with $u$ and $v$ exchanged. 

If two functions satisfy the sourced recurrences and these analytic conditions, their difference can be written as \begin{equation} F_r^{(1)}-F_r^{(2)} = M(u,v;q)\,e(u,v;q), \end{equation} where $e$ is entire and two-periodic in both internal labels. The mass-strip bounds give \begin{equation} |e(x+iy,v;q)| \le C|y|^{3-d}e^{2\pi|y|}. \end{equation} As in appendix~\ref{app:KL-uniqueness}, the Fourier modes with $|n|>2$ are excluded exponentially, while the modes $|n|=2$ vanish because $|y|^{3-d}\to0$. Evenness and compatibility between the two internal labels therefore reduce the difference to \begin{equation}\label{eq:app-homogeneous-modes} e(u,v;q) = A_0(q)+A_1(q)\cos(\pi u)\cos(\pi v). \end{equation} 

The defining integral also satisfies 
\begin{equation}\label{eq:app-diagonal-bound} F_r(i\mu,i\mu;q) = o(\mu^{d-3}), \qquad \mu\to+\infty. \end{equation} 
 To prove this, put
$\psi_{r,\mu}=(\sinh\rho)^h w_{i\mu}(-\cosh\rho+ri0)$.
It satisfies
\[
 \psi_{r,\mu}''+[\mu^2-h(h-1)\operatorname{csch}^2\rho]\psi_{r,\mu}=0.
\]
The large-distance connection coefficients and Stirling's formula
select $e^{-ri\mu\rho}$, with coefficient
$\mu^{h-1}(\beta_r+o(1))$, while the opposite wave is exponentially
suppressed. Since
$\int_\rho^\infty\operatorname{csch}^2t\,\dd t=\coth\rho-1$,
the Volterra equation bounds the correction to either free wave by
$C(\mu\rho)^{-1}e^{C/(\mu\rho)}$ for $\rho\ge R/\mu$.
After division by $\mu^{d-3}$, the transform on this region is
therefore the integral of the fixed amplitude
$\Omega_{d-1}\beta_r^2\varphi_q(\rho)$ against
$e^{-2ri\mu\rho}$, with error $O(R^{-1})+o(1)$. 
The amplitude is integrable, so the Riemann--Lebesgue lemma applies.
On $0<\rho<R/\mu$, the rescaled equation has Frobenius exponents
$h$ and $1-h$ and uniformly bounded rescaled data; its normalized
contribution is bounded by
$C_R\mu^{-1}\int_0^R s^{2-2h}\,\dd s$, which tends to zero
since $d<4$. Taking $\mu\to\infty$ first and then $R\to\infty$
proves eq.~\eqref{eq:app-diagonal-bound}, uniformly on compact
principal-series spectator sets.
Since 
\begin{equation} 
{M(i\mu,i\mu;q)\sim \frac{\Gamma((h+q)/2)^2\Gamma((h-q)/2)^2}{8\pi^{(d-1)/2}\Gamma(h)\Gamma(h+q)\Gamma(h-q)}\,\mu^{d-3}e^{-2\pi\mu}} 
\end{equation} 
and $\cosh^2(\pi\mu)\sim e^{2\pi\mu}/4$, the second term in eq.~\eqref{eq:app-homogeneous-modes} would be of order $\mu^{d-3}$. Hence $A_1(q)=0$. 

The simultaneous large-mass estimate therefore removes the
$\cos(\pi u)\cos(\pi v)$ mode, and the remaining homogeneous ambiguity
is $A_0(q)M(u,v;q)$. 

For fixed generic principal-series $u$ and $v$, the defining transform
is holomorphic for
\[
|\operatorname{Re}q|<h
\]
and satisfies
\begin{equation}\label{eq:A-q-strip-bound}
F_r(u,v;x+iy)=O(1),
\qquad
|x|\le1,
\qquad
|y|\to\infty,
\end{equation}
uniformly in $x$ and locally uniformly in the spectator labels. The spherical integral representation gives $|\varphi_{x+iy}(\rho)|\le\varphi_x(\rho)$;
uniformly for $|x|\le1$, the latter is bounded at the origin and is $O((1+\rho)e^{-(h-1)\rho})$ at infinity. The equal-branch radial
integrand is therefore bounded by $C\rho^{3-d}$ at the origin and by $C(1+\rho)^N e^{-(h-1)\rho}$ at infinity. These integrable bounds prove
eq.~\eqref{eq:A-q-strip-bound}. Replacing $1$ by any $b<h$ gives holomorphy on compact substrips $|\operatorname{Re}q|\le b$.

The normalised transform-label contiguity applied to the remaining
homogeneous ambiguity $A_0(q)M(u,v;q)$ gives
\begin{equation}\label{eq:A-antiperiodic}
A_0(q+1)=-A_0(q).
\end{equation}
Evenness and eq.~\eqref{eq:A-antiperiodic} imply $A_0(n+\tfrac12)=0$, so $G(q)=A_0(q)/\cos(\pi q)$ is entire and
one-periodic. The strip bounds and Stirling's formula give $G(x+iy)=O((1+|y|)^{4-d})$ on a fundamental strip. Translating the
Fourier-coefficient integral towards either end of the imaginary axis therefore eliminates every nonzero Fourier mode. Hence
\begin{equation}\label{eq:A-cosine-ambiguity}
 A_0(q)=A\cos(\pi q),
\end{equation}
where $A$ is independent of $q$.

It remains to determine $A$ from the ultraviolet normalisation. The
universal short-distance contribution may be evaluated using the
conformally coupled internal pair:
\begin{equation}\label{eq:A-conformal-reference-transform}
F_{r,c}(q)
=
e^{-ir\pi(d-2)}C_d
\frac{\Gamma(a+q)\Gamma(a-q)}
{\Gamma(\frac12+q)\Gamma(\frac12-q)},
\end{equation}
where
\begin{equation}\label{eq:A-conformal-reference-coefficient}
C_d
=
2^{-d}\pi^{-d/2}
\frac{
\Gamma(\frac{d-2}{2})^2
\Gamma(\frac{4-d}{2})
}{
\Gamma(d-2)
}.
\end{equation}
The difference between the general internal product and this conformal
reference has its leading Hadamard singularity cancelled. Green's
identity then gives
\begin{equation}\label{eq:A-ultraviolet-difference}
F_r(u,v;i\lambda)-F_{r,c}(i\lambda)
=
O(\lambda^{-2}),
\qquad
\lambda\to+\infty,
\end{equation}
and hence
\begin{equation}\label{eq:A-ultraviolet-leading}
F_r(u,v;i\lambda)
=
e^{-ir\pi(d-2)}C_d\lambda^{d-4}
+
o(\lambda^{d-4}).
\end{equation}
The full solution has the same leading coefficient, as established
in Proposition~\ref{prop:A-candidate-admissibility}. On the other hand,
$\cos(\pi q)M(u,v;q)$ has a nonzero contribution of order
$\lambda^{d-4}$ on the physical line $q=i\lambda$. Equality of the
ultraviolet coefficients therefore forces
\[
A=0.
\]
Thus the residual homogeneous ambiguity vanishes identically.

We have proved that the chronological and antichronological defining
transforms are the unique functions in the stated analytic class that
satisfy the sourced recurrences, the internal and transform-label
contiguity relations, the reflection and interchange symmetries, the
simultaneous large-mass estimate and the ultraviolet normalisation.
Consequently,
\begin{equation}\label{eq:A-sourced-identification}
F_r(u,v;q)
=
s_r(d)P(u,v;q)
+
e^{-ir\pi h}
\bigl[
\cos(\pi q)-irC_u(v,q)
\bigr]
M(u,v;q),
\qquad
r=\pm1,
\end{equation}
initially in the common convergence domain and, thereafter, by
meromorphic continuation.

\section{A branchwise contour theorem and dimensional subtraction}
\label{sec:independent-contour}
We establish the deformation of the bubble boundary diagram separately for each
Schwinger--Keldysh branch assignment. The contour argument is performed at fixed integer
dimension, with an analytic regulator at coincidence. At $d=4$ we then match its finite part
explicitly to the dimensionally subtracted result of section~\ref{sec:4d-limit}, so that no
contour argument in non-integer dimension is needed. Dimensionally continued quantities
below are those defined by the spectral representation of sections~\ref{sec:solution}
and~\ref{sec:attachment}.

\paragraph{Assumptions and prescription.}
The boundary directions $\xi_i=\xi(\bm k_i)$ are distinct, the internal
labels $u,v$ are generic principal-series labels, and the external
normalizations are finite. At fixed integer $d\ge3$, put $\delta_i={\rm Re}(\Delta_i)$, and assume
\begin{equation}\label{eq:cd-domain}
 \delta_1+\delta_2>h,\quad |\delta_1-\delta_2|<h,\qquad
 \delta_3+\delta_4>h,\quad |\delta_3-\delta_4|<h.
\end{equation}
These inequalities imply $\delta_i>0$. They include principal-series
external components and, at $d=4$, the slow conformal components.
Exceptional spectral parameters are treated subsequently by
meromorphic continuation, with any required contour residues retained.

Before deformation the branch contribution is formally
\begin{align}\label{eq:four-contributions}
 \mathcal C_{rs}^{\mathrm A}={}&-rs\int\!\dd\mu(X)\int\!\dd\mu(X')\,
 K_r^{\Delta_1}(X,\xi_1)K_r^{\Delta_2}(X,\xi_2)\nonumber\\
 &\times G^u_{rs}(X,X')\,G^v_{rs}(X,X')\,K_s^{\Delta_3}(X',\xi_3)K_s^{\Delta_4}(X',\xi_4),
\end{align}
where vertices are integrated over the expanding patch $\eta<0$, with
$\dd\mu(X)$ as in eq. \eqref{meapp} and $-rs$ is the product of the vertex factors $-ir$
and $-is$. Its precise definition is eq.~\eqref{eq:cd-zeta-domain} below; the coupling and
symmetry prefactors are restored after summing the branches.

\paragraph{The common rotation and the boundary values.}
For $t>0$ and $0<\alpha\le\pi/2$, define
\begin{equation}\label{eq:cd-homotopy}
 \eta_r=-t\,e^{-ir\alpha},\qquad Z_r(t,\bm x;\alpha)=X(\eta_r,\bm x),\qquad r=\pm1 .
\end{equation}
Thus $t=|\eta_r|$ is held fixed while the phase of the conformal time is rotated: at $\alpha=0$,
$t=-\eta$ is the real conformal time up to sign, and at $\alpha=\pi/2$, $\eta_r=rit$ and $t$ is the
radial coordinate $y$ of the cycle $\mathcal H_r$, eq.~\eqref{eq:H-coords}.
Writing $Y=Y(t,\bm x)\in H^d$ we get
\begin{align}\label{eq:cd-external}
 Z_r\cdot\xi(\bm k)
 &=-\frac t2e^{-ir\alpha}
   +\frac{|\bm x-\bm k|^2}{2t}e^{ir\alpha},
 &{\Im}(Z_r\cdot\xi)&=r\sin\alpha\,Y\cdot\xi.
\end{align}
The same sign holds at the point at infinity of the flat section.
 Thus $Z_r\in\mathcal T_r$ for all $0<\alpha\le\pi/2$: throughout the rotation, the plane waves of
the external lines remain in their tuboid of analyticity, and reduce to the boundary
values $K^\Delta_r$ as $\alpha\downarrow0$. Moreover,
\begin{equation}\label{eq:cd-leg-sandwich}
 \sin\alpha\,Y\cdot\xi\le |Z_r\cdot\xi|\le Y\cdot\xi.
\end{equation}
At the endpoint $Z_r=riY$ the external factor is
$e^{-ir\pi\Delta/2}(Y\cdot\xi)^{-\Delta}$. Including the reversal
of the $t$ orientation, the measure becomes
\begin{equation}\label{eq:cd-measure}
 \int_{-\infty}^{0}\frac{\dd\eta}{(-\eta)^d}F(\eta)
 \longrightarrow
 e^{ir(d-1)\alpha}\int_0^\infty\frac{\dd t}{t^d}F(\eta_r).
\end{equation}
Together with $-ir$, its endpoint phase is
$\varpi_r=(-ir)^{2-d}$, as in eq.~\eqref{eq:vertex-weight}.

Let $R=|\bm x-\bm x'|$ and
$\cosh\rho=(t^2+t'^2+R^2)/(2tt')$. The two internal invariants are
\begin{align}\label{eq:cd-invariants}
 z_{rr}&=-\frac{t^2+t'^2}{2tt'}+e^{2ir\alpha}\frac{R^2}{2tt'},
 \nonumber\\
 z_{r,-r}&=\frac{R^2-\cos(2\alpha)(t^2+t'^2)
                  +ir\sin(2\alpha)(t^2-t'^2)}{2tt'}.
\end{align}
For equal branches and $0<\alpha<\pi/2$, the invariant lies on side $r$ of the cut; it is real only for $\bm x=\bm x'$, where
the corresponding boundary value is used. Away from $z=-1$, that value extends locally to the sheet selected by the deformation.
At $\alpha=\pi/2$, $z_{rr}=-\cosh\rho$ for every pair, with the same boundary prescription. For mixed branches, a real invariant
in the open angular interval requires $t=t'$, when $z_{r,-r}>-1$; at the endpoint $z_{r,-r}=\cosh\rho$. Thus the only singularity
encountered is equal-branch coincidence. Consequently
\begin{equation}\label{eq:cd-final-sheets}
 z_{rr}\big|_{\alpha=\pi/2}=-\cosh\rho+ri0,
 \qquad
 z_{r,-r}\big|_{\alpha=\pi/2}=\cosh\rho.
\end{equation}
As $\alpha\downarrow0$, eq.~\eqref{eq:cd-invariants} gives exactly
the chronological, antichronological and mixed boundary values in
eqs.~\eqref{eq:SK-def} and~\eqref{eq:side}. 

\paragraph{Regulated amplitudes.} To regulate the coincidence singularity of the equal-branch
integrand, we multiply the product of the two internal propagators, $w_u(z_{rs})\,w_v(z_{rs})$,
by $(1+z_{rs})^\zeta$. For equal branches this power is, like the propagators, the boundary value
from the side $r$ of its cut. For mixed branches the integrand is regular and no regulator
is needed; we include the same factor for uniformity. Put
\begin{align}\label{eq:cd-tilted-amplitude}
& \mathcal C_{rs}(\alpha;d,\zeta)
 =-rs\int_0^\infty\!\dd t\int_0^\infty\!\dd t'
       \int\!\dd^{d-1}x\int\!\dd^{d-1}x'\,I_{rs},\nonumber\\
& I_{rs}
 =\frac{e^{i(r+s)(d-1)\alpha}}{t^dt'^d}
 \prod_{i=1,2}a_{\Delta_i}(Z_r\cdot\xi_i)^{-\Delta_i}
 (1+z_{rs})^\zeta w_u(z_{rs})w_v(z_{rs})
 \prod_{j=3,4}a_{\Delta_j}(Z_s'\cdot\xi_j)^{-\Delta_j}.
\end{align}
Initially
\begin{equation}\label{eq:cd-zeta-domain}
 \frac{d-4}{2}<{\rm Re}(\zeta)<h,\qquad
 \mathcal C_{rs}^{\mathrm A}(d,\zeta)
 :=\lim_{\alpha\downarrow0}\mathcal C_{rs}(\alpha;d,\zeta).
\end{equation}
The existence of this limit will follow from constancy in $\alpha$;
no interchange of the angle limit with an unregulated integral is
assumed. At $\alpha=\pi/2$ the amplitude is the cycle integral
\eqref{eq:def-oneloop}, with its internal product multiplied by
\begin{equation}\label{eq:cd-endpoint-regulator}
 \begin{cases}
 e^{ir\pi\zeta}(\cosh\rho-1)^\zeta,&s=r,\\
 (\cosh\rho+1)^\zeta,&s=-r.
 \end{cases}
\end{equation}

\paragraph{Uniform integrability at positive angle.}
Fix $\alpha_0>0$. All constants in this paragraph are uniform for
$\alpha\in[\alpha_0,\pi/2]$ and parameters in compact subsets of
the stated strict domain. Set
\[
 f_L^*(Y)=\prod_{i=1,2}(Y\cdot\xi_i)^{-\delta_i},\qquad
 f_R^*(Y')=\prod_{j=3,4}(Y'\cdot\xi_j)^{-\delta_j}.
\]
The external plane-waves are bounded by a constant times $f_L^*f_R^*$.
For the internal factors, eq.~\eqref{eq:cd-invariants} gives
\begin{align}\label{eq:cd-distance-bounds}
 \sin\alpha(\cosh\rho-1)&\le |1+z_{rr}|\le\cosh\rho-1,
 \nonumber\\
 \sin^2\alpha(\cosh\rho+1)&\le |1+z_{r,-r}|\le\cosh\rho+1.
\end{align}
Both invariants also obey $|z_{rs}|\ge\tfrac12\sin\alpha\cosh\rho$
when $\cosh\rho\ge4$. For example, with
$H=(t^2+t'^2)/(2tt')$, $B=R^2/(2tt')$ and $k=\sin^2\alpha$, the
mixed lower bound follows from
\[
 |1+z_{r,-r}|^2-k^2(H+B+1)^2
 =(1-k)\{(1+k)(H-B-1)^2+4k[H(B+1)-1]\}\ge0.
\]
Together with these bounds, the behaviour of $w_\nu$ near $z=-1$, eq.~\eqref{eq:hadamard}, and at
large distances, eq.~\eqref{eq:falloffs}, gives
\begin{equation}\label{eq:cd-kernel-bound}
 |(1+z_{rs})^\zeta w_u(z_{rs})w_v(z_{rs})|\le b_\zeta(\rho),
 \qquad
 b_\zeta(\rho)\le C
 \begin{cases}
 \rho^{4-2d+2{\rm Re}(\zeta)}(1+|\log\rho|)^N+1,&0<\rho<1,\\
 (1+\rho)^Ne^{-(2h-{\rm Re}(\zeta))\rho},&\rho\ge1.
 \end{cases}
\end{equation}
The logarithmic factor allows resonant dimensional parameters.

The function $\log(Y\cdot\xi)$ has a gradient of unit length in the hyperbolic metric, so it
changes by at most $\rho$ between two points at distance $\rho$:
$e^{-\rho}\le(Y'\cdot\xi)/(Y\cdot\xi)\le e^{\rho}$. Hence
$f_R^*(Y')\le e^{(\delta_3+\delta_4)\rho}f_R^*(Y)$, and for $\rho<1$ the external factor at $Y'$ can be
replaced by a constant times its value at $Y$. The near-diagonal integral is thus bounded by
\begin{equation}\label{eq:cd-contact-bound}
 C\int_0^1(\sinh\rho)^{d-1}b_\zeta(\rho)\,\dd\rho
   \int_{H^d}\dd\mu_H(Y)\,f_L^*(Y)f_R^*(Y)<\infty.
\end{equation}
Indeed, eq.~\eqref{eq:cd-domain} implies
$\sum_i\delta_i>2h$ and $\sum_j\delta_j-2\delta_i>0$ at each
insertion. These are exactly the convergence conditions of the
contact integral, including spatial infinity after compactifying
the boundary section.

In the far region, $\rho\ge1$, we compare the kernel with a function whose integral against
the external factors is already known: the Mehler--Fock kernel at $\lambda=0$, in the plane-wave
form of eq.~\eqref{eq:attachment-plane-waves},
\begin{equation}\label{eq:cd-Phi0}
 \varphi_0(Y,Y')=\frac1{\Omega_{d-1}}\int_{S^{d-1}}\!\dd\omega(\xi)\,(Y\cdot\xi)^{-h}(Y'\cdot\xi)^{-h},\qquad
 \Omega_{d-1}=\frac{2\pi^{d/2}}{\Gamma(d/2)}.
\end{equation}
It is positive, depends only on the distance $\rho$, and decays no faster than $e^{-h\rho}$:
placing $Y$ at the origin of the spherical section, where $Y\cdot\xi=1$ and $Y'\cdot\xi\le e^{\rho}$,
gives $\varphi_0(\rho)\ge e^{-h\rho}$. Since ${\rm Re}\,\zeta<h$, the bound $b_\zeta$ decays faster, so
$b_\zeta\le C\,\varphi_0$ for $\rho\ge1$. The integral of $f_L^*\,\varphi_0\,f_R^*$ is then the calculation of
section~\ref{sec:attachment} at $\lambda=0$, with real exponents. All factors being positive, the
integrations can be done in any order, and the two integrals over $H^d$ are the triangular
invariants~\eqref{eq:attachment-star-triangle} with exponents $(\delta_1,\delta_2,h)$ and
$(\delta_3,\delta_4,h)$, which converge in the domain~\eqref{eq:cd-domain}. The remaining
integral over the cone carries the four powers $(h\pm(\delta_1-\delta_2))/2$ and
$(h\pm(\delta_3-\delta_4))/2$, each between $0$ and $h$, with sum $2h$: it is independent of the
section and converges at the four distinct boundary points.
Altogether, for all $\alpha\in[\alpha_0,\pi/2]$,
\begin{equation}\label{eq:cd-majorant}
 |I_{rs}|\le\mathcal M(t,t',\bm x,\bm x')=\frac{C\,f_L^*(Y)\,f_R^*(Y')}{(t\,t')^d}\times
 \begin{cases}
  b_\zeta(\rho)+\varphi_0(\rho),&\rho<1,\\[2pt]
  \varphi_0(\rho),&\rho\ge1,
 \end{cases}
\end{equation}
where $\mathcal M$ is integrable and independent of $\alpha$. The bound holds, in particular,
when the two vertices approach the conformal boundary together.

\paragraph{The contour identity at fixed dimension.} The argument is Cauchy's theorem for
the two conformal times. The integrand depends on the angle only through
$\eta_r=-te^{-ir\alpha}$ and $\eta'_s=-t'e^{-is\alpha}$, and, away from coincidence, is locally holomorphic in both on the sheets selected by the
 deformation. Changing
$\alpha$ is therefore equivalent to a displacement along the two contours, since
$\partial_\alpha\eta_r=-ir\,t\,\partial_t\eta_r$, and the derivative of the integrand, with its
Jacobian, is a total derivative in the two times:
\begin{equation}\label{eq:cd-angular-derivative}
 \partial_\alpha I_{rs}=-i\big\{r\,\partial_t(t\,I_{rs})+s\,\partial_{t'}(t'I_{rs})\big\}.
\end{equation}
The derivative of the amplitude with respect to $\alpha$ is thus a sum of boundary terms:
from the ends of the two contours, $t,t'\to0$ and $t,t'\to\infty$, and, for equal branches,
from the coincidence point $Y=Y'$, where the integrand is singular. The spatial variables
are not rotated and give no boundary term. We show that both kinds of term vanish.

To this end we cut off the integral. We multiply it by $\varrho_L(t)\varrho_L(t')$, where
$\varrho_L(t)=\varrho(\log t/L)$ equals one for $e^{-L}<t<e^{L}$ and vanishes outside a larger
interval, so that $|t\,\partial_t\varrho_L|\le C/L$, and by spatial cutoffs of radius $M$; for
equal branches we also remove the ball $\rho\le\varepsilon$ around coincidence. Integrating
eq.~\eqref{eq:cd-angular-derivative} by parts, the cut-off amplitude satisfies
\begin{equation}\label{eq:cd-cutoff-bound}
 \big|\partial_\alpha\mathcal C_{rs}^{L,M,\varepsilon}\big|\le\frac CL\int\!\mathcal M
 +C\,\varepsilon^{4-d+2{\rm Re}\,\zeta}\,(1+|\log\varepsilon|)^N\int_{H^d}\!\dd\mu(Y)\,f_L^*(Y)\,f_R^*(Y).
\end{equation}
The first term comes from the ends of the contours: the derivative falls on the time
cutoffs, and what remains is bounded by the integral of $\mathcal M$,
eq.~\eqref{eq:cd-majorant}. The second term, present only for equal branches, is the flux
through the sphere $\rho=\varepsilon$. It is small for three reasons: the displacement
$V=r(t\partial_t+t'\partial_{t'})$ is almost tangent to the sphere,
\begin{equation}\label{eq:cd-tangency}
 V(\cosh\rho)=-r\,\frac{R^2}{tt'},\qquad |V\rho|\le\rho ,
\end{equation}
so its radial component is of order $\varepsilon$; the area of the sphere is of order
$\varepsilon^{d-1}$; and the integrand there is of order $\varepsilon^{4-2d+2{\rm Re}\,\zeta}$, up to
logarithms. Both terms vanish as $L\to\infty$ and $\varepsilon\to0$, the second because
${\rm Re}\,\zeta>(d-4)/2$. Removing the cutoffs with the bound~\eqref{eq:cd-majorant}, the amplitude
is therefore independent of the angle for $0<\alpha\le\pi/2$:
\begin{equation}\label{eq:cd-main-identity}
 \mathcal C_{rs}^{\mathrm A}(d,\zeta)=\mathcal C_{rs}(\alpha;d,\zeta)=\mathcal C_{rs}(\pi/2;d,\zeta),
\end{equation}
for $\zeta$ in the domain~\eqref{eq:cd-zeta-domain}, at fixed integer $d$. The first equality is
the definition~\eqref{eq:cd-zeta-domain}: since the amplitude does not depend on $\alpha$, its
limit as $\alpha\downarrow0$ exists. No estimate uniform down to $\alpha=0$ is needed. The
left-hand side is the diagram on the real contour, defined by its tilt, and the right-hand
side is the integral over the Lobachevskij cycles: their equality is thus proved for the
regulated amplitudes. At $d=3$ the regulator can be set to zero. At $d=4$, where $\zeta=0$ lies on
the boundary of the domain~\eqref{eq:cd-zeta-domain}, it is removed in the next paragraph.

\paragraph{Removing the regulator at four dimensions.} 

Fix $d=4$. The two sides of
eq.~\eqref{eq:cd-main-identity} are equal for $0<{\rm Re}\,\zeta<3/2$. We now show that both
extend to meromorphic functions of $\zeta$ around $\zeta=0$, with at most a simple pole there.
Being equal on an open set, they then have the same pole and the same finite part.

The pole comes from the coincidence point of the equal-branch integrands; for
principal-series internal labels it is the only one, since the large-distance region
converges for ${\rm Re}\,\zeta<3/2$. At $d=4$ the product of the two internal propagators near
$z=-1$ is $c^2(1+z)^{-2}$, with $c=1/(8\pi^2)$, whatever the masses, plus terms less singular by
one power of $1+z$, up to logarithms.

Near $Y'=Y$, let $\rho$ be the
hyperbolic distance and $\omega$ the direction of $Y'$ seen from $Y$. Then
$1+z_{rr}=\rho^2\,\kappa_{r,\alpha}(Y,\omega,\rho)$, where $\kappa_{r,\alpha}$ is smooth and, by
eq.~\eqref{eq:cd-distance-bounds}, bounded away from zero for $\alpha\ge\alpha_0$. With the
regulator $(1+z_{rr})^\zeta$ and the measure $\rho^3\dd\rho\,\dd\omega$, the integrand near coincidence is
\begin{equation}\label{eq:cd-local-zeta}
 \rho^{2\zeta-1}\,\big[A_{r,\alpha}(Y,\omega;\zeta)+O\big(\rho\,(1+|\log\rho|)^N\big)\big]\,\dd\rho\,\dd\omega,
\end{equation}
with $A_{r,\alpha}$ analytic in $\zeta$. This is the logarithmic divergence of the bubble: the first
term, integrated over $\rho<1$, gives $A_{r,\alpha}/(2\zeta)$, a simple pole at $\zeta=0$, while the
remainder is integrable, and analytic in $\zeta$, for ${\rm Re}\,\zeta>-1/2$.

The integrations over $\omega$ and over the position $Y$ of the pair do not spoil this. The
external factors and their first derivatives are bounded by constants times $f_L^*f_R^*$, by
eq.~\eqref{eq:cd-leg-sandwich}, and the integral of $f_L^*f_R^*$ over $Y$ converges, by
eq.~\eqref{eq:cd-contact-bound}; the region $\rho\ge1$ is controlled by $\varphi_0$ as before, for
${\rm Re}\,\zeta<3/2$. For mixed branches the integrand is regular, and the amplitude is analytic
at $\zeta=0$. Hence each side of eq.~\eqref{eq:cd-main-identity} is meromorphic near $\zeta=0$, with
a simple pole for equal branches only, and the equality extends there. In particular the
finite parts at $\zeta=0$, denoted by ${\rm FP}$, coincide:
\begin{equation}\label{eq:cd-finite-part-identity}
 \underset{\zeta=0}{\rm FP}\;\mathcal C_{rs}^{\mathrm A}(4,\zeta)
 =\underset{\zeta=0}{\rm FP}\;\mathcal C_{rs}(\pi/2;4,\zeta).
\end{equation}
We define $\mathcal B_{\rm an}$ as the sum of these four finite parts, multiplied by $Sg_Lg_R$: it is
the boundary correlator of the bubble in analytic regularisation. On the cycles the
regulator of the equal branches is $e^{ir\pi\zeta}(\cosh\rho-1)^\zeta$,
eq.~\eqref{eq:cd-endpoint-regulator}; its phase contributes to the finite part and is kept.

\paragraph{Relation to dimensional subtraction.} We have regulated the bubble by
continuing the dimension in section~\ref{sec:4d-limit}, whereas the contour identity has been
proved with the analytic regulator $\zeta$. The two finite results at $d=4$ differ by a finite
local term, which we now compute. Since it originates at coincidence, it does not involve
the external lines, and can be found by comparing the Mehler--Fock transforms of the
kernels in the two regularisations. Let
\[
 \varphi_{i\lambda,d}(\rho)
 =2^{d/2-1}\Gamma(d/2)(\sinh\rho)^{-(d-2)/2}
   P_{-1/2+i\lambda}^{-(d-2)/2}(\cosh\rho),
 \qquad \varphi_{i\lambda,d}(0)=1.
\]
In the normalization of eq.~\eqref{eq:transform-convention},
\begin{equation}\label{eq:cd-radial-transform}
 F_{r,d}(i\lambda)=\Omega_{d-1}\int_0^\infty
  (\sinh\rho)^{d-1}\varphi_{i\lambda,d}(\rho)
  \mathcal K_{r,d}(\rho)\,\dd\rho,
\end{equation}
initially for $3<{\rm Re}(d)<4$. The analytically regulated transform
$F_{r,\zeta}^{\rm an}$ is the same integral at $d=4$, multiplied
by $e^{ir\pi\zeta}(\cosh\rho-1)^\zeta$. Set
\begin{equation}\label{eq:cd-matching-constants}
 R_0=\frac1{16\pi^2},\qquad
 c_* =1+\gamma_{\rm E}+\log(2\pi).
\end{equation}
Here and below the curvature radius is one and no extra
dimension-dependent scale or normalization factor is introduced.

The dimensionally continued radial singular coefficient is
\begin{align}\label{eq:cd-local-coefficient}
 L_r(d)
 &=\Omega_{d-1}e^{-ir\pi(d-2)}
      \left[\frac{\Gamma(d/2-1)}{4\pi^{d/2}}\right]^2,\nonumber\\
 L_r(4-2\epsilon)
 &=2R_0\left[1+\epsilon
       (2ir\pi+\gamma_{\rm E}+\log\pi+1)+O(\epsilon^2)\right].
\end{align}
Subtract $L_r(d)\rho^{3-d}$ on $0<\rho<1$ in
eq.~\eqref{eq:cd-radial-transform}. The remainder has a common
limit $\overline F_r(i\lambda)$ as $d\to4$ and as $\zeta\to0$ in the
analytic regularization. Therefore
\begin{align}\label{eq:cd-two-laurent-expansions}
 F_{r,4-2\epsilon}(i\lambda)
 &=\frac{R_0}{\epsilon}+\overline F_r(i\lambda)
   +R_0(2ir\pi+\gamma_{\rm E}+\log\pi+1)+O(\epsilon),\nonumber\\
 F_{r,\zeta}^{\rm an}(i\lambda)
 &=\frac{R_0}{\zeta}+\overline F_r(i\lambda)
   +R_0(ir\pi-\log2)+O(\zeta).
\end{align}
In the second line the factor $2^{-\zeta}$ comes from
$\cosh\rho-1=\rho^2/2+O(\rho^4)$. It follows that
\begin{equation}\label{eq:cd-branch-matching}
 \lim_{\epsilon\to0}
 \left[F_{r,4-2\epsilon}(i\lambda)-\frac{R_0}{\epsilon}\right]
 =\operatorname{FP}_{\zeta=0}F_{r,\zeta}^{\rm an}(i\lambda)
    +R_0(c_*+ir\pi).
\end{equation}
Only the leading Hadamard term contributes to this difference. It is therefore
independent of $u,v$ and $\lambda$ and is a local delta term. Mixed branches have no such term.

The pole-subtracted transforms converge pointwise in $\lambda$ and are bounded by
$C(1+\lambda)^2$, uniformly for regulator parameters near zero. Indeed, $|\varphi_{i\lambda}|\le C\varphi_0$ and
$\varphi_{i\lambda,d}(\rho)-1=O\big((1+\lambda)^2\rho^2\big)$ near $\rho=0$, while the terms of
$w_\nu$ beyond the leading short-distance one are $O\big(\rho^{-\eta}(1+|\log\rho|)\big)$ for
$|d-4|<\eta<1/4$, so the remainders are integrable at $\rho=0$ uniformly in $d$; at large $\rho$
the integrand decays exponentially. This bound allows the limits to be taken under the spectral integral, against the function
$\mathcal A_\lambda$ of eq.~\eqref{eq:spectral-attachment}, which decays exponentially up to a polynomial factor by
eq.~\eqref{eq:attachment-large-lambda}.
\paragraph{The matching formula.} It remains to integrate against the external plane
waves. In analytic regularisation at $d=4$, the spectral
representation~\eqref{eq:attached-spectral} holds with $F$ replaced by the regulated
transforms. To justify the interchange of the spectral and vertex integrations, take
first $1<{\rm Re}\,\zeta<3/2$, with the lower and upper bounds ensuring
square integrability at coincidence and infinity, respectively,
and insert the heat factor $e^{-\tau(\lambda^2+h^2)}$, $\tau>0$. The interchange is then
absolutely convergent, by $|\varphi_{i\lambda}|\le\varphi_0$ and the positive factorisation
used for eq.~\eqref{eq:cd-majorant}. In position space this factor is the heat operator
$e^{\tau\LapH}$, which can be moved onto $f_L$, where it is harmless:
$e^{\tau\LapH}f_L^*\le e^{C\tau}f_L^*$, since $\nabla\log f_L^*$ and $\LapH f_L^*/f_L^*$ are bounded.
The limit $\tau\downarrow0$ is then dominated on both sides, by eq.~\eqref{eq:cd-majorant}
and by eq.~\eqref{eq:attachment-large-lambda}. The identity extends to $\zeta=0$ by
meromorphic continuation, and expresses $\mathcal B_{\rm an}$ as the spectral integral of the
finite parts of the regulated transforms. For a constant transform, the same argument
gives
\begin{equation}\label{eq:cd-contact-closure}
 \int_0^\infty\dd\lambda\,p_4(\lambda)\,\mathcal A_{\lambda,4}=\mathcal D_4 :
\end{equation}
a constant in $\lambda$ corresponds to the contact integral.

On the dimensional side, the limit $\epsilon\to0$ in eq.~\eqref{eq:ren-spectral} can be taken
under the spectral integral. The pole-subtracted transforms are bounded by $C(1+\lambda)^2$,
as shown above, and $\mathcal A_\lambda$ is analytic in $d$ near $d=4$, with a bound
$C(1+\lambda)^N e^{-\pi\lambda}$, for a fixed finite $N$, uniformly on compact parameter
 sets near $d=4$ within the strict external convergence domain and away from poles. Indeed, with one boundary point at infinity the other three
lie in a plane, and the cone integral~\eqref{eq:F4-def} involves a single transverse
vector, with measure
\begin{equation}\label{eq:cd-single-vector-measure}
 \frac{2\pi^{(d-3)/2}}{\Gamma\big(\frac{d-3}{2}\big)}\,\dd^2x\,s^{d-4}\,\dd s,\qquad 0<s<\infty :
\end{equation}
for ${\rm Re}\,d>3$ this is an absolutely convergent integral, whose integrand has a modulus
independent of $\lambda$, while the two coefficients $c_d$ supply exponential decay multiplied by
powers of $\lambda$. Together with the polynomial growth of $p_d$, these
bounds justify dominated convergence of the subtracted spectral integral.

By eq.~\eqref{eq:cd-branch-matching}, the two spectral representations differ by a constant
in $\lambda$, hence, by eq.~\eqref{eq:cd-contact-closure}, by a multiple of $\mathcal D_4$. Writing
$\dot\Theta=\partial_\epsilon\Theta(4-2\epsilon)|_{\epsilon=0}$ for the chosen continuation of the external
exponents, and keeping the derivative of the branch phases,
\begin{equation}\label{eq:cd-weighted-matching}
 \mathcal B_{\rm dim}-\mathcal B_{\rm an}=\frac{Sg_Lg_R}{8\pi^2}\,
 \big[c_*\cos\Theta_4-(\pi+\dot\Theta)\sin\Theta_4\big]\,\mathcal D_4 .
\end{equation}
For the fixed external exponents and coupling prefactor of
section~\ref{sec:4d-limit}, $\dot\Theta=-2\pi$, and
\begin{equation}\label{eq:cd-dim-matching}
 \mathcal B_{\rm dim}=\mathcal B_{\rm an}+\frac{Sg_Lg_R}{8\pi^2}\,
 \big[(1+\gamma_{\rm E}+\log2\pi)\cos\Theta_4+\pi\sin\Theta_4\big]\,\mathcal D_4 .
\end{equation}
The dimension dependence of $p_d$, of the external normalisations and of $\mathcal A_\lambda$ is
kept until the pole is subtracted, and generates no further term. For the subtraction of the
pole alone, eq.~\eqref{eq:finite-contact} gives instead
\begin{equation}\label{eq:cd-pole-matching}
 \mathcal B_{\rm pole}=\mathcal B_{\rm an}+\frac{Sg_Lg_R}{8\pi^2}\,
 \big[c_*\cos\Theta_4+\pi\sin\Theta_4\big]\,\mathcal D_4
 -\frac{Sg_Lg_R\cos\Theta_4}{4\pi^2}\,\mathcal D'_4 .
\end{equation}
These formulas hold for the conventions stated here: additional dimensional scale
factors, a different continuation of the external exponents or a different finite
counterterm change the local term accordingly.

{
\section{Comparison with recent literature} \label{app:gs} 
The purpose of this appendix is to compare the two-loop bubble-chain result of section~\ref{sec:twoloop} with the recent analysis of Ref.~\cite{GraefeSachs2026}, and to identify the origin of their discrepancy. We trace it to the delta-function contribution in the split representation of the equal-branch propagator, which is discarded in Ref.~\cite{GraefeSachs2026} as odd in the spectral variable, whereas on the full spectral line it is even. Once this contribution is restored, the equal-branch blocks acquire an additional term, which reproduces the term in $M_LM_R$ of eq.~\eqref{eq:chain-finite-conformal}.

For the purpose of the comparison, we temporarily adopt the notation of Ref.~\cite{GraefeSachs2026}. In particular, throughout this appendix $\db$ denotes the number of spatial dimensions. We retain the distinction between $\nu$ and $\tilde\nu$ used in that reference, with $\nu=i\tilde\nu$.

We first observe that the harmonic kernels $\Omega_\nu^\pm$ of Ref.~\cite{GraefeSachs2026} are nothing but the spatial Fourier transforms of the equal-tuboid two-point functions introduced in Section \ref{sametr}. Using the Fourier convention of eq.~\eqref{eq:plane-wave-Fourier-definition}, we may write 
\begin{align} 
W_{i\tilde\nu}^{++} \bigl(X(\eta,\bm x),X(\eta',\bm x')\bigr) &= -\frac{\pi}{\tilde\nu\sinh(\pi\tilde\nu)} \int_{\mathbb R^\db}\frac{\dd^\db p}{(2\pi)^\db}\, e^{-i\bm p\cdot(\bm x-\bm x')} \Omega_\nu^+(p;\eta,\eta'), \label{eq:GS-inverse-plus} \\ W_{i\tilde\nu}^{--} \bigl(X(\eta,\bm x),X(\eta',\bm x')\bigr) &= -\frac{\pi}{\tilde\nu\sinh(\pi\tilde\nu)} \int_{\mathbb R^\db}\frac{\dd^\db p}{(2\pi)^\db}\, e^{-i\bm p\cdot(\bm x-\bm x')} \Omega_\nu^-(p;\eta,\eta'). \label{eq:GS-inverse-minus} \end{align} 
The completeness relation proven in Ref.~\cite{GraefeSachs2026} is simply the reproducing-kernel property of the equal-tuboid two-point functions, analogous to the property established for the maximally analytic Wightman function by Bros, Epstein and Moschella~\cite{BEM2010}: 
\begin{align} \int_{dS_{\db+1}^{up}}\dd\mu(X'')\, W_{i\tilde\nu}^{++}(X,X'') W_{i\tilde\mu}^{++}(X'',X')  =\pi\, \frac{ \delta(\tilde\nu-\tilde\mu)+\delta(\tilde\nu+\tilde\mu)}{2\tilde\nu\sinh(\pi \tilde\nu)}
W_{i\tilde \nu}^{++}(X,X')  \cr  = \frac{\pi}{|\sinh(\pi\tilde\nu)|}\, \delta(\tilde\mu^2-\tilde\nu^2)\, W_{i\tilde\nu}^{++}(X,X'). \label{eq:equal-tuboid-reproducing} \end{align} Similarly 
\begin{equation} \int_{dS_{\db+1}^{up}}\dd\mu(X'')\, W_{i\tilde\nu}^{--}(X,X'') W_{i\tilde\mu}^{--}(X'',X') =  \frac{\pi}{|\sinh(\pi\tilde\nu)|}\, \delta(\tilde\mu^2-\tilde\nu^2)\, W_{i\tilde\nu}^{--}(X,X'). \label{eq:equal-tuboid-reproducing2} \end{equation}

These identities are understood in the distributional sense. The absolute value in the coefficient is essential, since it ensures invariance under reflection of the spectral parameter. The failure to preserve this symmetry is precisely the source of the discrepancy between our results and those of Ref.~\cite{GraefeSachs2026}.

In eqs.~(4.15)--(4.16) of Ref.~\cite{GraefeSachs2026},
the equal-branch propagator with mass label $\tilde\nu'$
is written as
\begin{align}
G^{\nu'}_{aa}
&=
ia
\int_{-\infty}^{\infty}
\dd\tilde\mu\,
\Ncal_\mu\,
\frac{
u^a_\mu(\eta_1)u^a_\mu(\eta_2)
}{
(\tilde\mu^2-\tilde\nu'^2)_{ia\epsilon}
},
\nonumber\\
\frac{1}{
(\tilde\mu^2-\tilde\nu'^2)_{ia\epsilon}
}
&=
\frac{1}{2\sinh(\pi\tilde\nu')}
\left[
\frac{
e^{\pi\tilde\nu'}
}{
\tilde\mu^2-\tilde\nu'^2+ia\epsilon
}
-
\frac{
e^{-\pi\tilde\nu'}
}{
\tilde\mu^2-\tilde\nu'^2-ia\epsilon
}
\right].
\label{eq:415}
\end{align}
In their eq.~(4.22), the kernel is decomposed as
\begin{equation}
\PV\frac{1}{\tilde\mu^2-\tilde\nu'^2}
-
i\pi a\coth(\pi\tilde\nu')\,
\delta(\tilde\mu^2-\tilde\nu'^2).
\end{equation}
The principal-value part of the kernel in
eq.~\eqref{eq:415} is even in $\tilde\nu'$,
whereas its delta-function part is odd.
The propagator itself, however, must be even in
the mass label, since the modes are such that 
$
u_\nu=u_{-\nu}.
$
Thus, the representation \eqref{eq:415}, with the
displayed prescription, applies only to the positive
spectral parameter used in its derivation.

When this representation is extended to the full spectral line in the K\"all\'en--Lehmann integrals of eqs.~(4.11) and (4.20) of Ref.~\cite{GraefeSachs2026}, the coefficient of the delta-function term must be $|\coth(\pi\tilde\nu')|$.
The resulting term is even in $\tilde\nu'$ and gives a nonzero contribution when integrated against the even spectral density $\rho^+_\nu$. 
Indeed, \begin{equation} -\pi\int_{-\infty}^{\infty} \dd\tilde\nu'\, \rho^+_\nu(\tilde\nu') |\coth(\pi\tilde\nu')| \delta(\tilde\nu^2-\tilde\nu'^2) = -\cosh(\pi\tilde\nu) \frac{\rho^+_\nu(\tilde\nu)}{\Ncal_\nu}, \qquad \tilde\nu>0. \end{equation} 
The diagonal spectral block of Ref.~\cite{GraefeSachs2026} should therefore be amended to \begin{equation} \tilde q_{++} = P_\nu-\cosh(\pi\tilde\nu)\,M, \qquad M=\Ncal_\nu^{-1}\rho^+_\nu. \label{eq:corrected420} \end{equation} 
The omission of the contribution of $M$ is the source of the discrepancy between their construction and ours.

A second correction is nevertheless necessary. The prefactor $ia$ in eq.~(4.15) of Ref.~\cite{GraefeSachs2026} must be replaced by $1$ (see Section~\ref{sec:phase}).  This amounts to multiplying the right-hand side of that equation by $-ia$, namely $-i$ for the $++$ branch and $+i$ for the $--$ branch. With both corrections, the spectral blocks become \begin{equation} \tilde q_{\pm\pm} = -iP_\nu\pm i\cosh(\pi\tilde\nu)\,M, \qquad \tilde q_{\pm\mp}=M, \label{eq:trueblocks} \end{equation} where $P_\nu$ denotes the principal-value contribution  \begin{equation} P_\nu(\lambda) = i\int_{-\infty}^{\infty}\dd t\, \rho^+(t)\, \PV\frac{1}{t^2-\lambda^2} = 2i\,\PV\!\int_0^\infty\dd t\, \frac{\rho^+(t)}{t^2-\lambda^2}, \label{eq:Adef} \end{equation}
and not the Legendre functions as in the main text.

The discrepancy is already present at one loop, although it is hidden for conformally coupled external fields in four spacetime dimensions. Indeed, the equal-branch blocks in eq.~\eqref{eq:trueblocks} carry the phases $e^{\mp i\pi d/2}$, which reduce to unity at $d=4$. The additional terms proportional to $\cosh(\pi\tilde\nu)M$ then cancel upon summing over Schwinger--Keldysh branches. For massive external fields, this cancellation does not occur in general. Even for conformally coupled external fields, however, the cancellation is incomplete in dimensional regularization. At $d=4-2\epsilon$, the residual contribution is $-2\sin(\pi\epsilon)\cosh(\pi\tilde\nu)M$, which can yield a finite term when combined with an ultraviolet $1/\epsilon$ pole in the bubble-chain expansion.

We now examine how the corrected spectral blocks affect the bubble-chain construction of Ref.~\cite{GraefeSachs2026}. In that reference, the branch sum is expressed in terms of the matrix \begin{equation} K= \begin{pmatrix} P_\nu&M\\ M&P_\nu \end{pmatrix}, \qquad e= \begin{pmatrix} 1\\ 1 \end{pmatrix}. \end{equation} Since $Ke=\Sigma_\nu e$, with $\Sigma_\nu=P_\nu+M$, one obtains \begin{equation} e^TK^Ne=2\Sigma_\nu^N. \end{equation} The corresponding geometric resummation gives \begin{equation} \sum_{N=0}^{\infty} (-\lambda_{\rm GS})^N e^TK^Ne = \frac{2}{1+\lambda_{\rm GS}\Sigma_\nu}, \end{equation} in agreement with their eq.~(4.34). Here we denote the coupling parameter of Ref.~\cite{GraefeSachs2026} by $\lambda_{\rm GS}$, to distinguish it from our spectral variable $\lambda$. 

With the corrected spectral blocks \eqref{eq:trueblocks}, the matrix $K$ instead takes the form \begin{equation} K=-i P_\nu\,\mathbb I+B, \qquad B= \begin{pmatrix} i\gamma&M\\ M&-i\gamma \end{pmatrix}, \qquad \gamma=\cosh(\pi\tilde\nu)\,M. \end{equation} The matrix $B$ satisfies \begin{equation} B^2=-s^2\mathbb I, \qquad s=\sinh(\pi\tilde\nu)\,M, \qquad e^TBe=2M. \end{equation} It follows that \begin{equation} e^TK^Ne = 2\operatorname{Re}\bigl[(-i P_\nu+is)^N\bigr] + \frac{2M}{s} \operatorname{Im}\bigl[(-i P_\nu+is)^N\bigr]. \label{eq:corrected-sum} \end{equation} In particular, \begin{equation} e^TKe=2(-i P_\nu+M), \qquad e^TK^2e = 2(-i P_\nu+M)^2 - 2\cosh^2(\pi\tilde\nu)M^2. \end{equation} The corrected geometric resummation is therefore 
\begin{equation} 
\sum_{N=0}^{\infty} (-g)^N e^TK^Ne = \frac{ 2\bigl(1-igP_\nu-gM\bigr) }{ (1-igP_\nu)^2 + g^2M^2\sinh^2(\pi\tilde\nu) }. \label{eq:corrected-resum-GS} 
\end{equation}
Here $g$ is a formal expansion parameter. The series converges when the spectral radius of $gK$ is less than one; the rational function gives its meromorphic continuation elsewhere. The identification of $g$ with $\lambda_{\rm GS}$ would require matching the propagator, split-vertex, external-line and subtraction conventions.

 Finally, letting $\lambda$ denote again the Mehler--Fock spectral variable, the corrected same-branch kernels are 
 \begin{equation} 
 \mathcal K_\pm(\lambda) = e^{\mp i\pi d/2} \left[-i P_\nu(\lambda) \pm i\cosh(\pi\lambda)\,M(\lambda) \right], \qquad M(\lambda) = \Ncal^{-1}_{i\lambda}\rho^+(\lambda). \label{eq:structure} 
 \end{equation} 
 For conformally coupled external fields in $dS_4$, the one-loop branch sum gives \begin{equation} 2M+\mathcal K_++\mathcal K_- = 2(-iP_\nu+M), \end{equation} recovering the result of Ref.~\cite{GraefeSachs2026} at this order.

In $dS_3$, the on-shell term does not cancel even at one loop and gives
\begin{equation}
2\bigl[1-\cosh(\pi\lambda)\bigr]M(\lambda).
\end{equation}

For the two-loop configuration in $dS_4$, the middle-vertex weight in our conventions is
\[
\varpi_\pm=(\mp i)^{-2}=-1
\]
for both branches, while the external weights are branch independent. The resulting branch sum is therefore
\begin{equation}
-e^T K^2e
=
-2(-iP_\nu+M)^2
+
2\cosh^2(\pi\lambda)M^2.
\label{eq:corrected-two-loop-sum}
\end{equation}
The second term agrees with
\[
\frac{
2(c-\sin\vartheta)^2
}{
\cos\vartheta
}
M^2
\]
at $\vartheta=0$ in eq.~\eqref{eq:chain-factorised}. 
For identical bubbles, $M_L=M_R=M$, the restored on-shell contribution therefore reproduces the $M_LM_R$ term of
eq.~\eqref{eq:chain-finite-conformal}. This comparison concerns the branch-summed spectral blocks: a comparison of fully
normalized renormalized correlators additionally requires matching the vertex, external-line and subtraction conventions.

\subsection{Check of the phase correction}\label{sec:phase}
For completeness, here we prove the necessity of the phase correction cited above.

Let $\db$ denote spatial dimension, $s,t>0$ principal-series masses, set the Hubble scale to one, and use normalized mode, \cite{GraefeSachs2026},
\[
u_s(\eta)=\frac{\sqrt\pi}{2}e^{-\pi s/2+i\pi/4}(-\eta)^{\db/2}H^{(1)}_{is}(-k\eta),
\qquad u_s^-=u_s,\quad u_s^+=u_s^*.
\]
The last equality applies on the real contour; complex-time values are analytic continuations. Put
\[
E(s,s')=\tfrac12[\delta(s-s')+\delta(s+s')],\qquad N_s=\frac{s\sinh\pi s}{\pi}.
\]
With the Bunch--Davies contour prescription, the Hankel-to-$K$ transformation \cite{DLMF} and Kontorovich--Lebedev orthogonality \cite{Bateman1954,YakubovichLuchko} give
\begin{equation}\label{eq:overlap}
\int_{-\infty}^0\frac{\dd\eta}{(-\eta)^{\db+1}}u_s^a(\eta)u_{s'}^a(\eta)
=\frac{ia}{N_s}E(s,s').
\end{equation}
These are spectral distribution identities, on the even signed-label space. For example,
\[
H^{(1)}_{is}(ix)=\frac{2}{\pi i}e^{\pi s/2}K_{is}(x),\qquad
\int_0^\infty\frac{\dd x}{x}K_{is}(x)K_{is'}(x)
=\frac{\pi^2}{2s\sinh\pi s}\delta(s-s')
\]
for positive labels. Rotation of the $--$ bilinear gives $-i$, and conjugation gives $+i$ for $++$. Endpoint limits are taken distributionally after spectral smearing. 

Define the distribution appearing in \cite{GraefeSachs2026}, (4.16), for positive $t$,
\begin{align}
D_a(s,t)&=\frac{1}{2\sinh\pi t}
\left[\frac{e^{\pi t}}{s^2-t^2+ia0}-\frac{e^{-\pi t}}{s^2-t^2-ia0}\right]\cr
&=\PV\frac1{s^2-t^2}-ia\pi\coth(\pi t)\delta(s^2-t^2).
\end{align}
The correctly normalized representation is
\begin{equation}\label{eq:correctG}
G^t_{aa}(\eta_1,\eta_2)=\int_{\mathbb R}\dd s\,N_s u_s^a(\eta_1)u_s^a(\eta_2)D_a(s,t).
\end{equation}
\cite{GraefeSachs2026}, (4.15), instead has an additional prefactor $ia$.
To check the normalization independently of notation, let us use the Fourier-space Klein--Gordon operator for mass $m_t^2=\db^2/4+t^2$,
\begin{equation}
 L_t=\eta_1^2\partial_{\eta_1}^2 -(\db-1)\eta_1\partial_{\eta_1}+k^2\eta_1^2+\frac{\db^2}{4}+t^2.
\end{equation}
From the Bessel equation, we get $L_tu_s^a=(t^2-s^2)u_s^a$. The completeness identity corresponding to \eqref{eq:overlap} is
\begin{equation}
 \int_{\mathbb R}\dd s\,N_su_s^a(\eta_1)u_s^a(\eta_2)=ia(-\eta_1)^{\db+1}\delta(\eta_1-\eta_2). \label{eq:GS-mode-completeness}
\end{equation}
Since $(t^2-s^2)D_a(s,t)=-1$, applying $L_t$ to \eqref{eq:correctG} and using \eqref{eq:GS-mode-completeness} gives
\[
 L_tG^t_{aa}=-ia(-\eta_1)^{\db+1}\delta(\eta_1-\eta_2),
\]
the source fixed by the Wronskian jump of the chronological or antichronological mode product. An additional factor $ia$ would give a real source instead. This establishes the required phase in the stated mode and contour conventions; the boundary prescription remains the one specified in $D_a$.

}


\begin{thebibliography}{99}

\bibitem{FOSW2025}
H.~Furugori, N.~Ogawa, S.~Sugishita and T.~Waki,
``Celestial holography meets dS/CFT,''
\href{https://arxiv.org/abs/2507.17558}
{arXiv:2507.17558 [hep-th]}.

\bibitem{BCM}
B.~A. Bassett, S.~L. Cacciatori and U.~Moschella,
``Finite spectral recurrences for loop integrals in de Sitter and
anti-de Sitter space,''
\href{https://arxiv.org/abs/2610.28183}
{arXiv:2610.28183 [hep-th]}.

\bibitem{SW}
R.~F. Streater and A.~S. Wightman,
\emph{PCT, Spin and Statistics, and All That},
Princeton Landmarks in Mathematics and Physics,
Princeton University Press, Princeton, 2000,
ISBN: 978-0-691-07062-9.

\bibitem{eg}
H.~Epstein and V.~Glaser,
``The role of locality in perturbation theory,''
\emph{Ann. Inst. H. Poincar\'e Sect. A}
\textbf{19} (1973), no.~3, 211--295,
\href{https://www.numdam.org/item/AIHPA_1973__19_3_211_0/}
{Numdam: AIHPA\_1973\_\_19\_3\_211\_0}.

\bibitem{bm}
J.~Bros and U.~Moschella,
``Two-point functions and quantum fields in de Sitter universe,''
\emph{Rev. Math. Phys.} \textbf{8} (1996) 327--392,
\href{https://doi.org/10.1142/S0129055X96000123}
{doi:10.1142/S0129055X96000123},
\href{https://arxiv.org/abs/gr-qc/9511019}
{arXiv:gr-qc/9511019}.

\bibitem{BD}
T.~S. Bunch and P.~C.~W. Davies,
``Quantum field theory in de Sitter space:
renormalization by point-splitting,''
\emph{Proc. Roy. Soc. Lond. A} \textbf{360} (1978) 117--134,
\href{https://doi.org/10.1098/rspa.1978.0060}
{doi:10.1098/rspa.1978.0060}.

\bibitem{GH}
G.~W. Gibbons and S.~W. Hawking,
``Cosmological event horizons, thermodynamics,
and particle creation,''
\emph{Phys. Rev. D} \textbf{15} (1977) 2738--2751,
\href{https://doi.org/10.1103/PhysRevD.15.2738}
{doi:10.1103/PhysRevD.15.2738}.

\bibitem{bgm}
J.~Bros, J.-P.~Gazeau and U.~Moschella,
``Quantum field theory in the de Sitter universe,''
\emph{Phys. Rev. Lett.} \textbf{73} (1994) 1746--1749,
\href{https://doi.org/10.1103/PhysRevLett.73.1746}
{doi:10.1103/PhysRevLett.73.1746}.

\bibitem{GGV5}
I.~M. Gel'fand, M.~I. Graev and N.~Ya. Vilenkin,
\emph{Generalized Functions}, Vol.~5:
\emph{Integral Geometry and Representation Theory},
translated by E.~Saletan,
Academic Press, New York and London, 1966.

\bibitem{SleightTaronna2021}
C.~Sleight and M.~Taronna,
``From dS to AdS and back,''
\emph{JHEP} \textbf{12} (2021) 074,
\href{https://doi.org/10.1007/JHEP12(2021)074}
{doi:10.1007/JHEP12(2021)074},
\href{https://arxiv.org/abs/2109.02725}
{arXiv:2109.02725 [hep-th]}.

\bibitem{KatsevichSun2026}
A.~Katsevich and Z.~Sun,
``Cosmological correlators from the Euclidean sphere,''
\href{https://arxiv.org/abs/2610.28516}
{arXiv:2610.28516 [hep-th]}.

\bibitem{LochabDenef2026}
V.~Lochab and F.~Denef,
``Higher-loop vacuum integrals on $S^D$,''
\href{https://arxiv.org/abs/2610.30489}
{arXiv:2610.30489 [hep-th]}.

\bibitem{HMM2011}
A.~Higuchi, D.~Marolf and I.~A. Morrison,
``On the equivalence between Euclidean and in-in formalisms in de Sitter
QFT,''
\emph{Phys. Rev. D} \textbf{83} (2011) 084029,
\href{https://doi.org/10.1103/PhysRevD.83.084029}
{doi:10.1103/PhysRevD.83.084029},
\href{https://arxiv.org/abs/1012.3415}
{arXiv:1012.3415 [gr-qc]}.

\bibitem{BEGMP2010}
J.~Bros, H.~Epstein, M.~Gaudin, U.~Moschella and V.~Pasquier,
``Triangular invariants, three-point functions and particle stability
on the de Sitter universe,''
\emph{Commun. Math. Phys.} \textbf{295} (2010) 261--288,
\href{https://doi.org/10.1007/s00220-009-0875-4}
{doi:10.1007/s00220-009-0875-4},
\href{https://arxiv.org/abs/0901.4223}
{arXiv:0901.4223 [hep-th]}.

\bibitem{NowinskiSachs2025}
M.~Nowinski and I.~Sachs,
``Resummation of cosmological correlators and their
UV regularization,''
\emph{JHEP} \textbf{02} (2026) 228,
\href{https://doi.org/10.1007/JHEP02(2026)228}
{doi:10.1007/JHEP02(2026)228},
\href{https://arxiv.org/abs/2507.21224}
{arXiv:2507.21224 [hep-th]}.

\bibitem{Zhang2026}
H.~Zhang,
``Dimensional regularization of bubble diagrams in de Sitter
spacetime,''
\emph{JHEP} \textbf{02} (2026) 119,
\href{https://doi.org/10.1007/JHEP02(2026)119}
{doi:10.1007/JHEP02(2026)119},
\href{https://arxiv.org/abs/2507.19318}
{arXiv:2507.19318 [hep-th]}.

\bibitem{Benincasa:2024ptf} 
P.~Benincasa, G.~Brunello, M.~K.~Mandal, P.~Mastrolia and F.~Vaz\~ao, 
``One-loop corrections to the Bunch--Davies wave function of the universe,'' 
\href{https://doi.org/10.1103/PhysRevD.111.085016} {\emph{Phys. Rev. D} \textbf{111} (2025) 085016}.

\bibitem{DiPietroGorbenkoKomatsu2022}
L.~Di Pietro, V.~Gorbenko and S.~Komatsu,
``Analyticity and unitarity for cosmological correlators,''
\emph{JHEP} \textbf{03} (2022) 023,
\href{https://doi.org/10.1007/JHEP03(2022)023}
{doi:10.1007/JHEP03(2022)023},
\href{https://arxiv.org/abs/2108.01695}
{arXiv:2108.01695 [hep-th]}.

\bibitem{XianyuZhang2023}
Z.-Z.~Xianyu and H.~Zhang,
``Bootstrapping one-loop inflation correlators with the spectral
decomposition,''
\emph{JHEP} \textbf{04} (2023) 103,
\href{https://doi.org/10.1007/JHEP04(2023)103}
{doi:10.1007/JHEP04(2023)103},
\href{https://arxiv.org/abs/2211.03810}
{arXiv:2211.03810 [hep-th]}.

\bibitem{GraefeSachs2026}
J.~Gr\"afe and I.~Sachs,
``Split representations and bubble resummation for massive de Sitter
correlators,''
\emph{JHEP} \textbf{07} (2026) 074,
\href{https://doi.org/10.1007/JHEP07(2026)074}
{doi:10.1007/JHEP07(2026)074},
\href{https://arxiv.org/abs/2602.09977}
{arXiv:2602.09977 [hep-th]}.

\bibitem{Bateman1}
A.~Erd\'elyi, W.~Magnus, F.~Oberhettinger and F.~G. Tricomi,
\emph{Higher Transcendental Functions}, Vol.~I,
Bateman Manuscript Project,
McGraw--Hill, New York, 1953, Chapter~III.

\bibitem{MoschellaAdS}
U.~Moschella,
``Two point functions and quantum fields in the anti-de Sitter
universe,''
\emph{JHEP} \textbf{05} (2026) 262,
\href{https://doi.org/10.1007/JHEP05(2026)262}
{doi:10.1007/JHEP05(2026)262},
\href{https://arxiv.org/abs/2512.18672}
{arXiv:2512.18672 [hep-th]}.

\bibitem{bem}
J.~Bros, H.~Epstein and U.~Moschella,
``Towards a general theory of quantized fields on the anti-de Sitter
space-time,''
\emph{Commun. Math. Phys.} \textbf{231} (2002) 481--528,
\href{https://doi.org/10.1007/s00220-002-0726-z}
{doi:10.1007/s00220-002-0726-z},
\href{https://arxiv.org/abs/hep-th/0111255}
{arXiv:hep-th/0111255}.

\bibitem{MST} U.~Moschella, C.~Sleight and M.~Taronna, work in preparation.

\bibitem{BEM2010}
J.~Bros, H.~Epstein and U.~Moschella,
``Particle decays and stability on the de Sitter universe,''
\emph{Ann. Henri Poincar\'e} \textbf{11} (2010) 611--658,
\href{https://doi.org/10.1007/s00023-010-0042-7}
{doi:10.1007/s00023-010-0042-7},
\href{https://arxiv.org/abs/0812.3513}
{arXiv:0812.3513 [hep-th]}.

\bibitem{BMS2013} 
A.~Bzowski, P.~McFadden and K.~Skenderis, 
``Implications of conformal invariance in momentum space,'' 
JHEP \textbf{03} (2014) 111,
\href{https://doi.org/10.1007/JHEP03(2014)111}{doi:10.1007/JHEP03(2014)111}.

\bibitem{DLMF}
NIST Digital Library of Mathematical Functions,
F.~W.~J. Olver et al., eds.,
\href{https://dlmf.nist.gov/}
{\texttt{https://dlmf.nist.gov/}}.

\bibitem{Bateman1954} 
A.~Erd\'elyi, W.~Magnus, F.~Oberhettinger and F.~G.~Tricomi, 
\textit{Tables of Integral Transforms}, 
Vol.~II, McGraw--Hill Book Company, New York, 1954.

\bibitem{YakubovichLuchko} 
S.~B.~Yakubovich and Yu.~F.~Luchko, 
\textit{The Hypergeometric Approach to Integral Transforms and Convolutions: Kernels and Applications}, 
Mathematics and Its Applications, Vol.~287, Kluwer Academic Publishers, Dordrecht, Boston, London, 1994.

\end{thebibliography}
\end{document}